\documentclass[useAMS,usenatbib]{mn2e}
\usepackage{graphicx,amssymb,color,stmaryrd}
\usepackage[normalem]{ulem}

\newcommand{\rev}{ }
\newcommand{\revrev}{ }

\title[Relations and criteria for transits]
{Explicit relations and criteria for eclipses, transits and occultations}

\author[Veras]{
Dimitri Veras$^{1,2}$\thanks{E-mail: d.veras@warwick.ac.uk}\thanks{STFC Ernest Rutherford Fellow}
\\
$^{1}$Centre for Exoplanets and Habitability, University of Warwick, Coventry CV4 7AL, UK
\\
$^{2}$Department of Physics, University of Warwick, Coventry CV4 7AL, UK
}

\begin{document}
\label{firstpage}
\pagerange{\pageref{firstpage}--\pageref{lastpage}}
\maketitle

\begin{abstract}
Solar system, exoplanet and stellar science rely on transits, eclipses and occultations for dynamical and physical insight. Often, the geometry of these configurations are modelled by assuming a particular viewpoint. Here, instead, I derive user-friendly formulae from first principles independent of viewpoint and in three dimensions. I generalise the results of \cite{verbre2017} by (i) characterising three-body systems which are in transit but {\revrev are} not necessarily perfectly aligned, and by (ii) incorporating motion. For a given snapshot in time, I derive explicit criteria to determine whether a system is in or out of transit, if an eclipse is total or annular, and expressions for the size of the shadow, including their extreme values and a condition for engulfment. {\rev These results are exact}. For orbital motion, {\rev I instead obtain approximate results.} By assuming fixed orbits, I derive a single implicit algebraic relation which can be solved to obtain the frequency and duration of transit events -- including ingresses and egresses -- for combinations of moons, planets and stars on arbitrarily inclined circular orbits; the eccentric case requires the solution of Kepler's equation but remains algebraic. I prove that a transit shadow -- whether umbral, antumbral or penumbral -- takes the shape of a parabolic cylinder, and finally present geometric constraints on Earth-based observers hoping to detect a three-body syzygy {\revrev (or perfect alignment)} -- {\rev either in extrasolar systems or within the solar system} -- potentially as a double annular eclipse.
\end{abstract}


\section{Introduction}

Rovers on Mars, humans on Earth, and artificial satellites in space all have
the capacity to view
transits, eclipses and occultations of moons, planets and/or stars.
In a similar vein, extrasolar observers may detect the Earth and other solar system
planets through photometric events {\rev \citep{brarag2016,helpud2016,weletal2018}}. These 
varied situations all feature the same basic geometry, independent of viewpoint: 
the intersection of a radiation cone with spheres \citep{cayley1870,rigge1924}. 
Consequently, a thorough exploration of this geometry might reveal widely-applicable results.

Previous investigations of this architecture have targeted specific groups 
of observers. For Earth-bound viewers of solar system-based transit phenomena,
the annual Astronomical Almanac \citep{alm2018} provides 
detailed numerical data. \cite{ismetal2015}
supplied an analytical alternative incorporating
Solar radiation pressure and Earth's oblateness, {\rev and \cite{kawetal2018} highlighted 
the importance of analysing Earth's transmission spectra during a lunar eclipse with both umbral
and penumbral data.} Ground-based observatories
on Earth have also utilised stellar occultations to constrain minor planet shape
\citep{ortetal2017} and ring particle size {\rev \citep[e.g.][]{coletal2018,menetal2018}},
but have difficulty distinguishing amongst grazing, partial and total eclipses in binary
star systems \citep{morris1999}. For the external observers of solar system planets,
\cite{weletal2018} instead derived non-grazing transit 
visibility zones, whereas \cite{helpud2016} 
focused on the Earth's transit visibility zone. 

In the field of extrasolar planets, \cite{cabsch2007} considered the cone-sphere 
intersection in the context of flux changes and probabilities from imaging
and reflected light. Exoplanet-based photometric transits include an extensive
literature, with analytic treatments provided by {\rev
\cite{schche1990},  \cite{seamal2003}, \cite{tinsac2005}, \cite{kipping2008}, \cite{kipping2010}, \cite{stegau2013}, 
\cite{winn2014}, \cite{martri2015}, \cite{lugetal2017}, \cite{martin2017a} and \cite{reaetal2017}.}  Related are transit
timing variations \citep[e.g.][]{foretal2011,agofab2017}, for which analytic work 
\citep{agoetal2005,nesmor2008,nesvorny2009,litetal2012,decago2015,agodec2016,hadlit2016} 
has also yielded fruitful results. {\rev Although exoplanetary literature rarely distinguishes 
umbral, penumbral, and antumbral cases from one another, recently \cite{beretal2018} 
have analysed these different cases
in flux and polarization phase curves of exoplanets with orbiting exomoons.}

\begin{figure*}
\includegraphics[width=15cm]{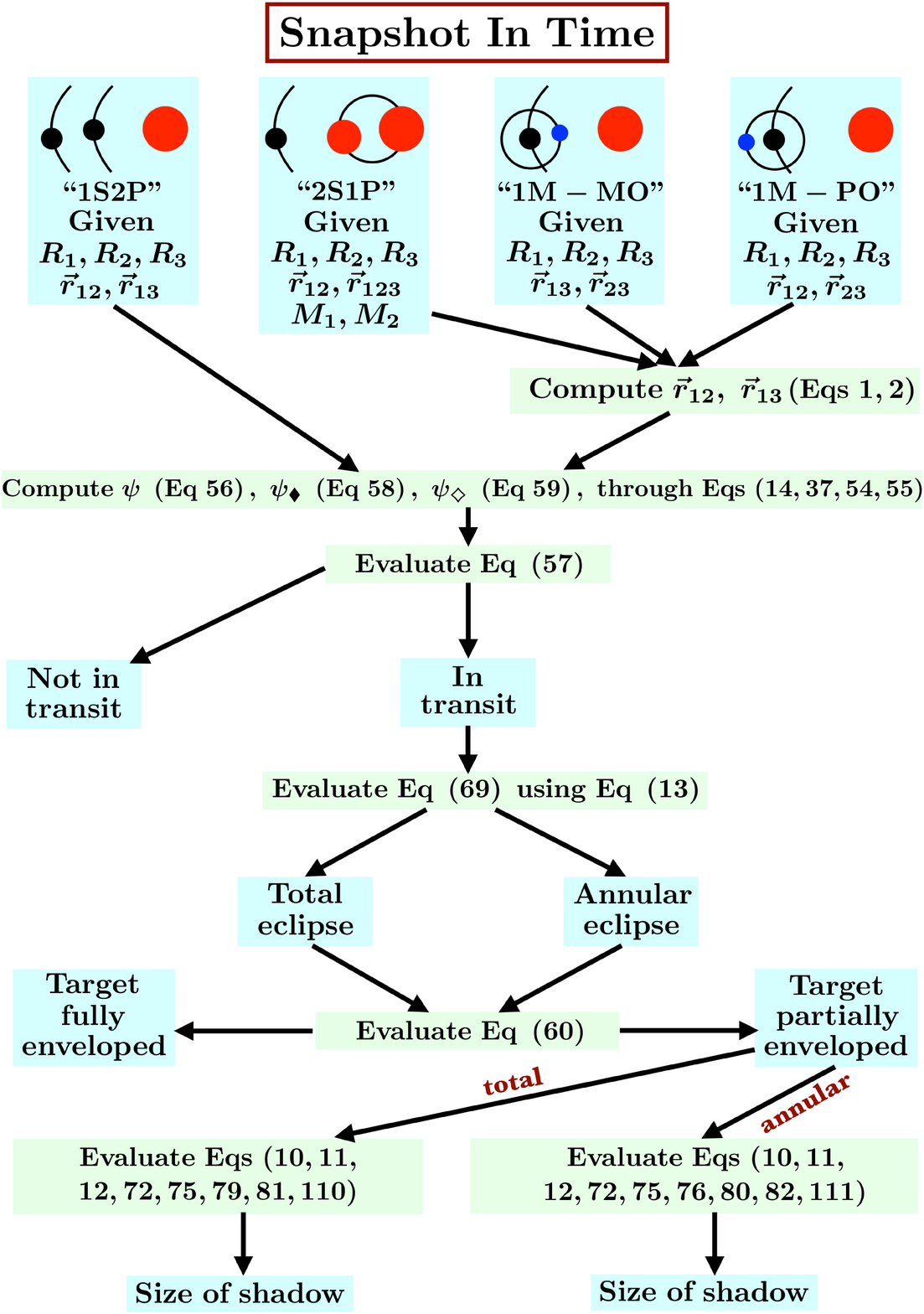}
\caption{
{\rev Procedure for obtaining results at snapshots in time}. The four configurations shown
are for two planets orbiting
one star (1S2P), one planet orbiting two stars (1P2S), and one moon, one planet
and one star with either the moon as the occulter (1M-MO) or the planet as
the occulter (1M-PO). This flowchart demonstrates how one can
determine with explicit expressions if three spherical bodies are in or out of
transit, whether the eclipse is total or annular, whether the target is fully or
partially engulfed in the shadow, and the size of the shadow. 
}
\label{bubbles1}
\end{figure*}

\begin{figure*}
\includegraphics[width=15cm]{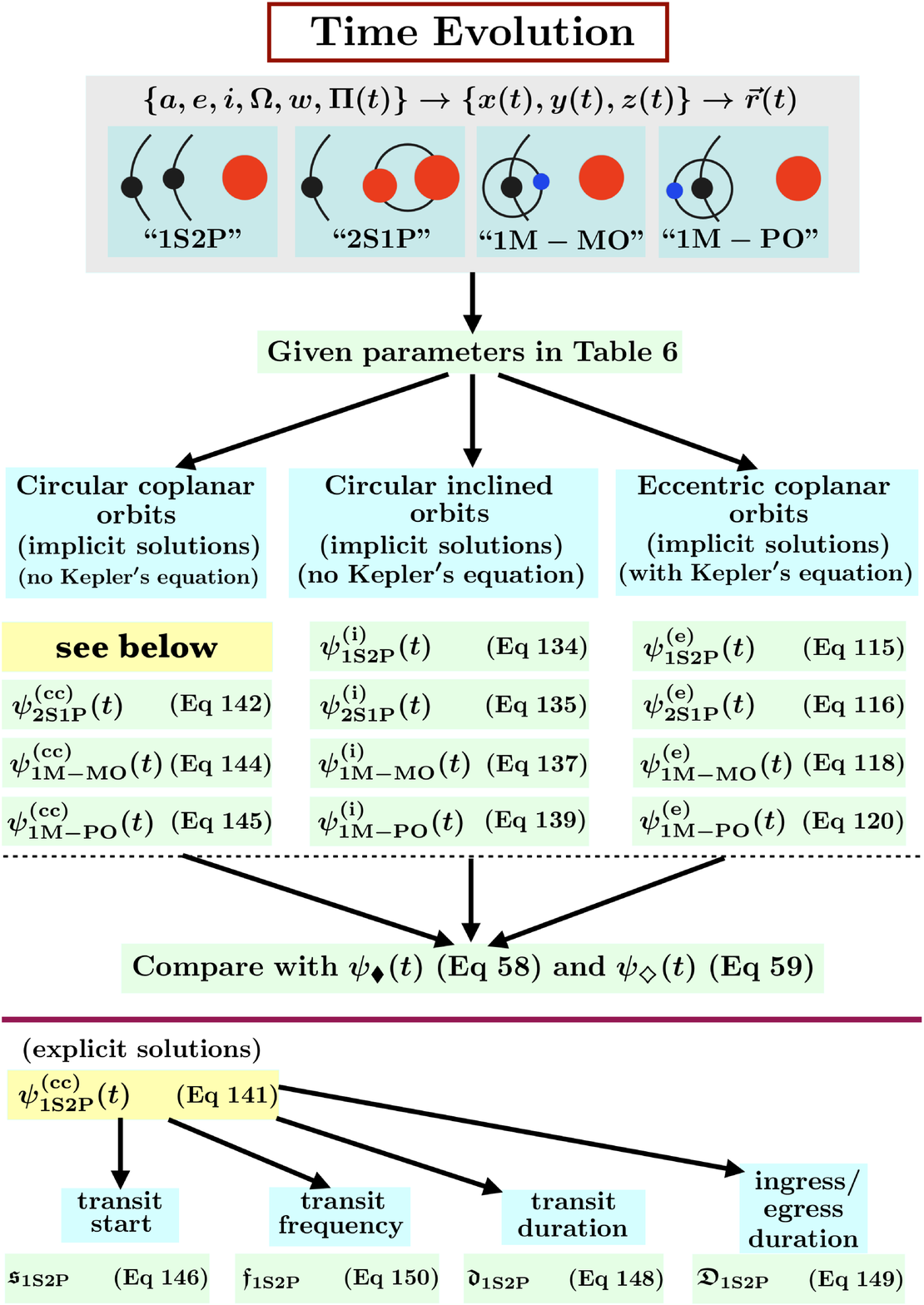}
\caption{
{\rev Procedure for obtaining results for motion along static orbits.} 
The four configurations shown are the same as in Fig. \ref{bubbles1}.
This flowchart illustrates how to obtain the frequency, duration
and start times of transits and ingresses/egresses given sets
of orbital elements.
}
\label{bubbles2}
\end{figure*}

\begin{figure*}
\includegraphics[width=15cm]{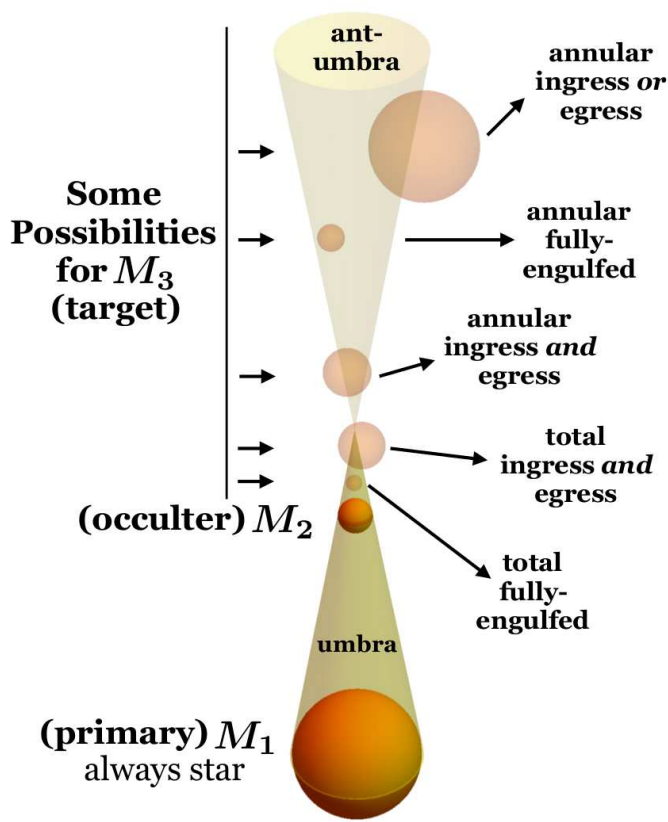}
\caption{
Potential configurations of three bodies in transit. The primary, which is always a star and the 
largest body, always forms a radiation cone with the occulter. If the target intersects this cone, then
the target is said to be ``in transit''. If the near side of the target intersects the bottom nappe,
the eclipse is total; otherwise, it is annular; {\rev the near side of the Earth, for example, lies coincidentally just
at the vertex of the cone, which is why there are both annular and total solar eclipses.} The entire target may or may not be completely
engulfed in shadow. When not engulfed, {\rev as is the case for solar eclipses as seen on Earth}, 
the target may be in ingress, egress or both, {\rev and the observability of the eclipse then depends on one's location
on the target surface.} The extension
to the penumbral case -- where a radiation cone is formed by internal tangent rays instead of external
tangent rays -- is covered in Section 9.
}
\label{difftarget}
\end{figure*}

\begin{figure}
\includegraphics[width=8cm]{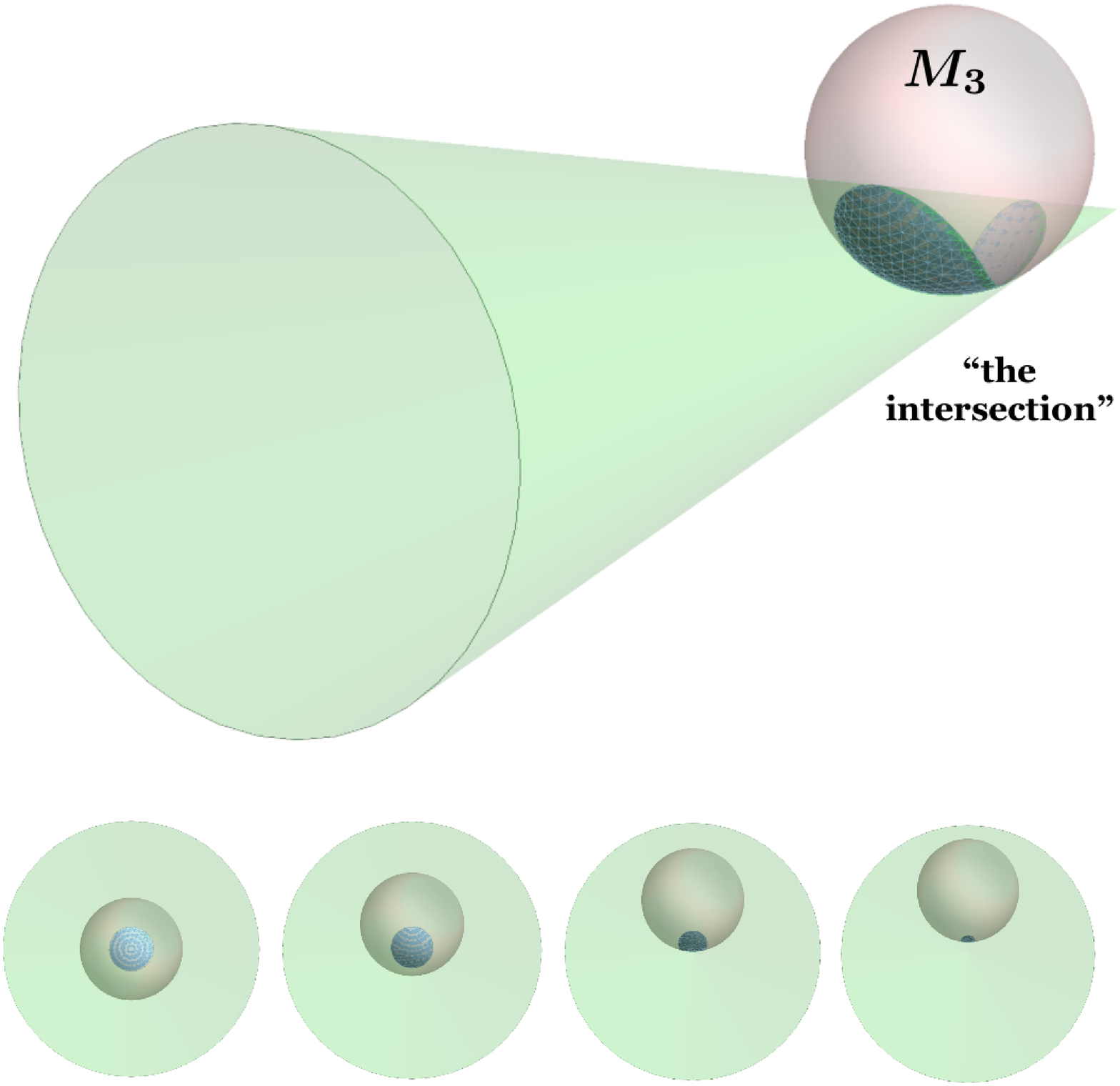}
\caption{
Visual representations of the intersection between the target and the radiation
cone. The intersection is the meshed surface, and the shadow is the dark
portion of that surface, which I prove is a parabolic cylinder in Section 4.
The bottom four cartoons are from the same point of
view, along the radiation cone axis, as the target gradually moves away from
syzygy (left to right). {\revrev These frames all demonstrate the case when the shadow does
not engulf the target, much like in a solar eclipse; an observer would see the eclipse
only by standing in the shadow. The true variation in the shadow size, however, can
be obscured by adopting the fixed perspective here} (see Section 7).
}
\label{intersec}
\end{figure}

Here I take a step back and make no assumptions about the observer. 
I generalise the results of \cite{verbre2017}
(hereafter Paper I), who characterised the geometry of syzygy -- a situation
when at least three bodies are co-linear. Now I consider non-syzygetic transits,
and henceforth for simplicity largely use the word ``transits'' to refer
also to eclipses and occultations\footnote{The Glossary of the Astronomical
Almanac \citep{alm2018} defines an {\it eclipse} as a ``the occultation of a celestial body caused by its passage through the shadow cast by another body'', 
a {\it transit}, in part, as ``the passage of one celestial body in front of another of greater apparent diameter'', an {\it occultation} as, in
part, ``the obscuration of one celestial body by another of greater apparent diameter'',
and a {\it syzygy}, as, in part, ``a configuration where three or more celestial bodies are positioned approximately
in a straight line in space''.}. 

In particular, my exploration seeks to establish what formulae 
can be derived from knowledge of only the radii and spatial locations of
the three bodies involved in the transit.
My assumptions are limited to (i) bodies being perfect spheres, and
(ii) that light does not bend. I consider the {\rev darkest central shadows -- the umbral and antumbral 
shadows -- first}, before moving onto {\rev the lighter peripheral} penumbral shadows. Further, 
I consider motion. By assuming fixed orbits, I show how to compute frequencies
and durations without having to solve differential equations.

{\rev

\subsection{Motivation for paper}

The motivation for this paper arises from the benefits of
a fully analytic treatment of eclipses, transits and occultations. The first
benefit is speed and convenience: although running multi-body numerical simulations 
can occasionally be faster than evaluations of analytic
treatments which involve long series expansions or differential equations, no series 
expansions nor differential equations are given in this paper. Further, ready-to-use 
algebraic equations do not require the setup time and output processing
involved when running multi-body simulations, particularly when objects are modelled as
solid bodies and ray tracing would be required to compute, for example, the condition
to be in or out of transit. The second benefit is a mathematical understanding of the results.
For example, the maximum shadow size on a solid body does not occur at a necessarily
obvious location. Also, a simple mathematical swap (shown later in the paper) allows
one to derive results for partial eclipses just as easily as for total or annular eclipses.
Without this knowledge, unnecessary effort might be expended in generating or
modifying a numerical module to process partial eclipses.

}

\subsection{Plan for paper}

Section 2 describes my basic setup. Sections 3-7 then explore what can be
deduced from a snapshot in time, whereas Section 8 covers time evolution.
Extensions are then presented in Sections 9-10 before the summary
in Section 11.

In particular, for snapshots in time, in Section 3, I derive the equation
of the intersection of a cone and a sphere. This equation, in combination
with the properties of quadrics {\rev (or quadratic curves)}, is then used to deduce 
the shape of the shadow (Section 4) and the condition to be 
in or out of transit (Section 5). Section 6 establishes a criteria to determine
whether the eclipse is annular or total, a direct extension of a similar
result from Paper I. Section 7 then extends the geometry of Paper I 
in order to determine the size of this shadow in all cases.

I introduce time evolution in Section 8 and in {\rev both Appendices A and B}, and consider transit
durations and frequencies in three specific cases: (i) arbitrarily eccentric,
coplanar orbits (Section 8.1), (ii) circular, arbitrarily inclined orbits
(Section 8.2) and (iii) circular coplanar orbits (Section 8.3).
For each case I provide three subcases:
when the transit includes (i) one star and two planets, (ii) two stars and
one planet, and (iii) one moon, one planet and one star. 

Section 9 discusses how all of these results can be applied to penumbral shadows
with relative ease. Section 10 then extends my off-syzygy results to four bodies in 
special cases where I do fix an observer.

Figures \ref{bubbles1} and \ref{bubbles2} provide a preview of the goals of the paper up through
Section 8, and summarises how to reach them. These figures may be used as a
convenient reference
and algorithm (allowing the reader to skip the details) for computing
eclipse-, transit- and occultation-related
quantities in planetary and stellar systems. The figures
also illustrate four different physical situations involving stars, planets
and moons for which my results are applicable.

All notation and geometry is consistent with that of Paper I, and I have taken
considerable care to avoid variable conflicts. Every variable in both papers
is identified for easy access in Tables \ref{TabRoman}-\ref{TabHebrew}.

\section{Physical and geometrical setup}

Throughout the paper I model three-body systems. Only
in Section 10 do I add a fourth body, and in a limited
capacity.

\subsection{The three bodies: radii and distances}

I consider all three bodies to be spheres, 
at least one of which is a star,
denoted the ``primary'' and with a radius $R_1$.
The other two bodies (the ``occulter'', and the ``target'')
are denoted respectively $R_2$ and $R_3$, with $R_1 > R_2$ and
$R_1 > R_3$. However, there are no constraints on the relative
sizes of the occulter and target.

The centre of the primary is taken to reside at the origin of an arbitrarily oriented 
orthogonal Cartesian $\left\lbrace xyz\right\rbrace$ coordinate system.
The radiation from the primary always forms an umbral and antumbral cone with
the occulter, as in Fig. \ref{difftarget}. Note that this cone has two
nappes ({\rev a nappe is one of the two sections of a double cone}) and can move and 
change size during a transit. The target may be anywhere in space.

Only when the target intersects either
or both nappes of the 
cone is a shadow formed. I refer to the side of the target facing 
the primary, where a shadow will be, as the ``near side''. The near side
of the target could represent slightly more or slightly less than half of its
total surface area (see e.g. equation A20 of Paper I).
Whether the shadow results in a total or annular eclipse depends on 
which nappe contains the intersection of the target's near side (see Section 6).

At a given moment in time, I consider the target to be either fully engulfed inside
of the cone, completely outside of the cone, or intersecting 
the cone. In this last case, if the surface of intersection occurs when the target
is entering/leaving the cone, I denote those cases as ingress/egress. The target
may be in both ingress and egress. Figure \ref{difftarget} illustrates many of 
the above possibilities. 

Common examples can be visualised with the figure. For example, a total
solar eclipse occurs when the Sun is the primary, the Moon is the occulter, the
Earth is the target, and the Earth intersects the umbral cone. In this case, the Earth
is usually is in both ingress and egress because of its relatively large size compared to 
the moon. Alternatively, an annular solar eclipse occurs when the Earth instead 
intersects the antumbral cone. For a lunar eclipse, the Earth becomes the occulter
and the moon becomes the target. 

Further, multiple planetary or stellar systems may 
be included: for example, an observer
on Earth (the target) could see a distant star (the primary) being occulted by a
Kuiper belt object (the occulter).
{\revrev For observing extrasolar planetary systems, one possibility is that the Earth is the target, the exoplanet
is the occulter, and the exoplanet host star is the primary. Another possibility -- when visualising what an extrasolar
observer would see within, for example, a circumbinary system -- is when the planet is the target,
the smaller star is the occulter, and the larger star is the primary}.
For more examples, see Paper I.

The distance vector $\vec{r}$ always begins at the origin. 
Specific time-dependent distances are 
the distance between the centres of the primary and occulter 
$\vec{r}_{12}(t) = \left\lbrace x_{12}(t), y_{12}(t), z_{12}(t) \right\rbrace$,
the primary and target $\vec{r}_{13}(t) = \left\lbrace x_{13}(t), y_{13}(t), z_{13}(t) \right\rbrace$
and the occulter and target 

\begin{equation}
\vec{r}_{23}(t) = \vec{r}_{13}(t) - \vec{r}_{12}(t)
.
\label{r23}
\end{equation}

Henceforth, I will drop the denotation of the time dependence of these variables.
As indicated in Fig. \ref{bubbles1}, for systems with two planets or stars, primarily $\vec{r}_{12}$ and  $\vec{r}_{13}$ are used,
whereas for systems with one moon, instead $\vec{r}_{13}$ and $\vec{r}_{23}$ are usually used.

Although the primary moves in time due to barycentric interactions, the origin of
my coordinate system is always fixed on the primary's centre. Because of this
barycentric motion, researchers often will define initial conditions in barycentric
or Jacobi coordinates, particularly in circumbinary planetary systems.
I denote the distance vector between the target and the
barycentre of the primary and occulter as $\vec{r}_{123}$ with

\begin{equation}
\vec{r}_{13} = \vec{r}_{123} + \left(\frac{M_1}{M_1 + M_2}\right)\vec{r}_{12}
.
\label{r123}
\end{equation}

\noindent{}In this expression, the masses of the primary and occulter are denoted as $M_1$ and $M_2$.  

{\rev Finally, depending on the particular setup, one may reasonably adopt certain approximations. One example is when the target is somewhere in the Solar System and the primary is another stellar system. From Earth, one may observe either an exoplanetary transit or an occultation of a distant star across Saturn's rings. In the former case, one might assume $r_{12} \gg r_{13}, r_{23}$, whereas in the latter case, one might assume $r_{12}, r_{13} \gg r_{23}$. Regardless, I emphasise that in order to keep the analytical treatment as general as possible, no such approximations are made throughout the paper.}

\subsection{Orbital elements}

In Paper I, only the radii and mutual distances
of the three bodies were needed to be given in order to obtain their results (see their figure 5). The same is also true here for snapshots in time. 

However, when I introduce motion, the number of 
initial parameters increases greatly: In that case, I also assume,
in addition to all of the radii, a given set of the following
orbital elements: $\left\lbrace a,e,i,\Omega,w,\Pi \right\rbrace$ for any two of the orbits.
These elements are, respectively, semimajor axis ($a$), eccentricity ($e$), inclination ($i$), 
longitude of ascending node ($\Omega$), argument of pericentre ($w$), and true anomaly ($\Pi$).
This set uniquely defines $r$ through:

\begin{equation}
r = \frac{a\left(1 - e^2\right)}
    {1 + e \cos{\Pi}}
\label{eqr}
\end{equation}

\noindent{}which is independent of $i, \Omega$, and $w$, and assumes
that the orbit is an ellipse. However,
the Cartesian components of $r$ (which are $x$, $y$, and $z$) are individually
dependent on $i, \Omega$ and $w$. Without loss of generality,
I assume that the $x$-axis lies along the major axis of the ellipse, with the orbital pericentre
being in the positive direction. Then

\begin{equation}
x = r \left[\cos{\Omega} \cos{\left(w + \Pi\right)} - \sin{\Omega} \sin{\left(w + \Pi\right) \cos{i}} \right]
,
\label{xr}
\end{equation}

\begin{equation}
y = r \left[\sin{\Omega} \cos{\left(w + \Pi\right)} + \cos{\Omega} \sin{\left(w + \Pi\right) \cos{i}} \right]
,
\label{yr}
\end{equation}

\begin{equation}
z = r \left[\sin{\left(w + \Pi\right)} \sin{i} \right]
.
\label{zr}
\end{equation}

Because the systems in this paper all contain more than two bodies, mutual
gravitational perturbations ensure that all of the orbital elements are time-dependent. Hence, the assumption of fixed orbits in this paper
becomes weaker over longer timescales. {\rev I quantify the breakdown of this assumption for a few representative cases in Appendix B.}

I will revisit motion {\rev along static orbits} in Section 8. However, for now, I treat the snapshot case.

\section{Expression for the Shadow}

\begin{table*}
 \centering
 \begin{minipage}{180mm}
  \centering
  \caption{Unstylised lowercase Roman variables used in this paper and in Paper I.}
  \label{TabRoman}
  \begin{tabular}{@{}lll@{}}
  \hline
   Variable & Explanation & Reference \\
   \hline
$a$ & Semimajor axis & \\[2pt]
$c$ & Auxiliary variable & Eq. A27 of Paper I \\[2pt]   
$d$ & Height of penumbral cone & Eq. B4 of Paper I \\[2pt]    
$e$ & Orbital eccentricity & \\[2pt]
$g$ & Fractional area of primary blocked out by occulter as seen by observer on target at syzygy & Eq. A30 of Paper I \\[2pt] 
$h$ & Height of umbral cone & equation (\ref{hexp}) \\[2pt]
$i$ & Orbital inclination & \\[2pt]
$j$ & Auxiliary variable & Eq. A28 of Paper I \\[2pt]    
$k$ & Perpendicular distance with which fourth body is offset from a syzygy & Section 10 only \\[2pt]
$k_{\forall}$ & Critical value of $k$ for which observer on fourth body cannot see a transit & equation (\ref{kall0})  \\[2pt]
$k_{\sqcup}$ & Critical value of $k$ for which target is entirely in the field of view of observer on fourth body & equation (\ref{ksqcup})  \\[2pt]
$k_{\parallel}$ & Critical value of $k$ for which occulter is entirely in the field of view of observer on fourth body & equation (\ref{kparallel})  \\[2pt]
$k_{\bullet}$ & Critical value of $k$ to produce a double annular eclipse for observer on fourth body & equations (\ref{l14eq}-\ref{l34low})  \\[2pt]  
$l$ & Auxiliary variable & equations (\ref{l14eq}-\ref{l34final}) \\[2pt]
$n$ & Distance between base of umbral cone and centre of primary & equation (\ref{hplusn}) \\[2pt]
$p$ & Transit number & equations (\ref{sta1S2P}-\ref{end1S2P})  \\[2pt]   
$\vec{r}_{12}$ & Distance vector between centres of primary and occulter & equations (\ref{r23}-\ref{r123})  \\[2pt]
$\vec{r}_{13}$ & Distance vector between centres of primary and target & equations (\ref{r23}-\ref{r123})  \\[2pt]
$\vec{r}_{23}$ & Distance vector between centres of occulter and target & equation (\ref{r23})  \\[2pt]
$\vec{r}_{123}$ & Distance vector between centre of target and barycentre of primary and occulter & equation (\ref{r123}) \\[2pt]
$\vec{r}_{14}$ & Distance vector between centres of primary and external body when &  \\[2pt]
 & the primary, occulter and target are in syzygy & Section 10 only \\[2pt]
$r_{23}^{\dagger}$ & Critical value of $r_{23}$ within which target is engulfed in a total eclipse at syzygy & Eq. A10 of Paper I \\[2pt]
$r_{23}^{\ddagger}$ & Critical value of $r_{23}$ beyond which target is engulfed in an annular eclipse at syzygy & Eq. A11 of Paper I \\[2pt]
$r_{23}^{\ominus}$ & Value of $r_{23}$ such that the umbral shadow radius is equal to $R_3$ at syzygy & Eq. A21 of Paper I \\[2pt]
$r_{23}^{\ast}$ & Critical value of $r_{23}$ beyond which target is engulfed in penumbral shadow at syzygy  & Eq. B8 of Paper I \\[2pt]
$r_{23}^{\uplus}$ & Critical value of $r_{23}$ beyond which target blocks primary's starlight  &  \\[2pt]
 & for observer on fourth body who is colinear with a syzygy  & equation (\ref{ruplus}) \\[2pt]   
$r_{23}^{\bullet}$ & Critical value of $r_{23}$ to produce a double annular eclipse for observer  & \\[2pt]
 & on fourth body who is offset with a syzygy  & equation (\ref{pythag}) \\[2pt]
$r_{\rm loc}$ & Distance between centre of primary and point on target surface & equation (\ref{rloc}) \\[2pt]
$t$ & Time & \\[2pt]
$u$ & Distance between base of penumbral cone and centre of primary & Eq. B6 of Paper I \\[2pt]  
$w$ & Argument of pericentre &   \\[2pt] 
$x$ & Cartesian component of $\vec{r}$ & equation (\ref{xr}) \\[2pt]
$x_{\rm loc}$ & Cartesian component of point location on target surface & equation (\ref{xloceq}) \\[2pt]
$y$ & Cartesian component of $\vec{r}$ & equation (\ref{yr}) \\[2pt]
$y_{\rm loc}$ & Cartesian component of point location on target surface & equation (\ref{yloceq}) \\[2pt]
$z$ & Cartesian component of $\vec{r}$ & equation (\ref{zr}) \\[2pt]
$z_{\rm loc}$ & Cartesian component of point location on target surface & equation (\ref{zloceq}) \\[2pt]
\hline
\end{tabular}
\end{minipage}
\end{table*}

\begin{table*}
 \centering
 \begin{minipage}{180mm}
  \centering
  \caption{Unstylised uppercase Roman variables used in this paper and in Paper I.}
  \label{TabRomanlower}
  \begin{tabular}{@{}lll@{}}
  \hline
   Variable & Explanation & Reference \\
   \hline
$A$ & Quadric coefficient of $x^2$ & equation (\ref{quadcone}) \\[2pt]
$B$ & Quadric coefficient of $y^2$ & equation (\ref{quadcone}) \\[2pt]
$C$ & Quadric coefficient of $z^2$ & equation (\ref{quadcone}) \\[2pt]
$D$ & Half of quadric coefficient of $xy$ & equation (\ref{quadcone}) \\[2pt]
$E$ & Half of quadric coefficient of $yz$ & equation (\ref{quadcone}) \\[2pt]
$F$ & Half of quadric coefficient of $xz$ & equation (\ref{quadcone}) \\[2pt]
$G$ & Half of quadric coefficient of $x$ & equation (\ref{quadcone}) \\[2pt]
$H$ & Half of quadric coefficient of $y$ & equation (\ref{quadcone}) \\[2pt]
$J$ & Half of quadric coefficient of $z$ & equation (\ref{quadcone}) \\[2pt]
$K$ & Quadric constant term & equation (\ref{quadcone}) \\[2pt]   
$M$ & Mass &   \\[2pt]
$R_1$ & Radius of primary (always a star) &  \\[2pt]
$R_2$ & Radius of occulter &  \\[2pt]
$R_3$ & Radius of target &  \\[2pt]
$R_4$ & Radius of fourth external body (only when other three are in syzygy) & Section 10 only  \\[2pt]
$R_{\rm c}$ & Radius of base of umbral cone & equation (\ref{Rcexp}) \\[2pt]
$R_{\rm d}$ & Radius of base of penumbral cone & Eq. B5 of Paper I \\[2pt]
$R_{\rm ant}$ & Antumbral shadow radius at syzygy & Eq. A23 of Paper I \\[2pt]   
$R_{\rm ant}^{\rm cen}$ & Auxiliary variable & equation (\ref{Rantcen}) \\[2pt]
$R_{\rm ant}^{\rm edge}$ & Auxiliary variable & equation (\ref{Rantedge}) \\[2pt]
$R_{\rm pen}$ & Penumbral shadow radius at syzygy & Eq. B13 of Paper I \\[2pt]
$R_{\rm umb}$ & Umbral shadow radius at syzygy & Eq. A19 of Paper I \\[2pt]   
$R_{\rm umb}^{\rm cen}$ & Auxiliary variable & equation (\ref{Rumbcen}) \\[2pt]
$R_{\rm umb}^{\rm edge}$ & Auxiliary variable & equation (\ref{Rumbedge}) \\[2pt]   
$S_{\rm ant}$ & Antumbral surface area at syzygy & Eq. A25 of Paper I \\[2pt]
$S_{\rm pen}$ & Penumbral surface area at syzygy & Eq. B14 of Paper I \\[2pt]   
$S_{\rm umb}$ & Umbral surface area at syzygy & Eq. A24 of Paper I \\[2pt]
\hline
\end{tabular}
\end{minipage}
\end{table*}

\begin{table*}
 \centering
 \begin{minipage}{180mm}
  \centering
  \caption{Greek variables used in this paper and in Paper I.}
  \label{TabGreek}
  \begin{tabular}{@{}lll@{}}
  \hline
   Variable & Explanation & Reference \\
   \hline
$\alpha$ & Semivertical angle of umbral and antumbral cones & equation (\ref{alpheq}) and Fig. \ref{zoomcart} \\[2pt]
$\beta$ & Auxiliary angle & equation (\ref{betaeq}) \\[2pt]
$\beta^{\vee}$ & Auxiliary angle & equation (\ref{betaeqvee}) \\[2pt]
$\gamma$ & Auxiliary angle & equation (\ref{gammaeq}) \\[2pt]
$\gamma^{\vee}$ & Auxiliary angle & equation (\ref{gammaveeeq}) \\[2pt]   
$\delta$ & Commonly-appearing auxiliary variable & equation (\ref{deltaeq}) \\[2pt]
$\Delta$ & Auxiliary variable & equation (\ref{Deltaequ}) \\[2pt]  
$\epsilon$ & Auxiliary angle & equation (\ref{epsiloneq}) \\[2pt]
$\epsilon^{\vee}$ & Auxiliary angle & equation (\ref{epsilonveeeq}) \\[2pt]    
$\zeta$ & Auxiliary angle & equation (\ref{earthfrac2}) \\[2pt]
$\eta_{\rm sha}$ & Angular diameter of shadow on primary as seen by observer on fourth body & equation (\ref{etasha}) \\[2pt]
$\eta_{14}$ & Angular diameter of primary as seen by observer on fourth body & equation (\ref{eta14arbk}) \\[2pt]
$\theta_{13}$ & Angular diameter of primary as seen by observer on target in penumbral shadow & Section B3 of Paper I \\[2pt]
$\theta_{23}$ & Angular diameter of occulter as seen by observer on target in penumbral shadow & Section B3 of Paper I \\[2pt]   
$\iota$ & Auxiliary angle & equation (\ref{earthfrac3}) \\[2pt]
$\kappa$ & Angle between $\vec{r}$ and $\vec{r}_{13}$ & equation (\ref{kappaeq}) \\[2pt]
$\nu_{\rm ant}^{\rm cen}$ & Auxiliary angle & equation (\ref{nuantcen}) \\[2pt]
$\nu_{\rm ant}^{\rm edge}$ & Auxiliary angle & equation (\ref{nuantedge}) \\[2pt]   
$\nu_{\rm umb}^{\rm cen}$ & Auxiliary angle & equation (\ref{nuumbcen}) \\[2pt]
$\nu_{\rm umb}^{\rm edge}$ & Auxiliary angle & equation (\ref{nuumbedge}) \\[2pt]
$\xi$ & Angle between $\vec{r}$ and $\vec{r}_{12}$ & equation (\ref{xieq}) \\[2pt]   
$\Pi$ & True anomaly &  \\[2pt]
$\sigma$ & Longitude of point on target & Section 5.2 \\[2pt]   
$\tau$ & Time of pericentre passage &  \\[2pt]  
$\phi_{13}$ & Angular diameter of primary as seen by observer on target in umbral shadow & Section A4 of Paper I \\[2pt] 
$\phi_{23}$ & Angular diameter of occulter as seen by observer on target in umbral shadow & Section A4 of Paper I \\[2pt]
$\chi$ & Auxiliary angle & equation (\ref{earthfrac1}) \\[2pt]
$\psi$ & Angle between $\vec{r}_{12}$ and $\vec{r}_{13}$ & equation (\ref{psieq}) \\[2pt]
$\psi_{\blacklozenge}$ & Value of $\psi$ where target is tangent and external to the cone & equation (\ref{lozdef}) \\[2pt]
$\psi_{\diamondsuit}$ & Value of $\psi$ where target is tangent and internal to the cone & equation (\ref{diadef}) \\[2pt]
$\psi_{\Updownarrow}$ & Value of $\psi$ where target is tangent to the cone axis & Section 7.2 \\[2pt]   
$\psi_{\rm loc}$ & Angle between $\vec{r}_{12}$ and a specific point on target surface & equation (\ref{psiloc}) \\[2pt]
$\omega$ & Auxiliary angle & equation (\ref{omegaeq}) \\[2pt]
$\omega^{\vee}$ & Auxiliary angle & equation (\ref{omegaveeeq}) \\[2pt]
$\Omega$ & Longitude of ascending node &  \\[2pt]
\hline
\end{tabular}
\end{minipage}
\end{table*}

\begin{table*}
 \centering
 \begin{minipage}{180mm}
  \centering
  \caption{Other variables used in this paper and in Paper I.}
  \label{TabHebrew}
  \begin{tabular}{@{}lll@{}}
  \hline
   Variable & Explanation & Reference \\
   \hline
$\mathfrak{x}$ & Component of transformed coordinate system & equation (\ref{mathfrakx}) \\[2pt]
$\mathfrak{x}_{\rm offset}$ & Offset from origin of transformed coordinate system & equation (\ref{mathfrakxoffset}) \\[2pt]
$\mathfrak{y}$ & Component of transformed coordinate system & equation (\ref{mathfraky}) \\[2pt]
$\mathfrak{y}_{\rm offset}$ & Offset from origin of transformed coordinate system & equation (\ref{mathfrakyoffset}) \\[2pt]
$\mathcal{C}$ & Auxiliary variables & equations (\ref{Cfirst}-\ref{Clast}) \\[2pt]
$\mathfrak{d}$ & Duration of transit & equations (\ref{dur1S2P}) \\[2pt]   
$\mathfrak{D}$ & Duration of ingress or egress only & equations (\ref{ing1S2P}) \\[2pt] 
$\mathfrak{e}$ & End time for transit & equation (\ref{end1S2P}) \\[2pt]
$\mathfrak{f}$ & Frequency of transits & equation (\ref{fre1S2P}) \\[2pt]     
$\mathcal{F}$ & Auxiliary variable & equation (\ref{calFeq}) \\[2pt]
$\mathcal{G}$ & Auxiliary variable & equation (\ref{calGeq}) \\[2pt]
$\mathcal{G}^{\vee}$ & Auxiliary variable & equation (\ref{calGeqvee}) \\[2pt]
$\mathfrak{G}$ & Gravitational constant & equation (\ref{meanmo}) \\[2pt]   
$\mathfrak{l}$ & Latitude for point on target & Section 5.2  \\[2pt]
$\mathcal{L}_{\rm ant}$ & Antumbral shadow length & equations (\ref{Lant}) and (\ref{Lantfinal}) \\[2pt]
$\mathcal{L}_{\rm ant}^{\rm syz}$ & Antumbral shadow length at syzygy $=2R_{\rm ant}$ & equation (\ref{Lantsyz}) \\[2pt]   
$\mathcal{L}_{\rm umb}$ & Umbral shadow length & equations (\ref{Lumb}) and (\ref{Lumbfinal}) \\[2pt]
$\mathcal{L}_{\rm umb}^{\rm syz}$ & Umbral shadow length at syzygy $=2R_{\rm umb}$ & equation (\ref{Lumbsyz}) \\[2pt]
$\mathcal{M}$ & Mean anomaly & equation (\ref{meananam}) \\[2pt]
$\mathfrak{n}$ & Mean motion & equation (\ref{meanmo}) \\[2pt]
$\mathcal{P}$ & Auxiliary variables & equations (\ref{Pfirst}-\ref{Plast}) \\[2pt]
$\mathfrak{s}$ & Start time for transit & equation (\ref{sta1S2P}) \\[2pt]   
$\mathcal{S}$ & Auxiliary variables & equations (\ref{Sfirst}-\ref{Slast}) \\[2pt] 
$\mathfrak{u}$ & Component in transformed coordinate system & Section 3.2 \\[2pt]
$\mathcal{U}$ & Auxiliary variables & equations (\ref{Ufirst}-\ref{Ulast}) \\[2pt]   
$\mathfrak{v}$ & Component in transformed coordinate system & Section 3.2 \\[2pt]
$\mathfrak{w}$ & Component in transformed coordinate system & Section 3.2 \\[2pt]
\hline
\end{tabular}
\end{minipage}
\end{table*}

A transiting system will produce an antumbral or umbral shadow on the target.
This shadow is defined as the intersection of the near
side of the target with the cone. Figure \ref{intersec} provides a cartoon
which illustrates the near side of the target
in shadow (the dark meshed region) and the far side of the target in the
intersection but not the shadow (the light meshed region).

In order to obtain the equation describing the shadow, I first need to
characterise the cone and sphere. The general Cartesian equations of both a
sphere and a cone have quadratic terms. Because both shapes are translated
from the origin, and the cone is arbitrarily oriented, these movements will introduce
cross terms in the equations. Therefore, I express all equations as
quadrics, also known as quadratic curves. These quadrics, whose properties
are described in \cite{smith1884}, \cite{coolidge1968}, \cite{zwillinger1996} and 
\cite{mccrea2006}, have the form:

\[
Ax^2 + By^2 + Cz^2 + 2Dxy + 2Eyz + 2Fxz 
\]

\begin{equation}
\ \ \ \ \ \ + 2Gx + 2Hy + 2Jz + K = 0
.
\label{quadcone}
\end{equation}

My immediate goal is to find expressions of the coefficients
$(A, B, C, D, E, F, G, H, J, K)$ in terms of 
$\left\lbrace R_1, R_2, R_3, x_{12}, y_{12}, z_{12}, x_{13},
y_{13}, z_{13} \right\rbrace$ (recall that the last
six of these variables can be combined to yield
$x_{23}, y_{23}, z_{23}, x_{123}, y_{123},$ and $z_{123}$).

\subsection{Cartesian equation of the target}

The general equation of the surface of the spherical target is 

\begin{equation}
\left(x - x_{13}\right)^2 + \left(y - y_{13}\right)^2 + \left(z - z_{13}\right)^2 = R_{3}^2
.
\label{eqtarget}
\end{equation}

\noindent{}When expressed as a quadric, this surface gives $A_{\rm tar}=B_{\rm tar}=C_{\rm tar}=1$, $D_{\rm tar}=E_{\rm tar}=F_{\rm tar}=0$,
$G_{\rm tar}=-x_{13}$, $H_{\rm tar}=-y_{13}$, $J_{\rm tar}=-z_{13}$ and $K_{\rm tar} = r_{13}^2 - R_{3}^2$. This equation does not distinguish
the near side from the far side of the target. I never have to consider the interior of this sphere.

\subsection{Cartesian equation of the cone}

\begin{figure}
\ \ \ \ \ \ \ \ \ \ \ \ \ \ \ \ \ \ \ \ \
\includegraphics[width=3.5cm]{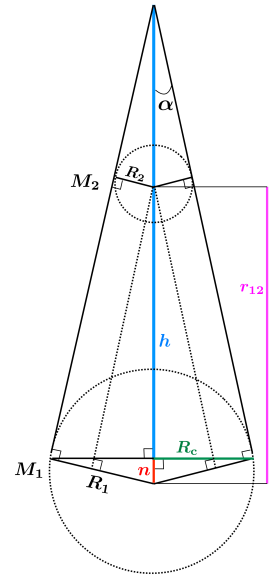}
\caption{
The radiation cone producing umbral and antumbral shadows; a reproduction of 
figure A1 from Paper I but with the inclusion of the semi-vertical angle $\alpha$.  
This figure helps illustrate the geometric meaning of $h$, $n$ and $\alpha$. Note
that no orientation with respect to $x$, $y$ or $z$ is assumed.
}
\label{zoomcart}
\end{figure}

A right circular cone with two nappes is always formed by the spherical primary and
the spherical occulter, such that the vertex is outside of both bodies. The centre
of the primary, which does not include the base of the cone,
lies at the origin of the coordinate system (see Fig. \ref{zoomcart}, which
is a duplicate of Fig. A1 of Paper I).

In order to construct the equation of this cone in 
Cartesian $(x,y,z)$ coordinates, consider first a different coordinate system
$(\mathfrak{u},\mathfrak{v},\mathfrak{w})$. In this new coordinate system, 
imagine a right circular cone with (i) the vertex at the origin, (ii) a base which is 
arbitrarily oriented, and (iii) a semi-vertical angle $\alpha$. 

The equation of this cone is

\begin{equation}
\left(\mathfrak{u}^2 + \mathfrak{v}^2 + \mathfrak{w}^2\right) \cos^2{\alpha} = \left(\frac{x_{12}}{r_{12}} \mathfrak{u} + \frac{y_{12}}{r_{12}} \mathfrak{v} + \frac{z_{12}}{r_{12}} \mathfrak{w} \right)^2
\label{rotcone}
\end{equation}

\noindent{}where I have inserted the direction cosines from my setup as the coefficients of $\mathfrak{u}$, $\mathfrak{v}$ and $\mathfrak{w}$. 
The semi-vertical angle $\alpha$, as shown in Fig. \ref{zoomcart}, is given by

\begin{equation}
\tan{\alpha} = \frac{R_{\rm c}}{h} 
\label{alpheq}
\end{equation}

\noindent{}where

\begin{equation}
h = R_1\left[ \frac{r_{12}^2 - \left(R_1 - R_2\right)^2 }{r_{12}\left(R_1 - R_2\right)} \right]
,
\label{hexp}
\end{equation}

\begin{equation}
R_{\rm c} = \frac{R_1}{r_{12}} \sqrt{r_{12}^2 - \left(R_1 - R_2\right)^2 }
.
\label{Rcexp}
\end{equation}

Equation (\ref{rotcone}) therefore reduces to a better result commonly found in textbooks
in the special case of the cone's axes coinciding
with the $z$-axis ($x_{12} = y_{12} = 0, z_{12} = r_{12}$). In Paper I (equation A17), this reduced
cone was also translated by a distance $h$, with an origin that was already translated by
a distance $n$.

Here the vertex is translated along the cone's axis by a distance of $h+n$,
where this sum is related to $R_1$, $R_2$ and $r_{12}$ through Fig. \ref{zoomcart} as 

\begin{equation}
h+n = \frac{R_1 r_{12}}{R_1 - R_2} = \frac{R_1}{\delta}
,
\label{hplusn}
\end{equation}

\noindent{}where

\begin{equation}
\delta \equiv \frac{R_1 - R_2}{r_{12}}
.
\label{deltaeq}
\end{equation}

\noindent{}$\delta$ is a particularly helpful auxiliary variable
representing a fundamental ratio in eclipse geometry, and is 
used throughout the paper. In fact, by utilising $\delta$
throughout the paper, I will be able to convert all of
my results from the antumbral case to the penumbral case
quickly in Section 9.

Introducing the correct translation and rotation finally yields the 
equation of the cone:

\[
\left( \frac{h^2}{h^2 + R_{\rm c}^2} \right)
\bigg[ 
  \left(x - \frac{x_{12} \left(h+n\right)}{r_{12}} \right)^2 
+ \left(y - \frac{y_{12} \left(h+n\right)}{r_{12}} \right)^2
\]

\[
+ \left(z - \frac{z_{12} \left(h+n\right)}{r_{12}} \right)^2
  \bigg]
= 
\bigg[
\frac{x_{12}}{r_{12}}
\left(x - \frac{x_{12} \left(h+n\right)}{r_{12}} \right)
\]

\begin{equation}
+
\frac{y_{12}}{r_{12}}
\left(y - \frac{y_{12} \left(h+n\right)}{r_{12}} \right)
+
\frac{z_{12}}{r_{12}}
\left(z - \frac{z_{12} \left(h+n\right)}{r_{12}} \right)
\bigg]^2
\end{equation}

\noindent{}where

\begin{equation}
r_{12} = \sqrt{x_{12}^2 + y_{12}^2 + z_{12}^2}
.
\end{equation}

The coefficients of the resulting quadric for the cone's surface -- which is
infinite in two directions -- can be expressed in the desired variables as

\begin{equation}
A_{\rm cone} = \frac{1}{r_{12}^2}\left[r_{12}^2 - x_{12}^2 - \left(R_1 - R_2\right)^2 \right]
           = 1 - \delta^2 - \left(\frac{x_{12}}{r_{12}}\right)^2
,
\end{equation}

\begin{equation}
B_{\rm cone} = \frac{1}{r_{12}^2}\left[r_{12}^2 - y_{12}^2 - \left(R_1 - R_2\right)^2 \right]
           = 1 - \delta^2 - \left(\frac{y_{12}}{r_{12}}\right)^2
,
\end{equation}

\begin{equation}
C_{\rm cone} = \frac{1}{r_{12}^2}\left[r_{12}^2 - z_{12}^2 - \left(R_1 - R_2\right)^2 \right]
           = 1 - \delta^2 - \left(\frac{z_{12}}{r_{12}}\right)^2
,
\end{equation}

\begin{equation}
D_{\rm cone} = -\frac{x_{12}y_{12}}{r_{12}^2}
,
\end{equation}

\begin{equation}
E_{\rm cone} = -\frac{y_{12}z_{12}}{r_{12}^2}
,
\end{equation}

\begin{equation}
F_{\rm cone} = -\frac{x_{12}z_{12}}{r_{12}^2}
,
\end{equation}

\begin{equation}
G_{\rm cone} = \frac{R_1\left(R_1 - R_2\right)x_{12}}{r_{12}^2}
           = R_1 \delta \left(\frac{x_{12}}{r_{12}} \right)
,
\end{equation}

\begin{equation}
H_{\rm cone} = \frac{R_1\left(R_1 - R_2\right)y_{12}}{r_{12}^2}
           = R_1 \delta \left(\frac{y_{12}}{r_{12}} \right)
,
\end{equation}

\begin{equation}
J_{\rm cone} = \frac{R_1\left(R_1 - R_2\right)z_{12}}{r_{12}^2}
           = R_1 \delta \left(\frac{z_{12}}{r_{12}} \right)
,
\end{equation}

\begin{equation}
K_{\rm cone} = -R_{1}^2
.
\end{equation}

\subsection{Cartesian equation of the intersection}

Now I construct the equation of the shadow by combining the results
of the equations for the cone and target. To do so I solve equation 
(\ref{eqtarget}) for $x^2 + y^2 + z^2$ and substitute the expression
into the quadric equation for the cone. The result is

\[
A_{\rm int}x^2 + B_{\rm int}y^2 + C_{\rm int}z^2 + 2D_{\rm int}xy + 2E_{\rm int}yz + 2F_{\rm int}xz 
\]

\begin{equation}
\ \ \ \ \ \ + 2G_{\rm int}x + 2H_{\rm int}y + 2J_{\rm int}z + K_{\rm int} = 0
\label{inteq}
\end{equation}

\noindent{}where ``int'' denotes intersection, and with

\begin{equation}
A_{\rm int} = \left(\frac{x_{12}}{r_{12}} \right)^2
,
\end{equation}

\begin{equation}
B_{\rm int} = \left(\frac{y_{12}}{r_{12}} \right)^2           
,
\end{equation}

\begin{equation}
C_{\rm int} = \left(\frac{z_{12}}{r_{12}} \right)^2
,
\end{equation}

\begin{equation}
D_{\rm int} = \frac{x_{12}y_{12}}{r_{12}^2}
,
\end{equation}

\begin{equation}
E_{\rm int} = \frac{y_{12}z_{12}}{r_{12}^2}
,
\end{equation}

\begin{equation}
F_{\rm int} = \frac{x_{12}z_{12}}{r_{12}^2}
,
\end{equation}

\begin{equation}
G_{\rm int} = -x_{13} \left(1 - \delta^2 \right) - \frac{\delta R_1 x_{12}}{r_{12}}
,
\end{equation}

\begin{equation}
H_{\rm int} = -y_{13} \left(1 - \delta^2 \right) - \frac{\delta R_1 y_{12}}{r_{12}}
,
\end{equation}

\begin{equation}
J_{\rm int} = -z_{13} \left(1 - \delta^2 \right) - \frac{\delta R_1 z_{12}}{r_{12}}
,
\label{Jinteq}
\end{equation}

\begin{equation}
K_{\rm int} = R_{1}^2 + \left(1 - \delta^2\right) \left(r_{13}^2 - R_{3}^2\right)
.
\label{TheKint}
\end{equation}

\noindent{}The surfaces of both the target and cone intersect only
when there exist $\left\lbrace x,y,z \right\rbrace$ which satisfy 
equation (\ref{inteq}).

\section{Shape of the shadow}

In order to determine the shape of the shadow, I utilise the properties of
quadrics \citep[Pgs. 316-319,][]{zwillinger1996}.
He showed that the shape defined by the quadric is determined by the following quantities, all using
the coefficients of the Cartesian equation of the intersection.
The first quantity is

\begin{equation} 
{\rm Rank} \left|
\begin{array}{ccc}
A_{\rm int} & D_{\rm int} & F_{\rm int} \\
D_{\rm int} & B_{\rm int} & E_{\rm int} \\
F_{\rm int} & E_{\rm int} & C_{\rm int} 
\end{array} 
\right|
= 1.
\label{rankA}
\end{equation}

\noindent{}and the next quantity is 

\begin{equation}
{\rm Rank} \left|
\begin{array}{cccc}
A_{\rm int} & D_{\rm int} & F_{\rm int} & G_{\rm int} \\
D_{\rm int} & B_{\rm int} & E_{\rm int} & H_{\rm int} \\
F_{\rm int} & E_{\rm int} & C_{\rm int} & J_{\rm int} \\
G_{\rm int} & H_{\rm int} & J_{\rm int} & K_{\rm int} 
\end{array} 
\right|
= 3.
\label{Bmatrix}
\end{equation}

These two values alone reveal that a {\it transit shadow is in the shape of a parabolic cylinder}.
For a visual representation of this shape, see Fig. \ref{intersec}. 

Another way of demonstrating the character of the shadow is by writing out and appropriately
transforming equation (\ref{inteq}). I note that the equation may be written as

\[
R_{1}^2 r_{12}^2 + \left(x x_{12} + y y_{12} + z z_{12}\right)^2 
- 2 \delta R_1 r_{12} \left(x x_{12} + y y_{12} + z z_{12} \right)
\]

\[
+\left(1-\delta^2\right)r_{12}^2 \left[r_{13}^2 - R_{3}^2 - 2 \left(x x_{13} + y y_{13} + z z_{13} \right) 
\right] = 0.
\]

\begin{equation}
\label{radialeq}
\end{equation}

\noindent{}I can rewrite this equation in the standard form for a parabolic cylinder as

\begin{equation}
\left(\mathfrak{x} - \mathfrak{x}_{\rm offset}\right)^2 + 2 \left(\mathfrak{y} - \mathfrak{y}_{\rm offset}\right) = 0
,
\label{stanform}
\end{equation}

\noindent{}by making the transformations 

\begin{equation}
\mathfrak{x} \rightarrow x x_{12} + y y_{12} + z z_{12}
\label{mathfrakx}
\end{equation}

\noindent{}and

\begin{equation}
\mathfrak{y} \rightarrow -\left(x x_{13} + y y_{13} + z z_{13}\right) r_{12}^2 \left(1 - \delta^2 \right)
,
\label{mathfraky}
\end{equation}

\noindent{}where 

\begin{equation}
\mathfrak{x}_{\rm offset} = \delta R_1 r_{12} 
\label{mathfrakxoffset}
\end{equation}

\noindent{}and

\begin{equation}
\mathfrak{y}_{\rm offset} = -\frac{1}{2}\left(1 - \delta^2\right) r_{12}^2 \left(r_{13}^2 + R_{1}^2 - R_{3}^2 \right).
\label{mathfrakyoffset}
\end{equation}

\section{Condition to be ``in transit''}

A natural follow-up to the last section is the determination of when a system would be in
transit in the first place. ``In transit'' can refer to a target that is partially or 
fully enveloped within the radiation cone. 

\subsection{Shadow anywhere on target}

The condition for these limiting 
cases may be derived from equation (\ref{radialeq}) by noting
that the Cartesian-based expressions there can all
be expressed as dot products. 

\subsubsection{Radial equation of the intersection}

I define $\xi$ as the angle
between $\vec{r}$ and $\vec{r}_{12}$ and $\kappa$ as the angle between 
$\vec{r}$ and $\vec{r}_{13}$. Then by assuming $r$ is the distance
to some point on the intersection,

\begin{equation}
\cos{\xi} = \frac{x x_{12} + y y_{12} + z z_{12}}{r r_{12}}
\label{xieq}
\end{equation}

\noindent{}and

\begin{equation}
\cos{\kappa} = \frac{x x_{13} + y y_{13} + z z_{13}}{r r_{13}}
.
\label{kappaeq}
\end{equation}

\noindent{}The equation of the intersection hence becomes

\[
r^2 \cos^2{\xi} + R_{1}^2 - 2 \delta r R_1 \cos{\xi}
\]

\begin{equation}
\ \ \ \ \ 
- \left(1 - \delta^2\right) \left(2 r r_{13} \cos{\kappa} - r_{13}^2 + R_{3}^2  \right)
= 0.
\end{equation}

\noindent{}Solving for $r$ yields

\[
r = \frac{1}{\cos^2{\xi}} 
\bigg[ 
\delta R_1 \cos{\xi} + r_{13} \cos{\kappa} \left(1 - \delta^2\right)
\]

\[
\ \ \ \ \pm \sqrt{1-\delta^2} 
\bigg\lbrace 2 \delta R_1 r_{13} \cos{\xi} \cos{\kappa} 
+ \left(1 - \delta^2\right) r_{13}^2 \cos^2{\kappa}
\]

\begin{equation}
\ \ \ \  - \cos^2{\xi}  
\left(r_{13}^2 + R_{1}^2 - R_{3}^2 \right)
\bigg \rbrace^{1/2}  
\bigg].
\label{requ}
\end{equation}

\noindent{}{\rev For a given $\xi$ and $\kappa$, the two signs in equation (\ref{requ}) refer to intersections occurring on the near side (negative sign) and far side (positive sign) of the target.}

\subsubsection{Target tangent to cone}

Equation (\ref{requ}) accounts for all possible values of $r$. These would, for
example, trace out the entire mesh in Fig. \ref{intersec}.
My concern here is to determine
the conditions for the target to be tangent to the cone, which will allow me to derive a transit
criterion for snapshots and other quantities when I introduce motion. If the target is tangent to the cone,
then $r$ can take one value only, denoted by $r_{\rm tan}$, requiring the determinant of equation (\ref{requ}) to be 
zero\footnote{I also consider
only positive values of $\cos{\xi}$ and $\cos{\kappa}$. The negative values would correspond
to the target intersecting the cone ``behind'' the primary, as the cone extends infinitely in
both directions. {\rev In exoplanetary science, this intersection would be referred to as a {\it secondary 
transit.} An analytical treatment of secondary transits may be carried out by proceeding with the analysis
in this section by assuming $\cos{\kappa} < 0$. This analysis may eventually yield, for example, an analytical expression
for the relative durations of the primary and secondary transits.}}. This requirement yields

\begin{equation}
\cos{\kappa_{\rm tan}} = \cos{\xi_{\rm tan}}
\left[
\frac
{\sqrt{\left(r_{13}^2 - R_{3}^2\right) \left(1 - \delta^2\right) + R_{1}^2} -\delta R_1}
{r_{13} \left(1 - \delta^2\right)} 
\right]
\end{equation}

\noindent{}and

\begin{equation}
r_{\rm tan} = 
\frac
{
\sqrt{\left(1 - \delta^2\right) \left(r_{13}^2 - R_{3}^2\right) + R_{1}^2  } 
}
{\cos{\xi_{\rm tan}}}
.
\end{equation}

In order to solve simultaneously for $r_{\rm tan}$, 
$\cos{\xi_{\rm tan}}$ and $\cos{\kappa_{\rm tan}}$
in terms of given variables, I need one more
equation: I use a relation which is obtained
by the triangle which connects the target (here tangent to the cone) 
and the centre of the primary:

\begin{equation}
R_{3}^2 = r_{\rm tan}^2 + r_{13}^2 - 2 r_{\rm tan} r_{13} \cos{\kappa_{\rm tan}}
.
\end{equation}

\noindent{}The final result is

\begin{equation}
r_{\rm tan} = \sqrt{ 
\frac
{K_{\rm int} + R_{1}^2 -2 \delta R_1 \sqrt{K_{\rm int}} }
{1 - \delta^2} 
},
\end{equation}

\begin{equation}
\cos{\xi_{\rm tan}} = \sqrt{ 
\frac
{K_{\rm int} \left(1 - \delta^2\right) }
{K_{\rm int} + R_{1}^2 -2 \delta R_1 \sqrt{K_{\rm int}}}
} = \frac{\sqrt{K_{\rm int}}}{r_{\rm tan}},
\end{equation}

\begin{equation}
\cos{\kappa_{\rm tan}} = 
\frac
{K_{\rm int}^{3/2} - \delta R_1 K_{\rm int}}
{r_{13}\sqrt{\left(1 - \delta^2\right) K_{\rm int} 
\left[ K_{\rm int} + R_{1}^2 -2 \delta R_1 \sqrt{K_{\rm int}} \right]   }}
.
\end{equation}

In order to develop an explicit criterion for transits, I wish to 
obtain a functional form in terms of the distance vectors to the centres
of the occulter and target (not the distance to the shadow itself).
So let $\psi$ represent the
angle between $\vec{r}_{12}$ and $\vec{r}_{13}$ such that

\begin{equation}
\cos{\psi} = \frac{x_{12}x_{13} + y_{12}y_{13} + z_{12}z_{13}}{r_{12}r_{13}}
.
\label{psieq}
\end{equation}

\noindent{}The angle $\psi$ also represents a crucial way to reduce the number
of degrees of freedom in the geometry, and will be applied repeatedly throughout
the paper.

Denote the limiting values of $\psi$ which correspond to when the target is
{\it tangent but external} to the cone as $\psi_{\blacklozenge}$: it is at these 
locations where the ingress begins and the egress ends. The criterion for a target to
be in transit is then:

\begin{equation}
\cos{\psi} \ge \cos{\psi_{\blacklozenge}}
,
\label{trancrit}
\end{equation}

\noindent{}where

\begin{equation}
\left| \psi_{\blacklozenge} \right| = \left|\xi_{\rm tan}\right| + \left|\kappa_{\rm tan}\right|,
\label{lozdef}
\end{equation}

\noindent{}which provides the basis for many
further results in this paper. The case where the target
is {\it tangent but internal} to the cone is also of interest,
for determining if, when and where the target is fully engulfed
in the shadow. Denote this critical angle as $\psi_{\diamondsuit}$,
and note that $\cos{\psi_{\diamondsuit}} \ge \cos{\psi_{\blacklozenge}}$.
Then

\begin{equation}
\left| \psi_{\diamondsuit} \right| = \left|\xi_{\rm tan} - \kappa_{\rm tan}\right|
.
\label{diadef}
\end{equation}

\noindent{}These tangent cases are illustrated graphically in Fig. \ref{bordcase}.

\subsubsection{Full engulfment in the shadow}

The figure demonstrates that a target which is tangent but internal to the cone
is not necessarily engulfed in shadow. The condition for the target to be fully 
engulfed in the shadow is

\begin{equation}
\cos{\psi} \ge \cos{\psi_{\diamondsuit}} \ \ \& \ \ \left|\xi_{\rm tan}\right| \ge \left|\kappa_{\rm tan}\right|
.
\label{engulfed}
\end{equation}

\subsubsection{Special case of syzygy}

\noindent{}I can perform a check by considering the critical target radius for engulfment
in the special case of syzygy. Here, 
$\xi_{\rm tan} = \kappa_{\rm tan}$, which yields,
for the umbral and antumbral cases respectively,

\begin{equation}
R_{3}^{\dagger} = R_1 - r_{13} \delta,
\end{equation}

\begin{equation}
R_{3}^{\ddagger} = r_{13} \delta - R_1
.
\end{equation}

\noindent{}These equations are equivalent to those of Eqs. A12-A13 of Paper I.
Note also 
that a target whose centre lies at the vertex of the radiation cone can never be engulfed in shadow. This
special case corresponds to $R_1 = r_{13} \delta$.

The physical meaning of the similar scenario 
where an observer on the target coincides with the cone's vertex is the following: the 
angular diameter of the occulter asymptotically would cover the primary's disc such that 
the eclipse would be considered just barely total. The size of the umbral shadow on the target
would be asymptotically zero at the observer's location. Other locations on the surface 
which are in shadow must be covered in the antumbral shadow.

\begin{figure}
\centering
\centerline{{\bf UMBRAL}}
\includegraphics[width=8cm]{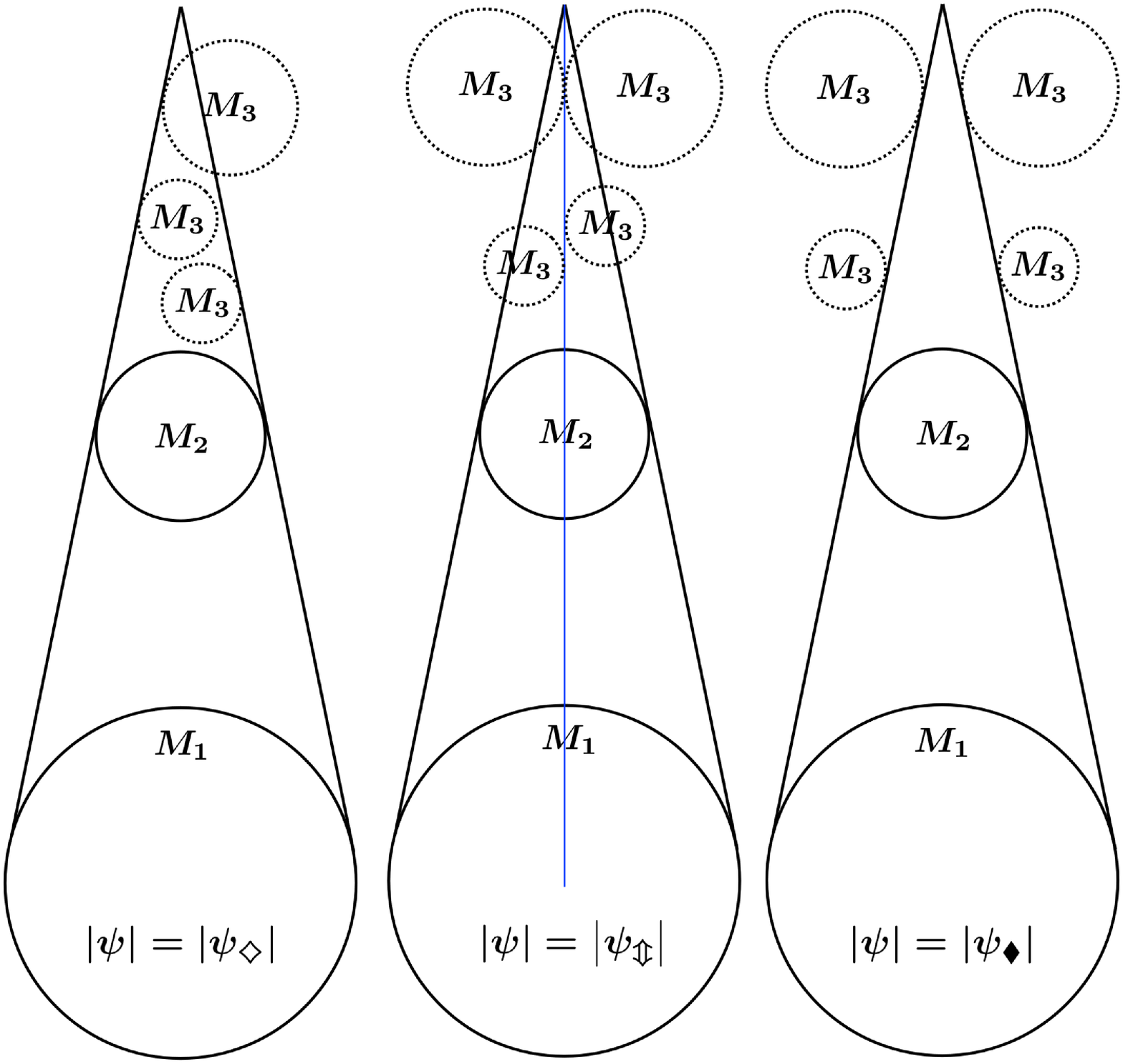}
---------------------------------------------------------------------------------
\centerline{{\bf ANTUMBRAL}}
\centerline{\phantom{}}
\includegraphics[width=8cm]{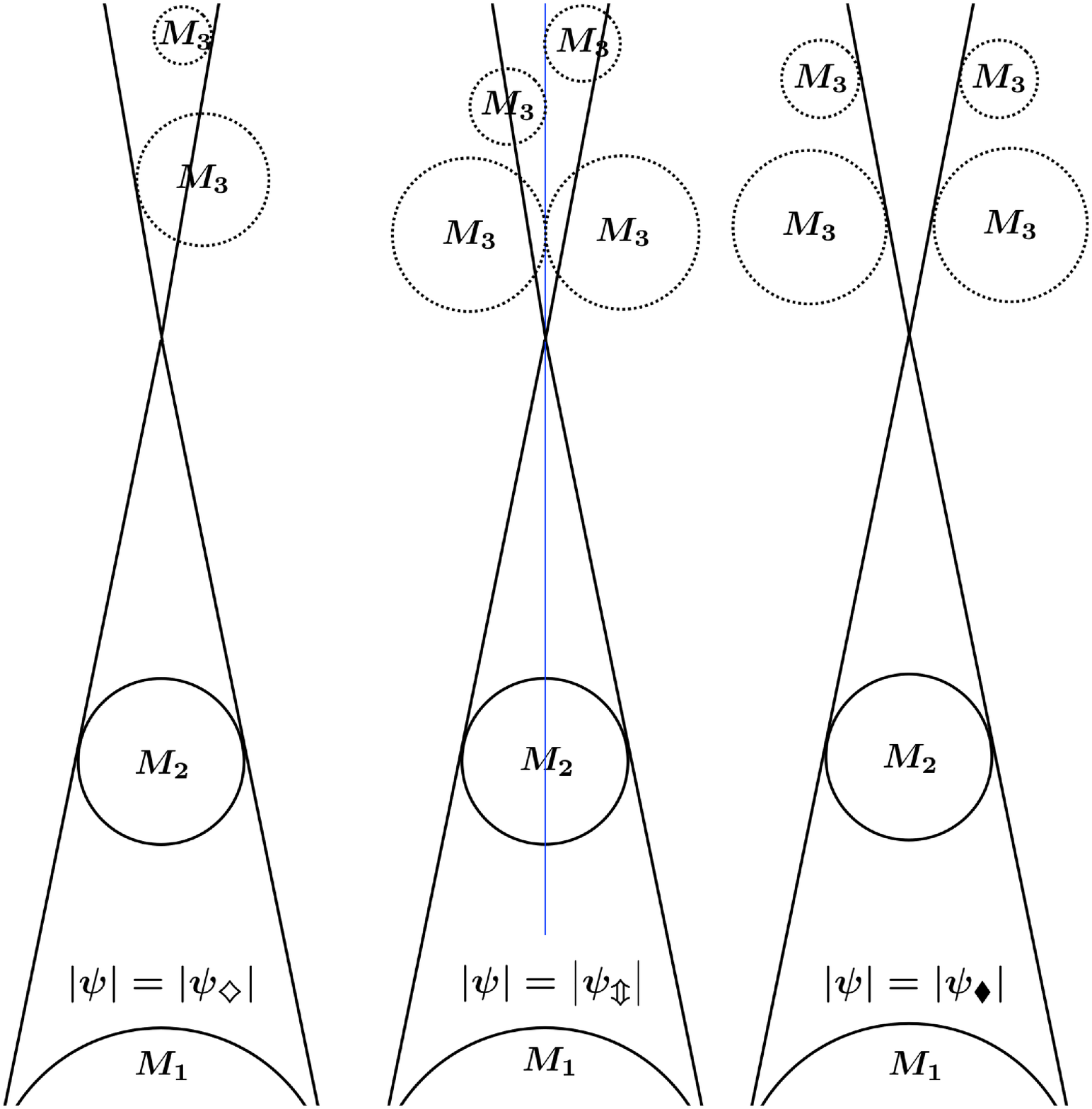}
\caption{
The geometric meaning of the critical angle values of $\psi_{\diamondsuit}$,
$\psi_{\blacklozenge}$, and $\psi_{\Updownarrow}$ for both
umbral cases (top panels) and antumbral cases
(bottom panels). Some different
possible locations of the target are illustrated with
dotted $M_3$ spheres. 
}
\label{bordcase}
\end{figure}

\subsection{Shadow on specific location on target}

The last paragraph illustrates the potential usefulness of a transit
criterion for
a specific point $(x_{\rm loc}, y_{\rm loc}, z_{\rm loc})$ on
the target's surface. In this section, I explore this scenario.

Assume that the point has a specific latitude $\mathfrak{l}$ and 
longitude $\sigma$, and
consider this point to be 
infinitesimal. Its Cartesian location is \citep[Pg. 205 of][]{roy2005}

\begin{eqnarray}
x_{\rm loc} &=& x_{13} - R_3 \cos{\sigma} \cos{\mathfrak{l}}
,
\label{xloceq}
\\
y_{\rm loc} &=& y_{13} - R_3 \sin{\sigma} \cos{\mathfrak{l}}
,
\label{yloceq}
\\
z_{\rm loc} &=& z_{13} - R_3 \sin{\mathfrak{l}}
.
\label{zloceq}
\end{eqnarray}

Then by assuming that this point is chosen to be on the near side of the target,
I can treat the point as representing the new target with
infinitesimal radius. Hence, the condition to be in transit is 

\begin{equation}
\cos{\psi_{\rm loc}} \ge \cos{\left[\psi_{\blacklozenge}\left(R_{3} = 0,r_{13}\rightarrow r_{\rm loc}\right)\right]}
,
\label{trancritloc}
\end{equation}

\noindent{}where

\begin{equation}
\cos{\psi}_{\rm loc} \equiv \frac
{x_{12} x_{\rm loc} + y_{12} y_{\rm loc} + z_{12}z_{\rm loc}}
{r_{12} r_{\rm loc}}
\label{psiloc}
\end{equation}

\noindent{}and

\begin{equation}
r_{\rm loc} = \sqrt{x_{\rm loc}^2 + y_{\rm loc}^2 + z_{\rm loc}^2}.
\label{rloc}
\end{equation}

Note that because this location is an infinitesimal point, it can never be partially engulfed in a shadow.

\section{Total vs. annular eclipse criterion}

In the last section, I determined the conditions necessary for a transit to exist.
Supposing it does, now I determine whether the transit produces a total or annular eclipse.
This exercise involves determining which nappe intersects the near side of the target at a given moment
in time. In rare cases, throughout a transit, the target's near side could at different times intersect both nappes, 
yielding a mixed or hybrid eclipse (here I consider just a snapshot in time).

\begin{figure}
\centering
\includegraphics[width=8cm]{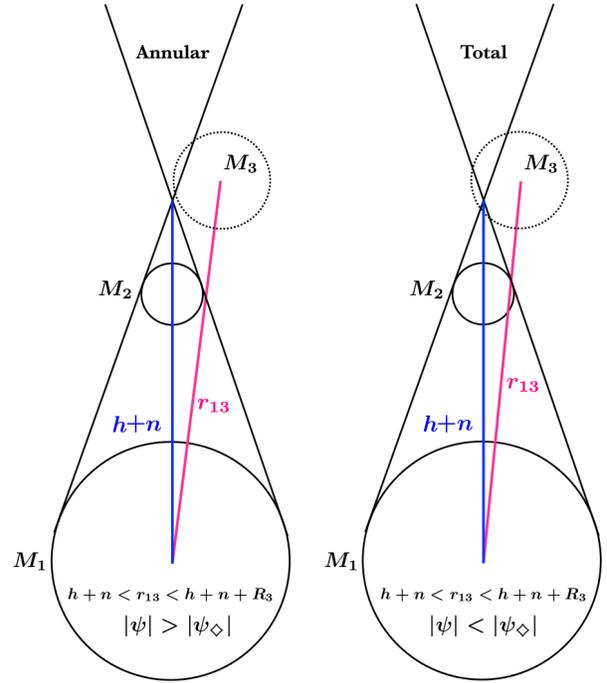}
\caption{
Determining whether an eclipse is total or annular. The figure illustrates
the two cases when $h+n<r_{13}<h+n+R_3$.
}
\label{anncon}
\end{figure}

For perspective, the criterion for
an annular eclipse to occur at syzygy (equation A7 of Paper I) is $r_{13} \ge h + n + R_3$, which
was derived by placing the target just above the cone's vertex at syzygy. The
criterion for a total eclipse to occur was obtained simply by switching the
sign.

Here, in the off-syzygy cases, there is a subtle difference. The same criterion
for an annular eclipse still holds, but is sufficient, not necessary. For a total
eclipse to occur, a sufficient
but not necessary condition is $r_{13} < h+n$,
because I consider only the near side of the target. Otherwise, when
$h+n < r_{13} < h+n+R_3$, the type of eclipse depends on $\psi$. Fig. \ref{anncon}
illustrates both possibilities: when $\left|\psi\right| >
\left|\psi_{\diamondsuit}\right|$, the eclipse is annular. Overall then,

\[
r_{13} \ge h+n+R_3,
\ \ \ \ \ \ \ \ \ \ \ \
\ \ \ \ \ \ \ \ \ \ \ \ \ \
\ \ \ \ \ \ \ \ \ \ 
{\rm annular}
\]

\[
h+n < r_{13} < h+n+R_3 \ \ \& 
\ \  
\left|\psi\right| \ge
\left|\psi_{\diamondsuit}\right|,
\ \ \ \ {\rm annular}
\]

\[
h+n < r_{13} < h+n+R_3 \ \ \&
\ \ 
\left|\psi\right| <
\left|\psi_{\diamondsuit}\right|,
\ \ \ \ {\rm total}
\]

\begin{equation}
r_{13} \le h+n,
\ \ \ \ \ \ \ \ \ \ \ \ \ \ \ \ \ \ \
\ \ \ \ \ \ \ \ \ \ \ \ \ \
\ \ \ \ \ \ \ \ \ \ \,
{\rm total}.
\label{newantot}
\end{equation}

\noindent{}I emphasise that this criterion applies only when the system is in transit in the first place (equation \ref{trancrit}).
In Table \ref{Tabcrit}, I summarise some important criteria so far listed.

At syzygy, $\psi=0$, and hence
$\left|\psi\right| < \left|\psi_{\diamondsuit}\right|$ is always true. Therefore,
in this case, the criterion from Paper I is recovered.

\begin{table*}
 \centering
 \begin{minipage}{180mm}
  \centering
  \caption{Summary of some important criteria}
  \label{Tabcrit}
  \begin{tabular}{@{}lll@{}}
  \hline
Criteria  &   & Meaning  \\
   \hline
   $\cos{\psi} < \cos{\psi_{\blacklozenge}}$
   & 
   & Not in transit
   \\[2pt]
   \hline
   $\cos{\psi} \ge \cos{\psi_{\blacklozenge}}$ :
   & 
   & In transit
   \\[6pt]
   & $\cos{\psi} \ge \cos{\psi_{\diamondsuit}} \ \ \& \ \ \left|\xi_{\rm tan}\right| \ge \left|\kappa_{\rm tan}\right|$
   & Engulfed in shadow
   \\[14pt]   
   & $r_{13} \ge h+n+R_3$
   & Annular eclipse
   \\[4pt]
   & $h+n < r_{13} < h+n+R_3$ \ \ \& \ \ $\left|\psi\right| \ge \left|\psi_{\diamondsuit}\right|$ 
   & Annular eclipse
   \\[14pt]
   & $h+n < r_{13} < h+n+R_3$ \ \ \& \ \ $\left|\psi\right| < \left|\psi_{\diamondsuit}\right|$ 
   & Total eclipse
   \\[6pt]    
   & $r_{13} \le h+n$
   & Total eclipse
   \\[14pt]
   & $\cos{\psi} = 1$ 
   & Syzygy
   \\[2pt]
   \hline
\end{tabular}
\end{minipage}
\end{table*}

\section{Size of shadow}

\begin{figure}
\centering
\includegraphics[width=8cm]{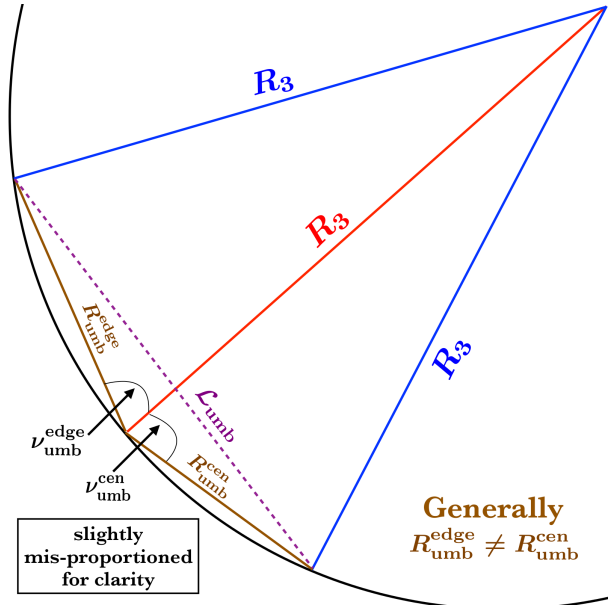}
\caption{
Defining the size of a shadow, with $\mathcal{L}$.
The endpoints of the dashed purple lines represent
locations where the radiation cone intersects the
shown target for both umbral and antumbral cases
(the ``umb'' subscripts shown here are just for
demonstration purposes, but could equally read
``ant'').
}
\label{Ldef}
\end{figure}

\begin{figure}
\centering
\includegraphics[width=5cm]{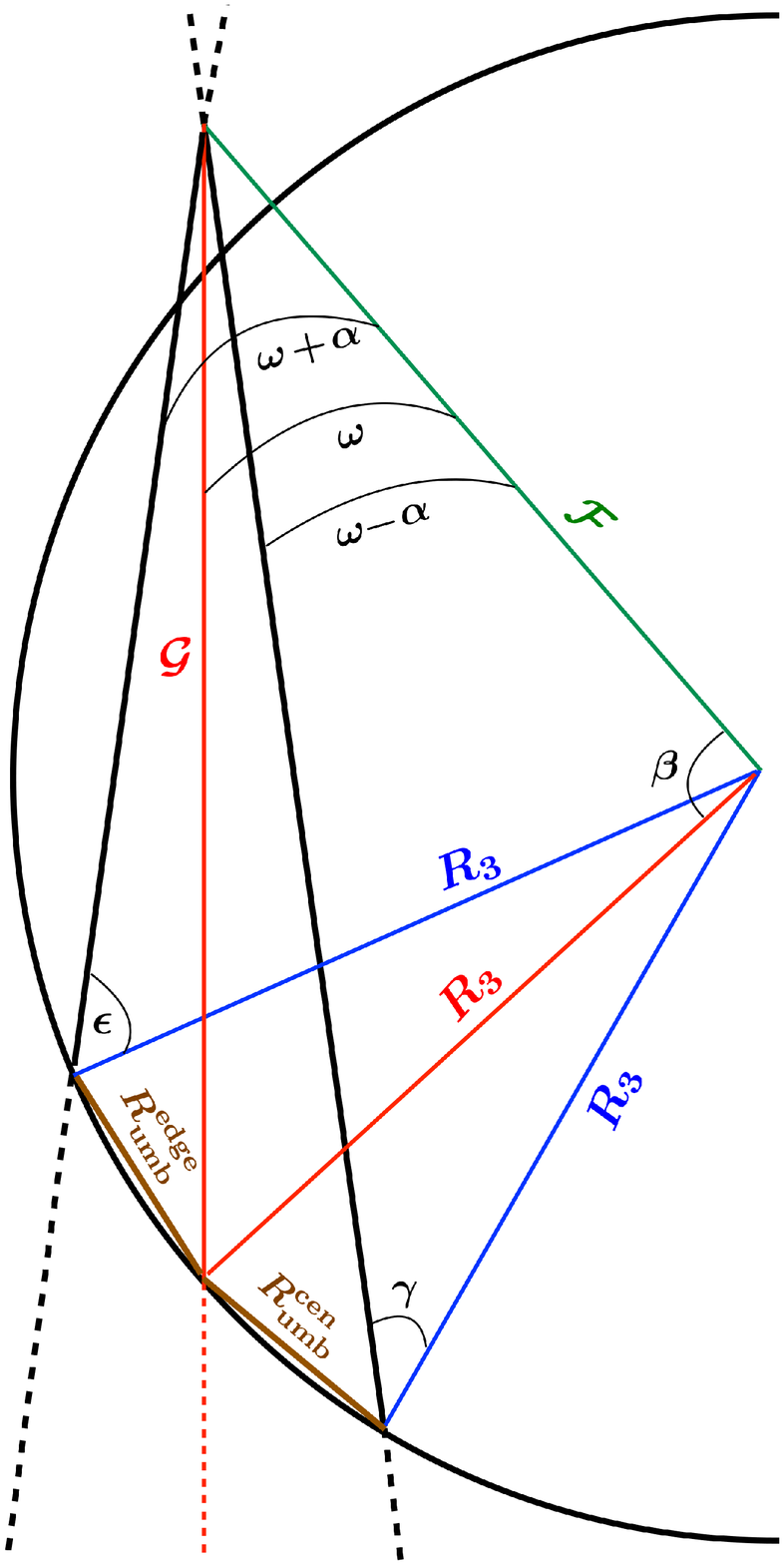}
\caption{
Geometry for determining size of an umbral shadow for simultaneous ingress and
egress. Shown is the target, with
the radiation cone bounded by diagonal black lines.
}
\label{umbsize}
\end{figure}

\begin{figure}
\centering
\includegraphics[width=8cm]{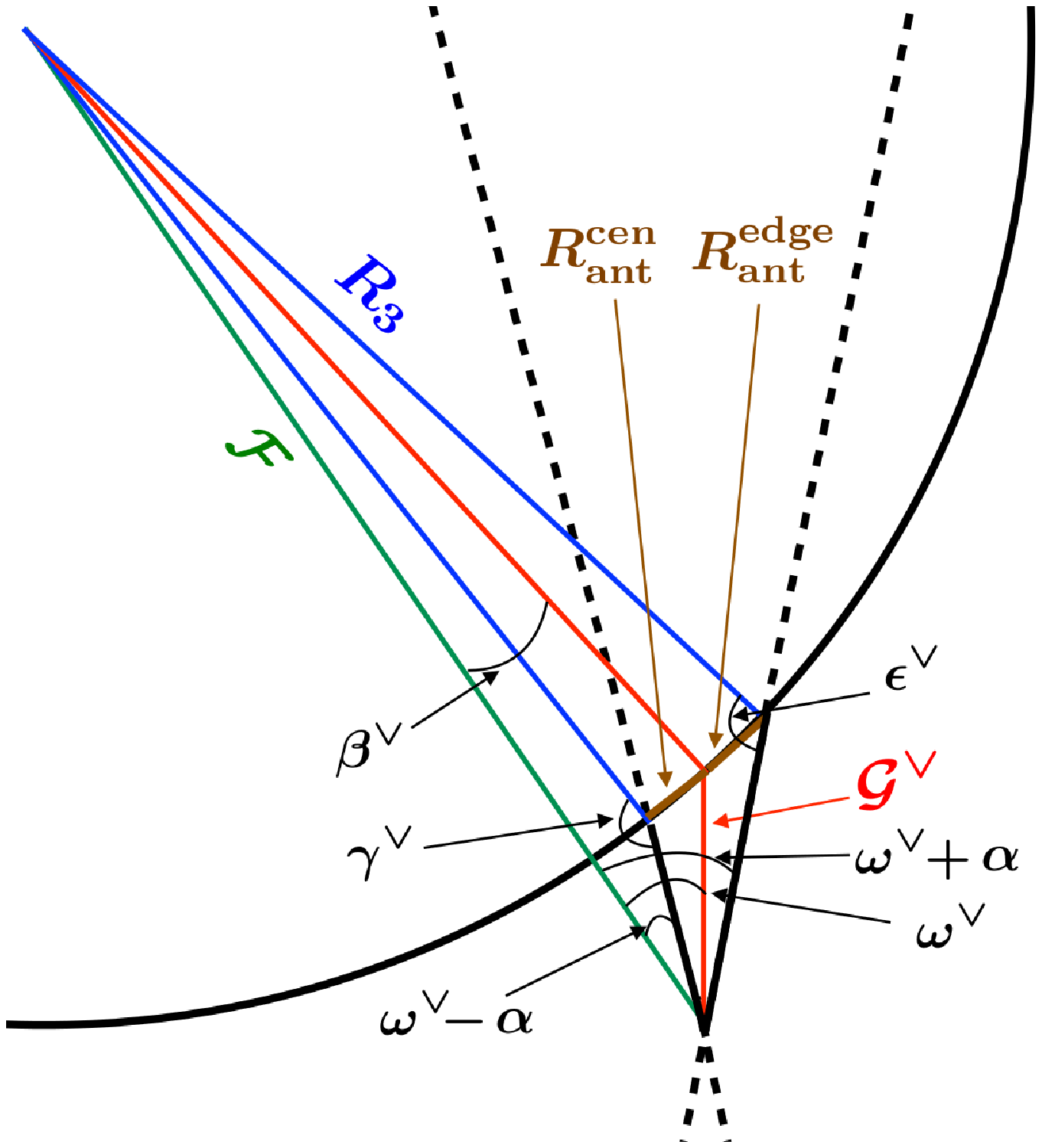}
\caption{
Geometry for determining size of an antumbral shadow for simultaneous ingress
and egress. Shown is the target, with
the radiation cone bounded by diagonal black lines.
}
\label{antsize}
\end{figure}

\begin{figure}
\centering
\includegraphics[width=6cm]{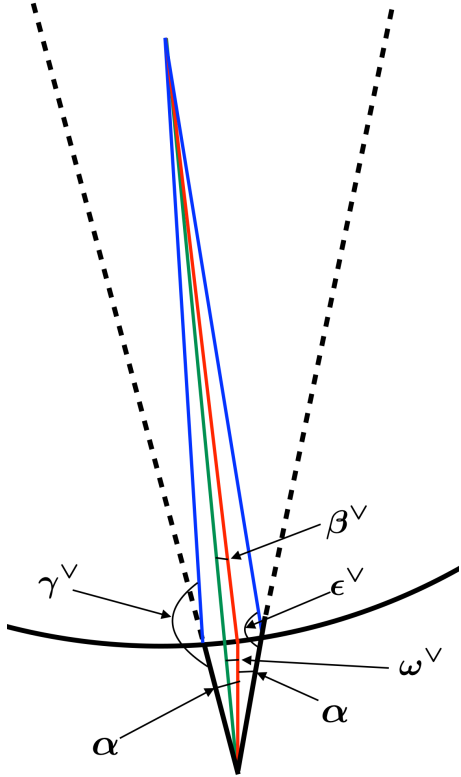}
\caption{
Geometry for determining size of an antumbral shadow for geometries
closer to syzygy than grazing. Despite the difference in relative location
of $\mathcal{F}$ (the green line) as compared to Fig. \ref{antsize}, 
the relevant geometric quantities in the main text remain unchanged.
}
\label{antsizeinvert}
\end{figure}

Now I pursue the task of determining the size of the shadow
when the target intersects the surface of the radiation cone,
using only the given variables
$\left\lbrace R_1, R_2, R_3, \vec{r}_{12}, \vec{r}_{13}\right\rbrace$.
To do so, I appeal to brute-force geometry 
from Figs. \ref{Ldef}-\ref{OutMostumb}, which cover both the umbral and antumbral
cases; the end result for each case is a single, compact piecewise function.

First, in Section 7.1, I consider the general case of the target being in both
ingress and egress
(see Fig. \ref{difftarget}), like the Earth is during nearly all of an
annular or total eclipse
of the Sun and Moon. Then in Section 7.2 I look at the geometry of
ingress only or egress only. In Section 7.3 I consider the case where the entire target
is engulfed, before collating the results in Section 7.4.

\subsection{Simultaneous ingress and egress}

\subsubsection{Size definition}

The shape and size of the shadow can vary considerably and non-monotonically during a transit, 
even under the assumption of a spherical target. Consequently, I seek a simple measure to quantify its size,
and denote the projected lengths $\mathcal{L}_{\rm umb}$ and $\mathcal{L}_{\rm ant}$ as the shadow
``sizes'' for the umbral and antumbral cases. For simultaneous ingress and egress, these 
lengths are split into two usually unequal
projected radii of the shadow ($R_{\rm umb}^{\rm edge}$ and $R_{\rm umb}^{\rm cen}$ for the umbral case 
and $R_{\rm ant}^{\rm edge}$ and $R_{\rm ant}^{\rm cen}$ for the antumbral case) as in Fig \ref{Ldef}. Hence,
through the law of cosines,

\[
\mathcal{L}_{\rm umb} = 
\]

\[
\sqrt{
\left(R_{\rm umb}^{\rm edge}\right)^2 + \left(R_{\rm umb}^{\rm cen}\right)^2
-2  R_{\rm umb}^{\rm edge}  R_{\rm umb}^{\rm cen} \cos{\left(\nu_{\rm umb}^{\rm edge} + \nu_{\rm umb}^{\rm cen}\right) }
},
\]

\begin{equation}
\label{Lumb}
\end{equation}

\[
\mathcal{L}_{\rm ant} =
\]

\[
\sqrt{
\left(R_{\rm ant}^{\rm edge}\right)^2 + \left(R_{\rm ant}^{\rm cen}\right)^2
-2  R_{\rm ant}^{\rm edge}  R_{\rm ant}^{\rm cen} \cos{\left(\nu_{\rm ant}^{\rm edge} + \nu_{\rm ant}^{\rm cen}\right) }
}.
\]

\begin{equation}
\label{Lant}
\end{equation}

I now set about determining all of the quantities on the right-hand-sides of 
equations (\ref{Lumb})-(\ref{Lant}).

\subsubsection{Auxiliary variables}

First consider Figs. \ref{umbsize}-\ref{antsizeinvert}, which cover the
umbral and antumbral cases in a general fashion. The auxiliary
variable $\mathcal{F}$ is common to both cases, and is the 
distance from the cone's vertex to the target
centre:

\[
\mathcal{F} =
\bigg[ 
  \left(x_{13} - \frac{x_{12} \left(h+n\right)}{r_{12}} \right)^2 
\]

\[
\ \ \ \ \ \ + \left(y_{13} - \frac{y_{12} \left(h+n\right)}{r_{12}} \right)^2
+ \left(z_{13} - \frac{z_{12} \left(h+n\right)}{r_{12}} \right)^2
  \bigg]^{1/2}
\]

\begin{equation}
\ \ \ = \sqrt{r_{13}^2 + \frac{R_{1}^2}{\delta^2} - \frac{2 R_1 r_{13}}{\delta} \cos{\psi} }
.
\label{calFeq}
\end{equation}

The other variables are not common to both the umbral and antumbral cases. All of the 
differences in the umbral and antumbral cases stem from the relative
location of the radiation cone's vertex to the near side of the target, which is
manifested through the lengths
$\mathcal{G}$ and $\mathcal{G}^{\vee}$:

\[
\mathcal{G} = h + n - \left( r_{13} \cos{\psi} - \sqrt{R_{3}^2 - r_{13}^2 \sin{\psi}} \right)
\]

\[
\ \ \ = \frac{R_1}{\delta} - \left( r_{13} \cos{\psi} - \sqrt{R_{3}^2 - r_{13}^2 \sin{\psi}} \right)
,
\]

\begin{equation}
\label{calGeq}
\end{equation}

\[
\mathcal{G}^{\vee} = \left( r_{13} \cos{\psi} + \sqrt{R_{3}^2 - r_{13}^2 \sin{\psi}} \right) - \left(h + n\right)
\]

\[
\ \ \ = \left( r_{13} \cos{\psi} + \sqrt{R_{3}^2 - r_{13}^2 \sin{\psi}} \right) - \frac{R_1}{\delta}
,
\]

\begin{equation}
\label{calGeqvee}
\end{equation}

\noindent{}and $\omega$ and $\omega^{\vee}$:

\begin{equation}
\omega = 
\arccos{
\left[
\frac
{\mathcal{F}^2 + \left(\frac{R_1}{\delta}\right)^2 - r_{13}^2}
{2 \mathcal{F} \left(\frac{R_1}{\delta}\right)}
\right]
}
,
\label{omegaeq}
\end{equation}

\begin{equation}
\omega^{\vee} = \pi - \omega
.
\label{omegaveeeq}
\end{equation}

\noindent{}The lengths $G$ and $G^{\vee}$ are derived by applying the law of cosines
to the triangle involving $R_3$, $r_{13}$, and the length
from the origin to the near side of the target along the cone's axis. The correct sign in front of the square root is obtained by choosing either the near or far side of the target.
For the 
derivation of $\omega$ and $\omega^{\vee}$, I use the law of cosines rather than the
law of sines. I do so in order to avoid a dependence on $G$ and $G^{\vee}$, because when $R_3 \le r_{13} \sin{\psi}$, then $G$ and $G^{\vee}$ will not exist.

Various angles now follow. The choice of whether to apply the law of cosines or sines
in the following depends on the quadrants in which these angles lie and in the interest
of avoiding ambiguity:

\begin{equation}
\beta = 
\arccos{
\left[
\frac
{\mathcal{F}^2 + R_{3}^2 - \mathcal{G}^2}
{2 \mathcal{F} R_3}
\right]
}
,
\label{betaeq}
\end{equation}

\begin{equation}
\beta^{\vee} = 
\arccos{
\left[
\frac
{\mathcal{F}^2 + R_{3}^2 - \left(\mathcal{G}^{\vee}\right)^2}
{2 \mathcal{F} R_3}
\right]
}
,
\label{betaeqvee}
\end{equation}

\begin{equation}
\epsilon = 
\arcsin{
\left[
\frac{\mathcal{F}}{R_3}
\sin{\left(\omega + \alpha\right)}
\right]
,
}
\label{epsiloneq}
\end{equation}

\begin{equation}
\epsilon^{\vee} = 
\pi -
\arcsin{
\left[
\frac{\mathcal{F}}{R_3}
\sin{\left(\omega^{\vee} + \alpha\right)}
\right]
,
}
\label{epsilonveeeq}
\end{equation}

\begin{equation}
\gamma = 
\arcsin{
\left[
\frac{\mathcal{F}}{R_3}
\sin{\left(\omega - \alpha\right)}
\right]
,
}
\label{gammaeq}
\end{equation}

\begin{equation}
\gamma^{\vee} = 
\pi
-
\arcsin{
\left[
\frac{\mathcal{F}}{R_3}
\sin{\left(\omega^{\vee} - \alpha\right)}
\right]
.
}
\label{gammaveeeq}
\end{equation}

\noindent{}Recall that $\alpha$ is the opening angle of the cone from equation (\ref{alpheq})
and Fig. \ref{zoomcart}, and is independent of the target.

For the umbral case, $\beta$ can lie in the first or second quadrants,
and $\epsilon$ and $\gamma$ always lie within
the first quadrant. However, for the antumbral case, although $\beta^{\vee}$
always lies within the first quadrant, $\epsilon^{\vee}$
always lies within the second quadrant, and $\gamma^{\vee}$ can lie
in either the second quadrant (Fig. \ref{antsize}) or third quadrant
(Fig. \ref{antsizeinvert}).

\subsubsection{Final solution}

I can now derive expressions for $R_{\rm umb}^{\rm edge}$, $R_{\rm umb}^{\rm cen}$, 
$R_{\rm ant}^{\rm edge}$, $R_{\rm ant}^{\rm cen}$ and the correspondingly labelled expressions
for $\nu$ through the isosceles triangles formed by these variables (Fig. \ref{Ldef}). I obtain

\begin{equation}
\nu_{\rm umb}^{\rm edge} = \pi - \frac{\left(\beta + \epsilon + \omega + \alpha \right)}{2}
,
\label{nuumbedge}
\end{equation}

\begin{equation}
\nu_{\rm umb}^{\rm cen} = \frac{\beta + \gamma + \omega - \alpha}{2}
,
\label{nuumbcen}
\end{equation}

\begin{equation}
\nu_{\rm ant}^{\rm edge} = \frac{\beta^{\vee} + \epsilon^{\vee} + \omega^{\vee} + \alpha}{2}
,
\label{nuantedge}
\end{equation}

\begin{equation}
\nu_{\rm ant}^{\rm cen} = \pi - \frac{\left(\beta^{\vee} + \gamma^{\vee} + \omega^{\vee} - \alpha \right)}{2}
,
\label{nuantcen}
\end{equation}

\noindent{}and, hence, through the law of sines,

\[
R_{\rm umb}^{\rm edge} = R_3 
\frac
{\sin{\left[-\left(\beta + \epsilon + \omega + \alpha\right)  \right]} }
{\sin{\left[\nu_{\rm umb}^{\rm edge}\right]}  } 
\]

\begin{equation}
\ \ \ \ \ \ \ \,
=
-2 R_3 \cos{
\left[
\frac{1}{2}
\left(
\alpha + \beta + \epsilon + \omega
\right)
\right]
}
,
\label{Rumbedge}
\end{equation}

\[
R_{\rm umb}^{\rm cen} = R_3
\frac
{\sin{\left[\beta + \gamma + \omega - \alpha  \right]} }
{\sin{\left[\nu_{\rm umb}^{\rm cen}\right]}  } 
\]

\begin{equation}
\ \ \ \ \ \ \ \,
=
2 R_3 \cos{
\left[
\frac{1}{2}
\left(
\alpha - \beta - \gamma - \omega
\right)
\right]
}
,
\label{Rumbcen}
\end{equation}

\[
R_{\rm ant}^{\rm edge} = R_3
\frac
{\sin{\left[\beta^{\vee} + \epsilon^{\vee} + \omega^{\vee} + \alpha  \right]} }
{\sin{\left[\nu_{\rm ant}^{\rm edge}\right]}  } 
\]

\begin{equation}
\ \ \ \ \ \ \ \,
=
2 R_3 \cos{
\left[
\frac{1}{2}
\left(
\alpha + \beta^{\vee} + \epsilon^{\vee} + \omega^{\vee}
\right)
\right]
}
,
\label{Rantedge}
\end{equation}

\[
R_{\rm ant}^{\rm cen} = R_3
\frac
{\sin{\left[-\left(\beta^{\vee} + \gamma^{\vee} + \omega^{\vee} - \alpha\right)  \right]} }
{\sin{\left[\nu_{\rm ant}^{\rm cen}\right]}  } 
\]

\begin{equation}
\ \ \ \ \ \ \ \,
=
-2 R_3 \cos{
\left[
\frac{1}{2}
\left(
\alpha - \beta^{\vee} - \gamma^{\vee} - \omega^{\vee}
\right)
\right]
}
.
\label{Rantcen}
\end{equation}

\noindent{}Inserting these expressions (equations \ref{nuumbedge}-\ref{Rantcen}) 
into equations (\ref{Lumb}) and (\ref{Lant}) 
produces cancellations in $\beta$, $\beta^{\vee}$, $\omega$, and $\omega^{\vee}$.
Taking care to use the correct (physically realistic) root finally yields the
following compact expressions

\begin{equation}
\mathcal{L}_{\rm umb} = 2 R_{3} \sin{\left[\alpha + \frac{\epsilon}{2} - \frac{\gamma}{2}  \right]}
,
\label{newLumb}
\end{equation}

\begin{equation}
\mathcal{L}_{\rm ant} = 2 R_{3} \sin{\left[- \left( \alpha + \frac{\epsilon^{\vee}}{2} - \frac{\gamma^{\vee}}{2} \right)\right]}
.
\label{newLant}
\end{equation}

For the umbral shadow, when the target experiences both ingress
and egress (see Fig. \ref{difftarget}),
the final answer is given by equation (\ref{newLumb}), 
which is solved by first computing variables in the following order from
equations (\ref{hexp}), (\ref{Rcexp}), (\ref{alpheq}), (\ref{deltaeq}), (\ref{calFeq}),
(\ref{omegaeq}),
(\ref{epsiloneq}), and (\ref{gammaeq}). 

Similarly, for the antumbral shadow, the final answer is given
by equation (\ref{newLant}), 
which is solved by first computing variables in the following order from
equations (\ref{hexp}), (\ref{Rcexp}), (\ref{alpheq}), (\ref{deltaeq}), (\ref{calFeq}),
(\ref{omegaeq}), (\ref{omegaveeeq}),
(\ref{epsilonveeeq}), and (\ref{gammaveeeq}).

\subsubsection{At syzygy}

I can check these solutions with those from the syzygy case,
where $\psi = 0$ and hence $\mathcal{F} = \left| R_1 - \delta r_{13} \right|/\delta$. The value of subsequent auxiliary variables depends on whether $R_1$ is greater than $\delta r_{13}$.

Regardless, the umbral shadow size is given by

\begin{equation}
\mathcal{L}_{\rm umb}^{\rm syz} = 2 R_3 \sin{\left\lbrace \alpha - 
\arcsin{\left[  
\frac{\left(r_{13} - \frac{R_1}{\delta}  \right) \sin{\alpha} }{R_3}
\right]}  \right\rbrace}.
\label{Lumbsyz}
\end{equation}

\noindent{}This expression checks out because it is equivalent to twice the umbral syzygetic radius
from equation (A19) of Paper I, just in a more compact form.

Now consider the antumbral syzygy. Here,

\begin{equation}
\mathcal{L}_{\rm ant}^{\rm syz} = 2 R_3 \sin{\left\lbrace -\alpha + 
\arcsin{\left[  
\frac{\left(r_{13} - \frac{R_1}{\delta}  \right) \sin{\alpha} }{R_3}
\right]}  \right\rbrace}.
\label{Lantsyz}
\end{equation}

\noindent{}Similarly, this expression checks out because it is equivalent to twice the antumbral syzygetic radius from equation (A23) of Paper I.

\subsubsection{Extreme shadow sizes}

The shadow sizes range from zero when the target is tangent to the radiation cone,
to a maximum value some time soon after the start of ingress, and then a local minimum
at syzygy. Determining these quantities is facilitated by the compact forms of 
$\mathcal{L}_{\rm umb}$ and $\mathcal{L}_{\rm ant}$ from equations 
(\ref{newLumb}) and (\ref{newLant}). They show that I need only consider
the differences $(\epsilon-\gamma)$ and $(\epsilon^{\vee} - \gamma^{\vee})$
when determining the extremes.

By inspection, I see that when both ingress and egress occur, the minimum value is
achieved when $\epsilon = -\gamma$ and $\epsilon^{\vee} = -\gamma^{\vee}$, giving

\begin{equation}
{\rm local \ min}\left(\mathcal{L}_{\rm umb}\right) = \mathcal{L}_{\rm umb}^{\rm syz}
,
\label{minshaumb}
\end{equation}

\begin{equation}
{\rm local \ min}\left(\mathcal{L}_{\rm ant}\right) = \mathcal{L}_{\rm ant}^{\rm syz}
.
\label{minshaant}
\end{equation}

\begin{figure}
\centering
\includegraphics[width=5cm]{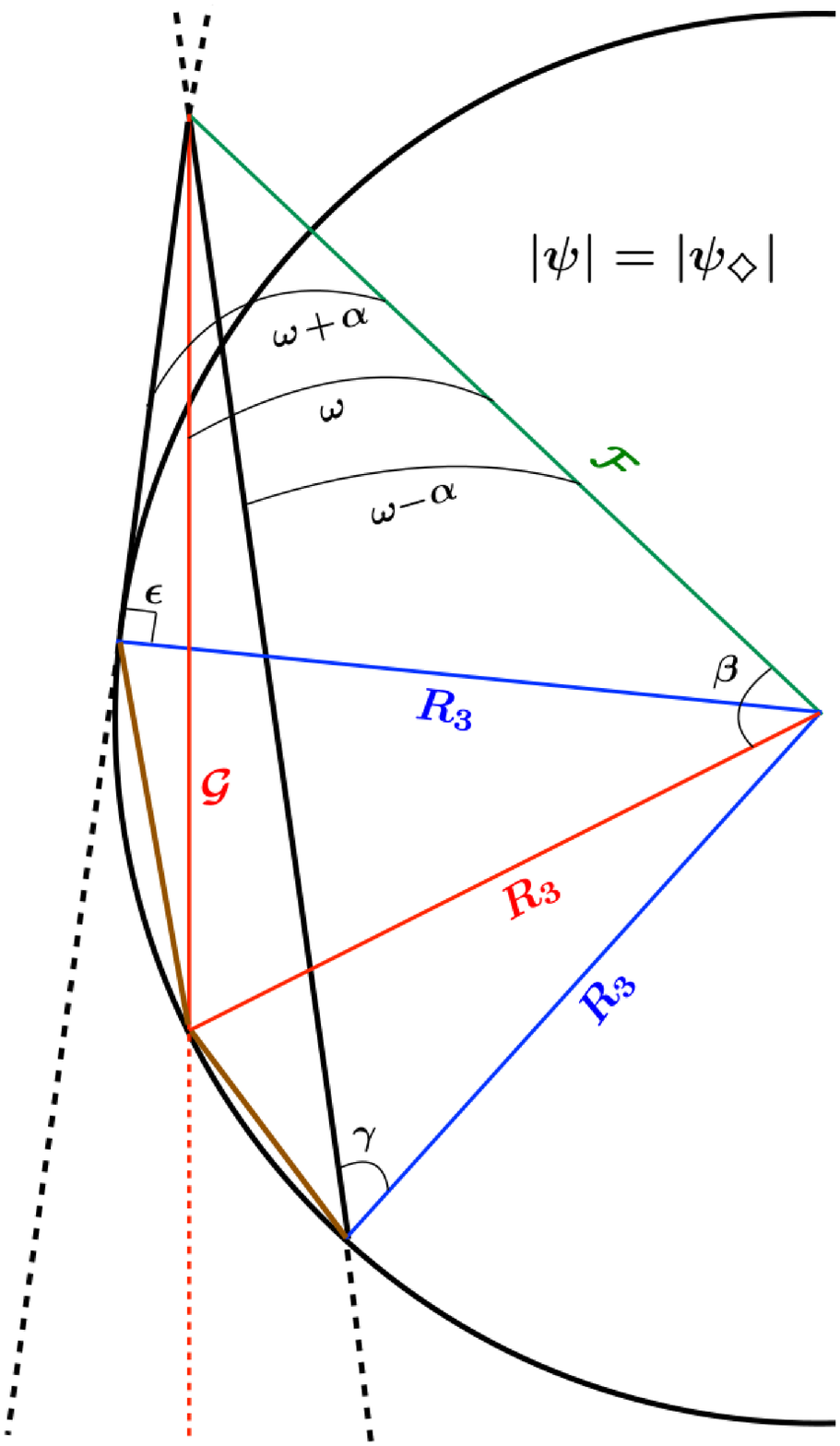}
\caption{
The configuration which gives the maximum size of the umbral shadow,
where $\epsilon = \pi/2$. 
}
\label{tansize}
\end{figure}

\begin{figure}
\centering
\includegraphics[width=5cm]{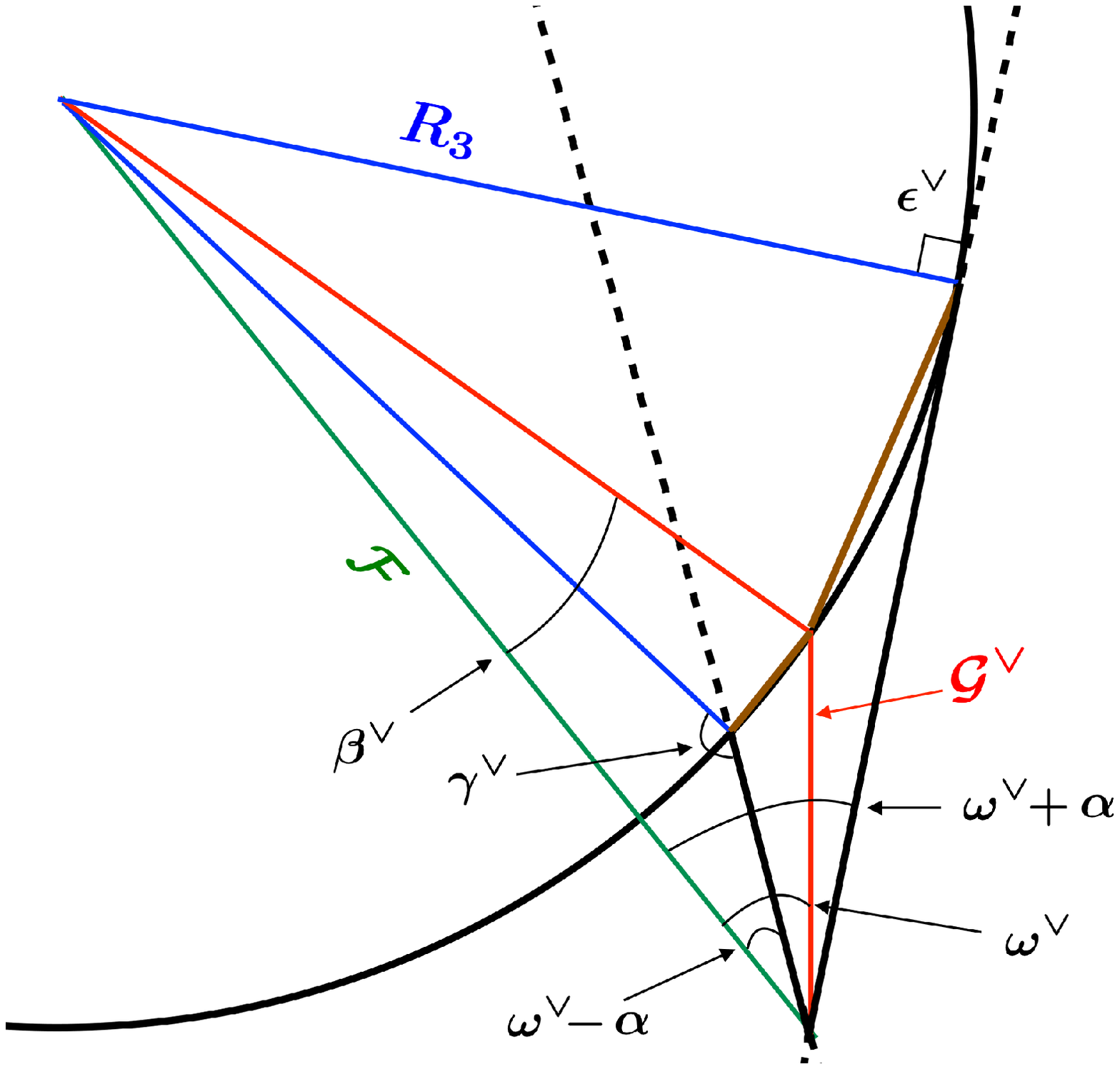}
\caption{
The configuration which gives the maximum size of the antumbral shadow,
where $\epsilon^{\vee} = \pi/2$. 
}
\label{anttansize}
\end{figure}

In order to determine the maximum shadow size, consider the allowable
range of $\psi$. The maximum value of $\psi$ for which the target is
in both ingress and egress occurs when $\epsilon = \pi/2$ or $\epsilon^{\vee} = \pi/2$,
as in Figs. \ref{tansize}-\ref{anttansize}.

The value of $\psi$ which corresponds to this situation is $\psi_{\diamondsuit}$.
The maximum shadow size must given by $\psi_{\diamondsuit}$ because both that angle
is the largest possible value of $\psi$ and the size of the shadow scales positively
with $\psi$, as shown by the functional dependence of $(\epsilon-\gamma)$ and
$(\epsilon^{\vee} - \gamma^{\vee})$ on $\psi$
through $\mathcal{F}$, $\omega$ and $\omega^{\vee}$. Therefore,

\begin{equation}
{\rm max}\left(\mathcal{L}_{\rm umb}\right) = 
2 R_3 \sin{\left[\alpha + \frac{\pi}{4} - \frac{\gamma \left(\psi = \psi_{\diamondsuit}\right)}{2}\right]}
,
\label{maxshaumb}
\end{equation}

\begin{equation}
{\rm max}\left(\mathcal{L}_{\rm ant}\right) =
2 R_3 \sin{\left[ -\left( \alpha + \frac{\pi}{4} - \frac{\gamma^{\vee} \left(\psi = \psi_{\diamondsuit}\right)}{2} \right) \right]}
.
\label{maxshaant}
\end{equation}

\subsection{Ingress only or egress only}

Now consider the case
$\left|\psi_{\diamondsuit}\right| \le \left| \psi \right| < \left|\psi_{\blacklozenge}\right|$,
when the target will still be in shadow and either ingress or egress will be occurring
(but not both).

\begin{figure}
\centering
\includegraphics[width=5cm]{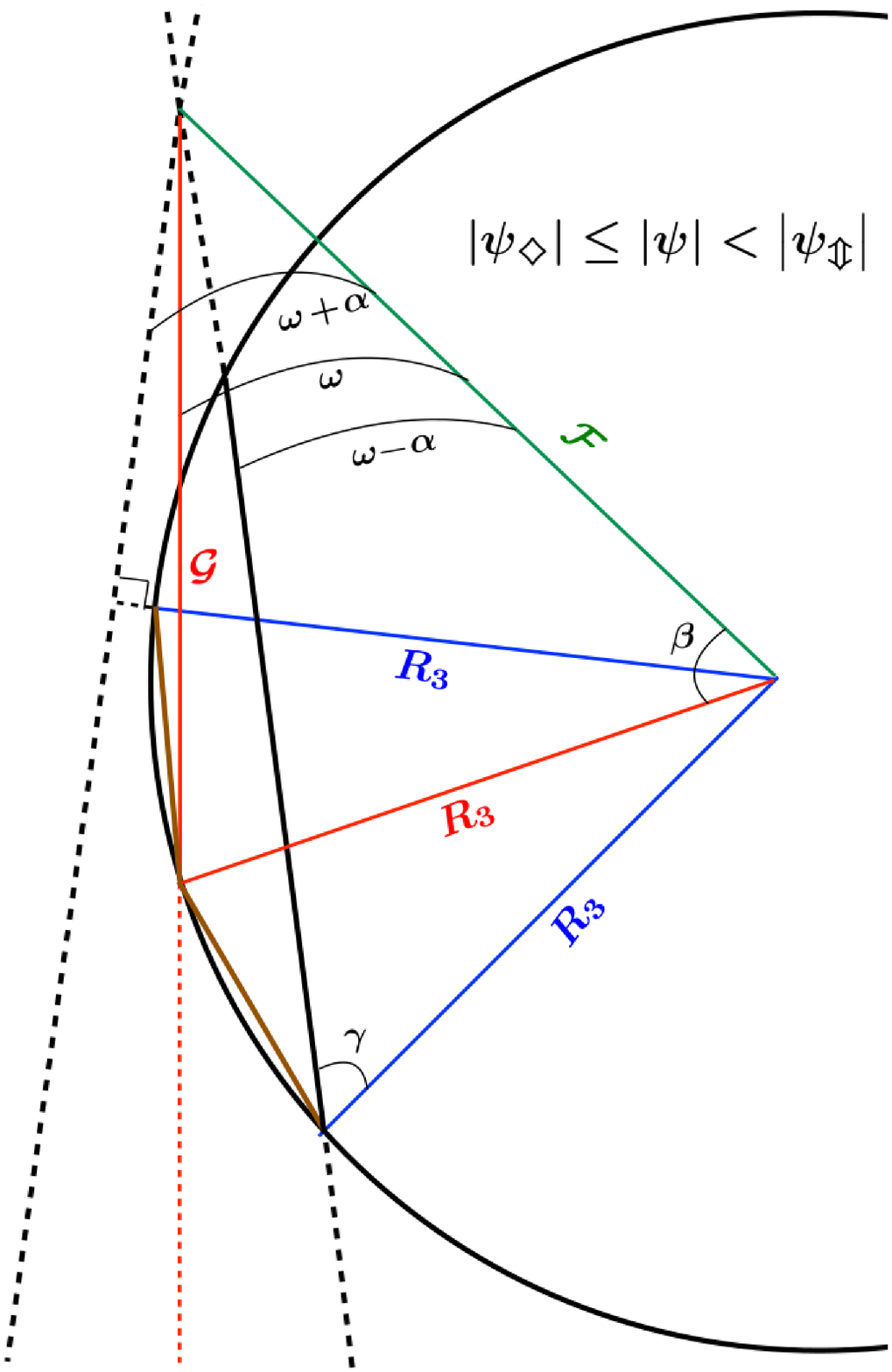}
\caption{
The umbral shadow during ingress only or egress only, when the cone
axis still intersects the target.
}
\label{stillGumb}
\end{figure}

\begin{figure}
\centering
\includegraphics[width=6.5cm]{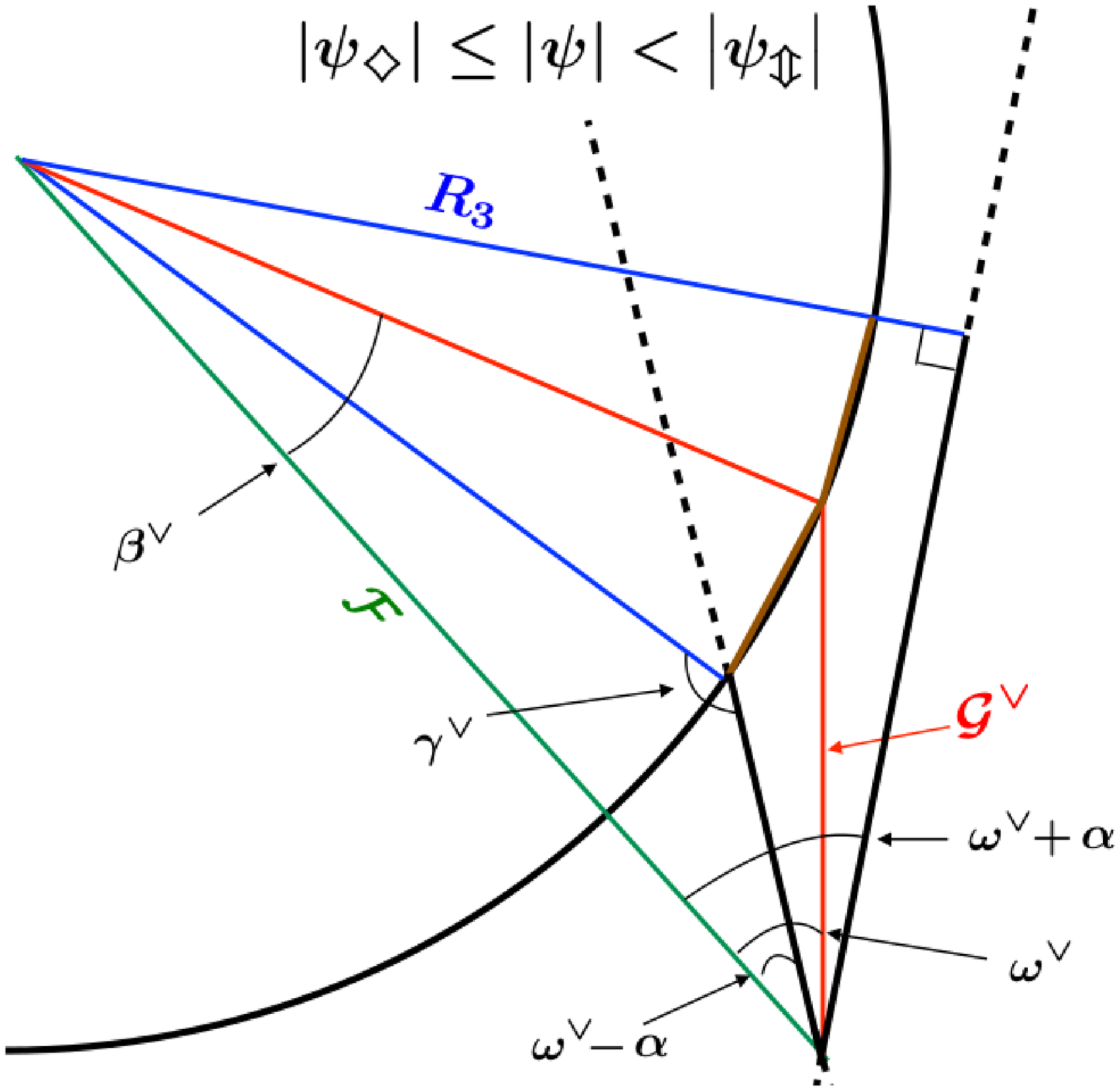}
\caption{
The antumbral shadow during ingress only or egress only, when the cone
axis still intersects the target.
}
\label{stillGant}
\end{figure}

\begin{figure}
\centering
\includegraphics[width=6.5cm]{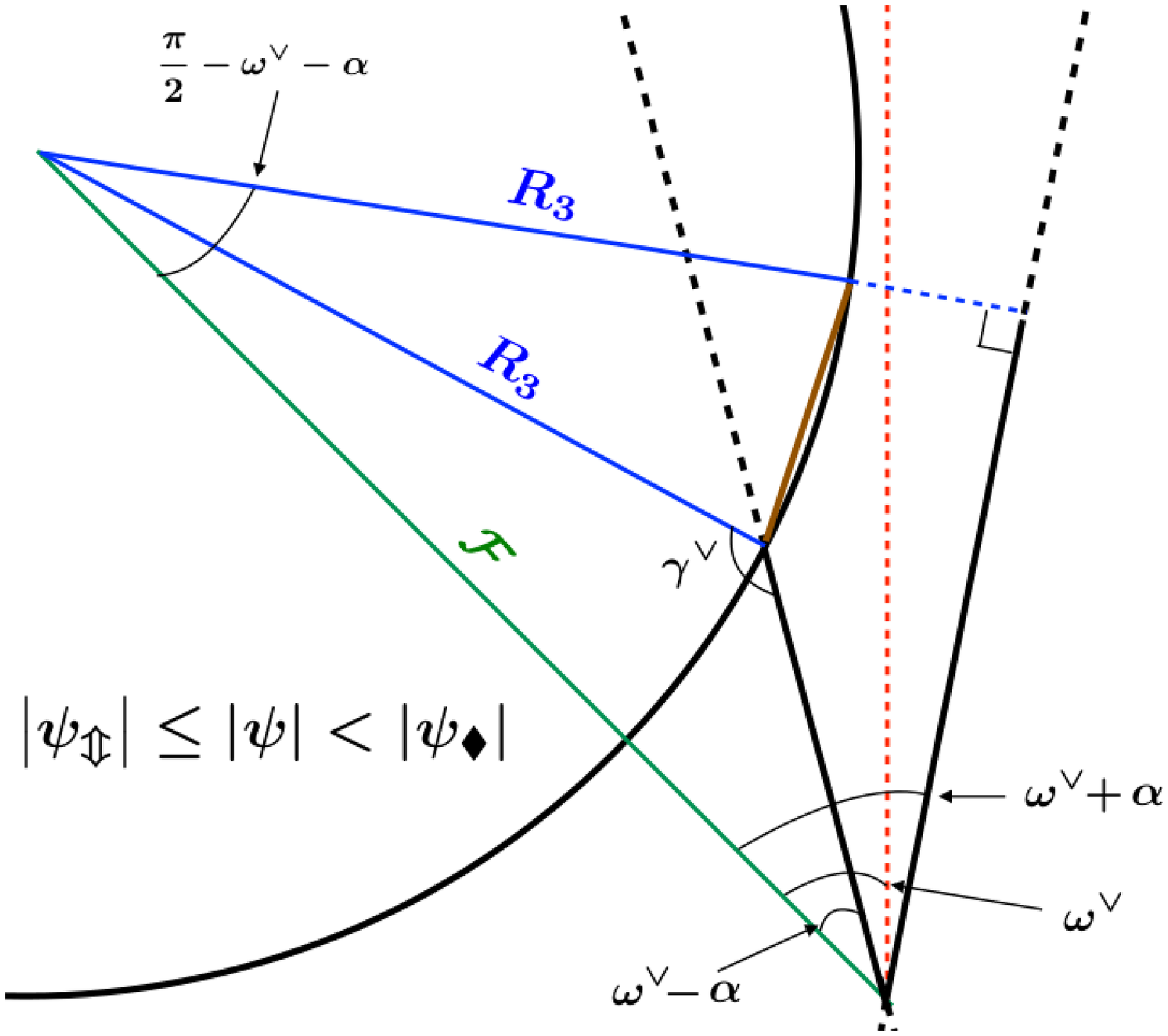}
\caption{
The antumbral shadow during ingress only or egress only, when the cone
axis does not intersect the target.
}
\label{Outant}
\end{figure}

\begin{figure}
\centering
\includegraphics[width=5cm]{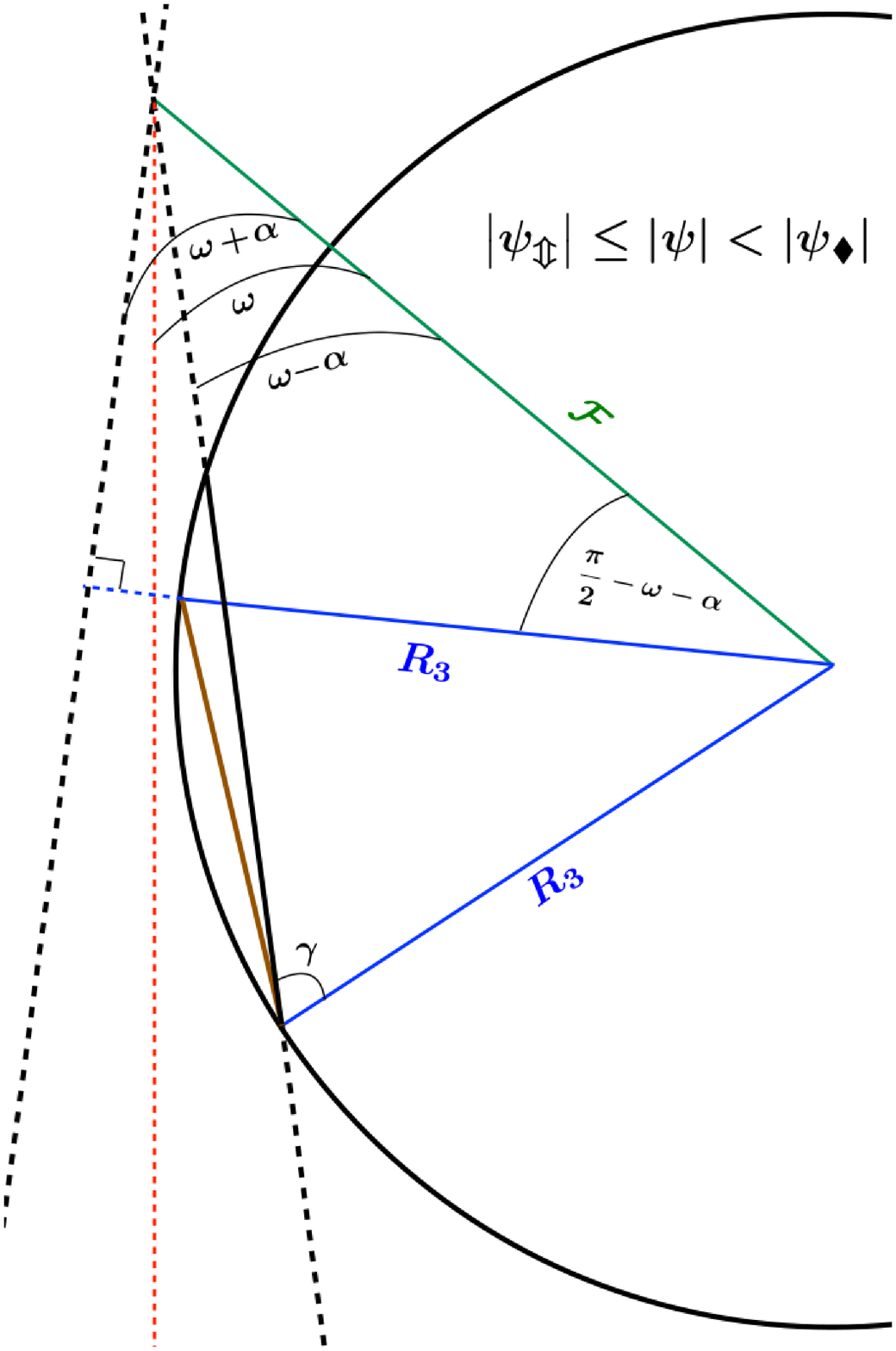}
\caption{
The umbral shadow during ingress only or egress only, when the cone
axis does not intersect the target and when the far side of the target
contains part of the cone surface-target intersection.
}
\label{Outumb}
\end{figure}

\begin{figure}
\centering
\includegraphics[width=5cm]{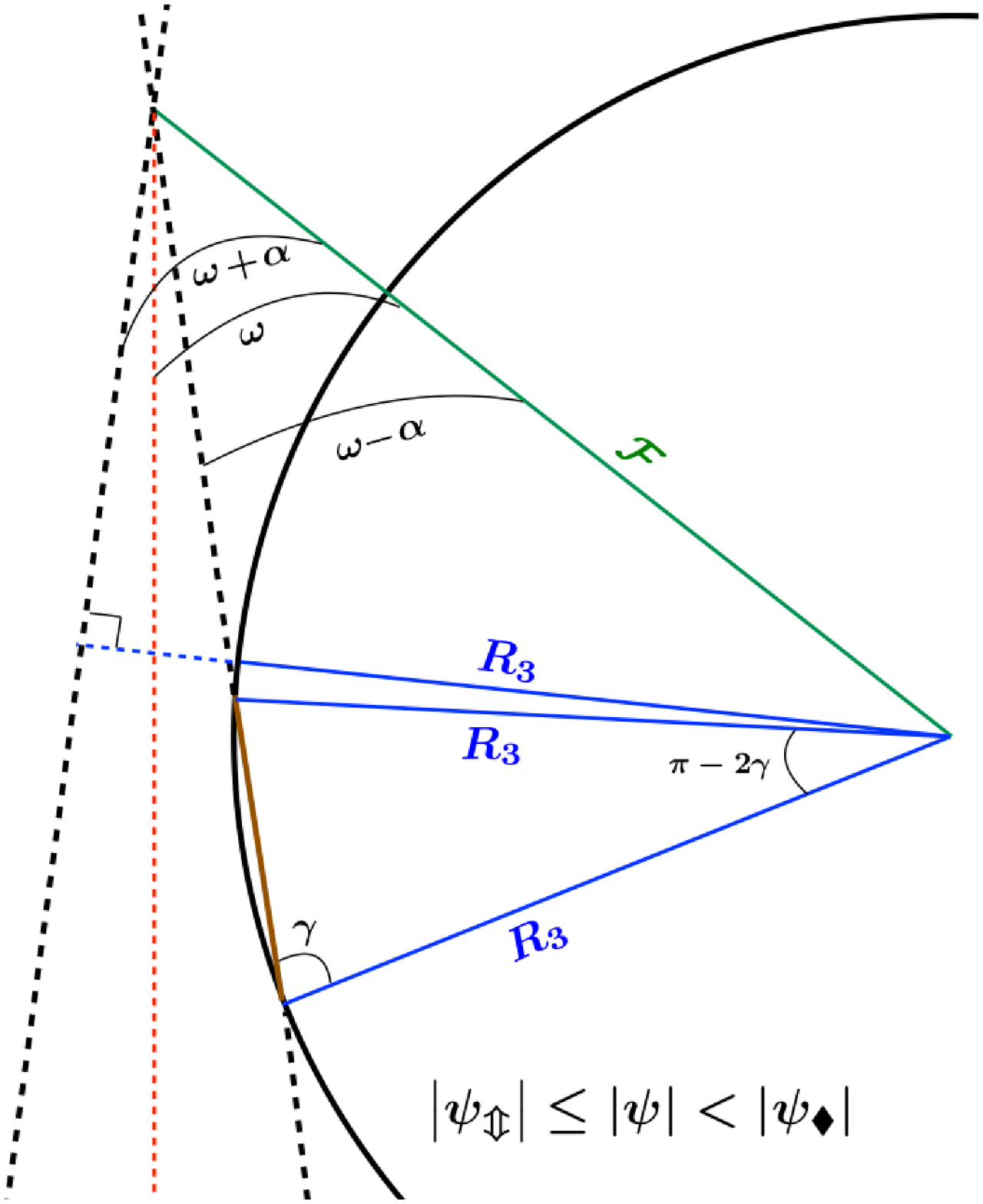}
\caption{
The umbral shadow during ingress only or egress only, when the cone
axis does not intersect the target and when only the near side of the target
contains part of the cone surface-target intersection.
}
\label{OutMostumb}
\end{figure}

As $\left|\psi\right|$ increases from $\left|\psi_{\diamondsuit}\right|$ to
$\left|\psi_{\blacklozenge}\right|$, the shadow 
will decrease in size until disappearing at $\psi_{\blacklozenge}$.
This transition comes in two parts, the first being
$\left|\psi_{\diamondsuit}\right| \le \left| \psi \right| < \left|\psi_{\Updownarrow}\right|$
and the second being
$\left|\psi_{\Updownarrow}\right| \le \left| \psi \right| < \left|\psi_{\blacklozenge}\right|$,
where $\psi_{\Updownarrow}$ is the value of $\psi$ for which the target is tangent
to the cone's axis (see Fig. \ref{bordcase}).

\subsubsection{Cone axis intersecting target}

This subsection describes the first part of this transition
($\left|\psi_{\diamondsuit}\right| \le \left| \psi \right| < \left|\psi_{\Updownarrow}\right|$), with geometries where the axis of the cone still intersects the target
(see Figs. \ref{stillGumb}-\ref{stillGant}). In this case, $R_{\rm umb}^{\rm cen}$, 
$R_{\rm ant}^{\rm cen}$, $\nu_{\rm umb}^{\rm cen}$ and $\nu_{\rm ant}^{\rm cen}$ are computed
in exactly the same way, as in equations (\ref{nuumbcen}), (\ref{nuantcen}), (\ref{Rumbcen}) 
and (\ref{Rantcen}). However, $R_{\rm umb}^{\rm edge}$ and $R_{\rm ant}^{\rm edge}$ are not, as now

\begin{equation}
R_{\rm umb}^{\rm edge} = -2R_3 \cos{\left[ \frac{1}{2} \left(\alpha + \beta + \frac{\pi}{2}  +\omega \right) \right]}
,
\end{equation}

\begin{equation}
R_{\rm ant}^{\rm edge} = 2R_3 \cos{\left[ \frac{1}{2} \left(\alpha + \beta^{\vee} + \frac{\pi}{2}  +\omega^{\vee} \right) \right]}
.
\end{equation}

\noindent{}Consequently, for $\mathcal{L}_{\rm umb}$ and $\mathcal{L}_{\rm ant}$, equations (\ref{newLumb}-\ref{newLant})
cannot be used. Instead equations (\ref{Lumb}-\ref{Lant}) must be used, along with

\begin{equation}
\nu_{\rm umb}^{\rm edge} = \pi - \frac{\beta + \frac{\pi}{2} + \omega + \alpha}{2}
,
\end{equation}

\begin{equation}
\nu_{\rm ant}^{\rm edge} = \frac{\beta^{\vee} + \frac{\pi}{2} + \omega^{\vee} + \alpha}{2}
.
\end{equation}

\noindent{}Fortuitously, the final result is exactly the same as equations (\ref{newLumb}) and (\ref{newLant})
but with $\epsilon = \epsilon^{\vee} = \pi/2$ such that

\begin{equation}
\mathcal{L}_{\rm umb} = 
2 R_3 \sin{\left[\alpha + \frac{\pi}{4} - \frac{\gamma}{2}\right]}
,
\label{Lumbpi}
\end{equation}

\begin{equation}
\mathcal{L}_{\rm ant} = 
2 R_3 \sin{\left[-\left(\alpha + \frac{\pi}{4} - \frac{\gamma^{\vee}}{2}\right)\right]}
.
\end{equation}

\subsubsection{Cone axis not intersecting target}

The next part of the transition, when the cone axis does not intersect the target 
(Figs. \ref{Outant}-\ref{OutMostumb}), occurs when 

\begin{equation}
\left|\psi_{\Updownarrow}\right|
\le 
\left| \psi \right|
<
\left|\psi_{\blacklozenge}\right|
.
\end{equation}

\noindent{}Here,

\begin{equation}
R_{\rm umb}^{\rm edge} = R_{\rm ant}^{\rm edge} = 0
,
\end{equation}

\noindent{}and $\omega$, $\omega^{\vee}$, $\gamma$ and $\gamma^{\vee}$ are still 
defined. 

For the antumbral case, the result is obtained from the geometry in Fig. \ref{Outant}, with
another fortuitous result

\[
\mathcal{L}_{\rm ant} = R_{\rm ant}^{\rm cen}
=
R_3 
\frac
{\sin{\left[ -\frac{\pi}{2} - 2 \alpha + \gamma^{\vee} \right]}  }
{\sin{\left[ \frac{3\pi}{4} +   \alpha - \frac{\gamma^{\vee}}{2} \right]}  }
\]

\begin{equation}
\ \ \ \ \ \ = 2 R_3 \sin{\left[-\left(\alpha + \frac{\pi}{4} - \frac{\gamma^{\vee}}{2}\right)\right]}
.
\end{equation}

\noindent{}Note that the shadow finally disappears when 
$\gamma^{\vee}/2 - \alpha = \pi/4$, even though
the radiation cone still intersects the target
on the ``far'' side, which would not be illuminated
anyway. 

The umbral case, however, is trickier, because two subcases
must be considered, as shown in Figs. \ref{Outumb} and \ref{OutMostumb}.
In Fig. \ref{Outumb}, where the far side of the target is contained
within the intersection, I obtain

\[
\mathcal{L}_{\rm umb} = R_{\rm umb}^{\rm cen}
=
R_3 
\frac
{\sin{\left[ \frac{\pi}{2} + 2 \alpha - \gamma \right]}  }
{\sin{\left[ \frac{\pi}{4} -   \alpha + \frac{\gamma}{2} \right]}  }
=2 R_3 \sin{\left[\alpha + \frac{\pi}{4} - \frac{\alpha}{2} \right]}
,
\]

\begin{equation}
\end{equation}

\noindent{}which is the same result as before (equation \ref{Lumbpi}). 
However, in Fig. \ref{OutMostumb}, where only the near side of
the target is contained within the intersection, I have

\begin{equation}
\mathcal{L}_{\rm umb} = R_{\rm umb}^{\rm cen} 
= 
R_3 \frac
{\sin{\left[ \pi - 2\gamma \right]}  }
{\sin{\left[ \gamma \right]}  }
=
2 R_3 \cos{\gamma}
\end{equation}

\noindent{}such that the shadow vanishes at $\gamma = \pi/2$. The transition
point between the two subcases occurs when $\gamma = \pi/2 - 2\alpha$.

\subsection{Engulfment in shadow}

When the target is sufficiently small and distant (equation \ref{engulfed}), the 
entire target may be engulfed. In this case, the size of the shadow may become
ambiguous depending on one's definition of shadow. Paper I avoided this issue
by not defining shadow surface areas at syzygy. My definition of shadow size
here provides a convenient solution. The maximum possible value of both
${\rm max}\left(\mathcal{L}_{\rm umb}\right)$ and ${\rm max}\left(\mathcal{L}_{\rm ant}\right)$
(equations \ref{maxshaumb}-\ref{maxshaant}) is $2R_3$, when for the umbra
$\gamma = 2 \alpha - \pi/2$ and for the antumbra $\gamma = 2 \alpha + 3\pi/2$.

\subsection{Summary}

All of the cases in this section can be compressed into the
following final expressions for the umbra

\[
\mathcal{L}_{\rm umb} = 2 R_3
, \ \ \ \ \ \ \ \ \ \ \ \ \ \ \ \ \ \ \ \ \ \ \ \,
\left|\psi\right| < \left|\psi_{\diamondsuit}\right| 
\]

\[
\ \ \ \ \ \ \ \ \ \ \ \ \ \ \ \ \
\ \ \ \ \ \
\ \ \ \ \ \ \ \ \
\ \ \ \ \ \ \ \ \ \ \ 
\&
\ \ \
\left|\xi_{\rm tan}\right| \ge \left|\kappa_{\rm tan}\right|
\]

\[
\ \ \ \ \ \ \ \ = 2 R_3 \sin{\left[\alpha + \frac{\epsilon}{2} - \frac{\gamma}{2} \right]}
, \ \  
\left|\psi\right| < \left|\psi_{\diamondsuit}\right| 
\]

\[
\ \ \ \ \ \ \ \ \ \ \ \ \ \ \ \ \
\ \ \ \ \ \
\ \ \ \ \ \ \ \ \
\ \ \ \ \ \ \ \ \ \ \
\&
\ \ \ 
\left|\xi_{\rm tan}\right| < \left|\kappa_{\rm tan}\right|
\]

\[
\ \ \ \ \ \ \ \, =
2 R_3 \sin{\left[\alpha + \frac{\pi}{4} - \frac{\gamma}{2} \right]}
, \ \ 
\left|\psi_{\diamondsuit} \right|
\le
\left|\psi\right|
<
\left|\psi_{\blacklozenge} \right|
\]

\[
\ \ \ \ \ \ \ \ \ \ \ \ \ \ \ \ \
\ \ \ \ \ \
\ \ \ \ \ \ \ \ \
\ \ \ \ \ \ \ \ \ \ \
\&
\ \ \ 
\gamma < \frac{\pi}{2} - 2\alpha
\]

\[
\ \ \ \ \ \ \ \, = 2 R_3 \cos{\gamma},  \ \ \ \ \ \ \ \ \ \ \ \ \ \ \ \ \ \,  
\left|\psi_{\diamondsuit}\right| 
\le
\left|\psi\right|
<
\left|\psi_{\blacklozenge} \right|
\]

\[
\ \ \ \ \ \ \ \ \ \ \ \ \ \ \ \ \
\ \ \ \ \ \ 
\ \ \ \ \ \ \ \ \
\ \ \ \ \ \ \ \ \ \ \
\&
\ \ \ 
\gamma \ge \frac{\pi}{2} - 2\alpha
\]

\[
\ \ \ \ \ \ \ \, = 0, \ \ \ \ \ \ \ \ \ \ \ \ \ \ \ \ \ \ \ \ 
\ \ \ \ \ \ \  \
\left|\psi\right| \ge \left|\psi_{\blacklozenge}\right|
,
\]

\begin{equation}
\label{Lumbfinal}
\end{equation}

\noindent{}and for the antumbra

\[
\mathcal{L}_{\rm ant} = 2 R_3
, \ \ \ \ \ \ \ \ \ \ \ \ \ \ \ \ \ \ \ \ \ \ \ \ \ \ \ \ \ 
\ \ \ \ \
\left|\psi\right| < \left|\psi_{\diamondsuit}\right| 
\]

\[
\ \ \ \ \ \ \ \ \ \ \ \ \ \ \ \ \
\ \ \ \ \ \ \ \ \ \ \ \ \ \ \ \ \
\ \ \ \ \ \ \ \ \
\ \ \ \ \ \ \ \ \ 
\&
\ \ \ 
\left|\xi_{\rm tan}\right| \ge \left|\kappa_{\rm tan}\right|
\]

\[
\ \ \ \ \ \ \, = 2 R_3 \sin{\left[-\left(\alpha + \frac{\epsilon^{\vee}}{2} - \frac{\gamma^{\vee}}{2}\right) \right]}
, \ \  
\left|\psi\right| < \left|\psi_{\diamondsuit}\right| 
\]

\[
\ \ \ \ \ \ \ \ \ \ \ \ \ \ \ \ \
\ \ \ \ \ \ \ \ \ \ \ \ \ \ \ \ \
\ \ \ \ \ \ \ \ \
\ \ \ \ \ \ \ \ \ 
\&
\ \ \ 
\left|\xi_{\rm tan}\right| < \left|\kappa_{\rm tan}\right|
\]

\[
\ \ \ \ \ \ \, = 2 R_3 \sin{\left[-\left(\alpha + \frac{\pi}{4} - \frac{\gamma^{\vee}}{2}\right) \right]}
, \ \ \ \, 
\left|\psi_{\diamondsuit}\right| \le \left|\psi\right| < \left|\psi_{\blacklozenge}\right|
\]

\[
\ \ \ \ \ \  = 0, \, \ \ \ \ \ \ \ \ \ \ \ \ \ \ \ \ \ \ \ \ \ 
\ \ \ \ \ \ \ \ \ \ \ \ \ \ \ \ \ 
\left|\psi\right| \ge \left|\psi_{\blacklozenge}\right|
.
\]

\begin{equation}
\label{Lantfinal}
\end{equation}

\noindent{}which are all independent of $\psi_{\Updownarrow}$,
$\mathcal{G}$, $\mathcal{G}^{\vee}$, 
$\beta$, and $\beta^{\vee}$.

\section{Time variation}

In Sections 3-7, I have characterised transits at specific snapshots in time. Now, I remove
that restriction, and consider {\rev the evolution of bodies moving along fixed orbits}, 
as described in Section 2.2. This assumption removes the three-body considerations
of stability and transit timing variations, which have both become substantial research
fields by themselves; {\rev in Appendix B I quantify the goodness of this approximation for
a few cases}. Nevertheless, as shown below, even the assumption of fixed orbits
is not sufficiently simple to provide explicit analytical results in nearly all cases.

I assume that the masses and radii of the primary, occulter and target are all known,
as well as some of their orbital parameters, depending on the specific case considered.
Table \ref{Tabassum} provides a list of the orbital parameters assumed for each case.

\begin{table*}
 \centering
 \begin{minipage}{180mm}
  \centering
  \caption{Orbital elements which are assumed to be given for the various cases in Section 8 and Appendix A: 1S2P $=$ One Star, Two Planets; 2S1P $=$ Two Stars, One Planet; 1M-MO $=$ One Moon, One Planet, One Star, Moon occulter; 1M-PO $=$ One Moon, One Planet, One Star, Planet occulter. The radii and masses of the primary, occulter and target are always assumed to be given.}
  \label{Tabassum}
  \begin{tabular}{@{}lllll@{}}
  \hline
Restrictions                   & 1S2P & 2S1P & 1M-MO & 1M-PO \\
   \hline
   Arbitrarily eccentric coplanar
   & $a_{12}$, $a_{13}$, $e_{12}$, $e_{13}$,
   & $a_{12}$, $a_{123}$, $e_{12}$, $e_{123}$,
   & $a_{23}$, $a_{13}$, $e_{23}$, $e_{13}$,
   & $a_{12}$, $a_{23}$, $e_{12}$, $e_{23}$,
   \\[2pt]
   & $\Pi_{12}$, $\Pi_{13}$, $\tau_{12}$, $\tau_{13}$
   & $\Pi_{12}$, $\Pi_{123}$, $\tau_{12}$, $\tau_{123}$
   & $\Pi_{23}$, $\Pi_{13}$, $\tau_{23}$, $\tau_{13}$
   & $\Pi_{12}$, $\Pi_{23}$, $\tau_{12}$, $\tau_{23}$
   \\[2pt]
   \hline
   Circular, arbitrarily inclined
   & $a_{12}$, $a_{13}$, $i_{12}$, $i_{13}$, $\Omega_{12}$,
   & $a_{12}$, $a_{123}$, $i_{12}$, $i_{123}$, $\Omega_{12}$,
   & $a_{23}$, $a_{13}$, $i_{23}$, $i_{13}$, $\Omega_{23}$, 
   & $a_{12}$, $a_{23}$, $i_{12}$, $i_{23}$, $\Omega_{12}$, 
   \\[2pt]
   & $\Omega_{13}$, $w_{12}$, $w_{13}$, $\tau_{12}$, $\tau_{13}$
   & $\Omega_{123}$, $w_{12}$, $w_{123}$, $\tau_{12}$, $\tau_{123}$
   & $\Omega_{13}$, $w_{23}$, $w_{13}$, $\tau_{23}$, $\tau_{13}$
   & $\Omega_{23}$, $w_{12}$, $w_{23}$, $\tau_{12}$, $\tau_{23}$
   \\[2pt]
   \hline
   Circular, coplanar
   & $a_{12}$, $a_{13}$, $\tau_{12}$, $\tau_{13}$
   & $a_{12}$, $a_{123}$, $\tau_{12}$, $\tau_{123}$
   & $a_{23}$, $a_{13}$, $\tau_{23}$, $\tau_{13}$
   & $a_{12}$, $a_{23}$, $\tau_{12}$, $\tau_{23}$
   \\[2pt]
   \hline
\end{tabular}
\end{minipage}
\end{table*}

My goal is to determine eclipse frequencies and durations for various
geometries. My procedure is to (i) find
$(x_{12}, y_{12}, z_{12}, x_{13}, y_{13}, z_{13})$
in terms of the given orbital elements (Table \ref{Tabassum}) and time
(equations \ref{r23}-\ref{zr}), 
(ii) then compute $\psi(t)$ from equation (\ref{psieq}),
and (iii) finally compare $\psi(t)$ to $\psi_{\blacklozenge}(t)$
(equation \ref{trancrit}) and $\psi_{\diamondsuit}(t)$
(equation \ref{engulfed}).
The complexity of the time dependence determines the ability
to carry out these tasks. 
The first task is completed in Appendix A, and the second task
is completed in this section. The third task yields an explicit
solution only in a special case, which is also presented in this section.

I perform these tasks in 12 different scenarios split according to whether the orbits are
arbitrarily eccentric and coplanar (Section 8.1), circular and arbitrarily inclined (Section 8.2),
or circular and coplanar (Section 8.3). Each of these scenarios are respectively denoted by
the superscripts of (e), (i) and (cc).

I split each of these subsections into four cases of interest. These cases {\rev are}:
(a) One Star, Two Planets (1S2P)\footnote{This case is also equivalent to two moons 
orbiting a sufficiently luminous planet, {\rev such as Phobos and Deimos orbiting Mars (when
all other solar system bodies are ignored).}}, 
(b) Two Stars, One Planet (2S1P) {\rev -- where the planet
is assumed to be a circumbinary planet --} (c) One Star, One Planet, One Moon
-- Moon occulter (1M-MO), and (d) One Star, One Planet, One Moon
-- Planet occulter (1M-PO). I will use these abbreviations throughout the section for clarity.  
This split is useful because each of these scenarios typically utilise different initial conditions,
as described in Section 2.1 and Table \ref{Tabassum}. 

{\rev These scenarios also represent known or suspected planetary systems.
The 1S2P case is particularly common. In fact, roughly 500 known planets
reside in 1S2P systems (from the Exoplanet Orbit Database, at exoplanets.org, as of October 2018). A couple dozen
planets reside in 2S1P systems, including, for example, PSR B1620-26, which features a planet orbiting 
both a millisecond pulsar and a white dwarf \citep{sigurdsson1993,thoetal1993,sigetal2003}, and Kepler-16,
which features a planet orbiting both a K-type star and an M-type star \citep{doyetal2011}. The 1M-MO and 1M-PO cases
can be represented, for example, by the Sun, Earth and Moon, when neglecting everything else in the
solar system. In exoplanetary systems, no exo-moon has yet been confirmed, although \cite{teakip2018}
presented tantalizing evidence for one in the Kepler-1625 system. The discovery of exomoons will likely
motivate additional eclipse studies.}

\subsection{Key parameters}

The key parameter in this
section, time ($t$), is propagated through the mean anomaly, 
$\mathcal{M}$, explicitly as follows:

\begin{equation}
\mathcal{M} = \mathfrak{n} \left(t - \tau\right)
.
\label{meananam}
\end{equation}

\noindent{}The variable $\tau$ is the time of pericentre passage,
a crucial parameter for determinations of transit times.
The proportionality constant is the mean motion,
$\mathfrak{n}$, which is a function of only
masses\footnote{If all of the masses are not known,
then one may
neglect $M_{\rm orbiter}$ in the computation of $\mathfrak{n}$ with a
corresponding loss of accuracy.}
and semimajor axis:

\begin{equation}
\mathfrak{n} = \sqrt{\frac{\mathfrak{G}\left(M_{\rm orbited} + M_{\rm orbiter} \right)}{a^3}}
,
\label{meanmo}
\end{equation}

\noindent{}where $\mathfrak{G}$ is the gravitational constant.

However, as demonstrated by equations (\ref{eqr}-\ref{zr}), positions
along an orbit are not generally given explicitly through the mean anomaly
$\mathcal{M}$, but rather the true anomaly $\Pi$. These anomalies are
related through Kepler's equation as

\begin{equation}
  \mathcal{M} =
\arctan{\left[
\frac{\sqrt{1-e^2}\sin{\Pi}}
{e + \cos{\Pi}}\right]}
-
\frac{e\sqrt{1-e^2} \sin{\Pi}}
     {1+e \cos{\Pi}}
     ,
\end{equation}

\noindent{}which is an implicit equation for $\Pi$ in terms of $\mathcal{M}$.
The result is that the comparison of $\psi$ to $\psi_{\blacklozenge}$ and
$\psi_{\diamondsuit}$ are treated differently in the three subsections below:

\begin{itemize}

\item  
For eccentric and coplanar orbits (Section 8.2), $\psi$,
$\psi_{\blacklozenge}$ and $\psi_{\diamondsuit}$ are all functions of
true anomaly, and hence
comparisons at each moment in time throughout the orbits
require an implicit solution for
time through Kepler's equation;

\item
For inclined and circular orbits (Section 8.3), 
$\Pi = \mathcal{M} =\mathfrak{n} \left(t - \tau\right)$.
Therefore, $\psi_{\blacklozenge}$ and $\psi_{\diamondsuit}$ 
are no longer functions of true anomaly and instead
are explicit functions of time (through $\delta$ and $K_{\rm int}$),
and $\psi$ is
a different explicit function of time, depending on architecture.
Hence Kepler's equation need not be solved, and the resulting
relations require just a single implicit solution for time.

\item
For circular and coplanar orbits (Section 8.4),
$\Pi = \mathcal{M} =\mathfrak{n} \left(t - \tau\right)$,
and the situation is the same for inclined, circular orbits
with one exception: For the 2P1S case,
$\psi_{\blacklozenge}$ and $\psi_{\diamondsuit}$ are constants,
enabling explicit solutions for time.
\end{itemize}

\subsection{Arbitrarily eccentric, coplanar orbits}

\subsubsection{Expressions for $\psi$}

By using the Cartesian elements $(x_{12}, y_{12}, z_{12}, x_{13}, y_{13}, z_{13})$
computed from the equations in
Appendix A, I now obtain expressions
for $\psi$ through equation (\ref{psieq}). The 1S2P case simplifies
to:

\begin{equation}
\psi_{\rm 1S2P}^{(\rm e)} = 
\left[\mathfrak{n}_{12} \left(t-\tau_{12} \right) 
    - \mathfrak{n}_{13} \left(t-\tau_{13} \right)
\right]
.
\label{phisimp}
\end{equation}

\noindent{}The 2S1P case does not feature such fortunate
cancellations, as
the resulting expression for $\psi$ becomes a function
of both time and true anomaly through $r_{12}(\Pi)$ and $r_{123}(\Pi)$:

\phantom{need space here}

\[
\cos{\psi_{\rm 2S1P}^{(\rm e)}} = 
\]

\[
\frac
    {
      \frac{M_1}{M_1 + M_2} r_{12,{\rm 2S1P}}^{(\rm e)}
      +
      r_{123,{\rm 2S1P}}^{(\rm e)}
      \cos{\left[
\mathfrak{n}_{12} \left(t - \tau_{12}\right)
-
\mathfrak{n}_{123} \left(t - \tau_{123}\right)
\right]}
    }
     {r_{13,{\rm 2S1P}}^{(\rm e)}}
,
\]

\begin{equation}
\label{psistar}
\end{equation}

\noindent{}where

\[
  \left(r_{13,{\rm 2S1P}}^{(\rm e)}\right)^2 =
  \left(\frac{M_1}{M_1 + M_2}\right)^2 \left(r_{12,{\rm 2S1P}}^{(\rm e)}\right)^2
  + \left(r_{123,{\rm 2S1P}}^{(\rm e)}\right)^2
\]

\[
  +
  \frac{M_1 r_{12,{\rm 2S1P}}^{(\rm e)} r_{123,{\rm 2S1P}}^{(\rm e)}
  \cos{\left[
\mathfrak{n}_{12} \left(t - \tau_{12}\right)
-
\mathfrak{n}_{123} \left(t - \tau_{123}\right)
\right]}
  }
  {M_1 + M_2}
  .
\]

\begin{equation} 
\end{equation}

\noindent{}The two cases which include moons are more compact but feature the same dependencies
on time and true anomaly through, for the moon
occulter, $r_{13}(\Pi)$ and
$r_{23}(\Pi)$:

\[
\cos{\psi_{\rm 1M-MO}^{(\rm e)}} =
\]

\[
\frac
    {r_{13,{\rm 1M-MO}}^{(\rm e)} - r_{23,{\rm 1M-MO}}^{(\rm e)}
\cos{\left[
\mathfrak{n}_{13} \left(t - \tau_{13}\right)
-
\mathfrak{n}_{23} \left(t - \tau_{23}\right)
\right]}
    }
    {r_{12,{\rm 1M-MO}}^{(\rm e)} }
,
\]

\begin{equation}
\label{psimoonMO}
\end{equation} 

\noindent{}where

\[
  \left(r_{12,{\rm 1M-MO}}^{(\rm e)}\right)^2 =
  \left(r_{13,{\rm 1M-MO}}^{(\rm e)}\right)^2
+ \left(r_{23,{\rm 1M-MO}}^{(\rm e)}\right)^2
\]

\[
  -
  2 r_{13,{\rm 1M-MO}}^{(\rm e)} r_{23,{\rm 1M-MO}}^{(\rm e)}
  \cos{\left[
\mathfrak{n}_{13} \left(t - \tau_{13}\right)
-
\mathfrak{n}_{23} \left(t - \tau_{23}\right)
\right]}
  ,
\]

\begin{equation} 
\end{equation}

\noindent{}and for the planet occulter,
$r_{12}(\Pi)$ and $r_{23}(\Pi)$:

\[
\cos{\psi_{\rm 1M-PO}^{(\rm e)}} = 
\]

\[
\frac
    {r_{12,{\rm 1M-PO}}^{(\rm e)} + r_{23,{\rm 1M-PO}}^{(\rm e)}
\cos{\left[
\mathfrak{n}_{12} \left(t - \tau_{12}\right)
-
\mathfrak{n}_{23} \left(t - \tau_{23}\right)
\right]}
    }
    {r_{13,{\rm 1M-PO}}^{(\rm e)} }
,
\]

\begin{equation}
\label{psimoonPO}
\end{equation} 

\noindent{}where

\[
  \left(r_{13,{\rm 1M-PO}}^{(\rm e)}\right)^2 =
  \left(r_{12,{\rm 1M-PO}}^{(\rm e)}\right)^2
+ \left(r_{23,{\rm 1M-PO}}^{(\rm e)}\right)^2
\]

\[
  +
  2 r_{12,{\rm 1M-PO}}^{(\rm e)} r_{23,{\rm 1M-PO}}^{(\rm e)}
  \cos{\left[
\mathfrak{n}_{12} \left(t - \tau_{12}\right)
-
\mathfrak{n}_{23} \left(t - \tau_{23}\right)
\right]}
  .
\]

\begin{equation} 
\end{equation}

\subsubsection{Solving for time}

Regardless if the angle $\psi$ is given by
equations (\ref{phisimp}), (\ref{psistar}), (\ref{psimoonMO}) or (\ref{psimoonPO}),
in the arbitrarily eccentric case, comparison with equations (\ref{trancrit})
and (\ref{engulfed})
will yield additional dependencies on true anomaly.
In these instances, Kepler's (implicit) equation must be solved at every
instance along the orbits until the angles are found to overlap.

\subsection{Circular, arbitrarily inclined orbits}

In order to aid readability, many of the expressions in this section can be simplified by utilising the following auxiliary variables

\begin{eqnarray}
\mathcal{S}_{12} &\equiv& \sin{\left[\mathfrak{n}_{12} \left(t - \tau_{12} \right) + w_{12} \right]} 
,
\label{Sfirst}
\\ 
\mathcal{S}_{13} &\equiv& \sin{\left[\mathfrak{n}_{13} \left(t - \tau_{13} \right) + w_{13} \right]} 
,
\\ 
\mathcal{S}_{23} &\equiv& \sin{\left[\mathfrak{n}_{23} \left(t - \tau_{23} \right) + w_{23} \right]} 
,
\\ 
\mathcal{S}_{123} &\equiv& \sin{\left[\mathfrak{n}_{123} \left(t - \tau_{123} \right) + w_{123} \right]}
,
\label{Slast}
\\
\mathcal{C}_{12} &\equiv& \cos{\left[\mathfrak{n}_{12} \left(t - \tau_{12} \right) + w_{12} \right]} 
,
\label{Cfirst}
\\
\mathcal{C}_{13} &\equiv& \cos{\left[\mathfrak{n}_{13} \left(t - \tau_{13} \right) + w_{13} \right]} 
,
\\
\mathcal{C}_{23} &\equiv& \cos{\left[\mathfrak{n}_{23} \left(t - \tau_{23} \right) + w_{23} \right]} 
,
\\
\mathcal{C}_{123} &\equiv& \cos{\left[\mathfrak{n}_{123} \left(t - \tau_{123} \right) + w_{123} \right]}
\label{Clast}
,
\\
\mathcal{P}_{12} &\equiv& \mathcal{C}_{12} \sin{\Omega_{12}} + \mathcal{S}_{12} \cos{\Omega_{12}} \cos{i_{12}}
\label{Pfirst}
,
\\
\mathcal{P}_{123} &\equiv& \mathcal{C}_{123} \sin{\Omega_{123}} + \mathcal{S}_{123} \cos{\Omega_{123}} \cos{i_{123}}
,
\label{Plast}
\\
\mathcal{U}_{12} &\equiv& \mathcal{C}_{12} \cos{\Omega_{12}} - \mathcal{S}_{12} \sin{\Omega_{12}} \cos{i_{12}}
,
\label{Ufirst}
\\
\mathcal{U}_{123} &\equiv& \mathcal{C}_{123} \cos{\Omega_{123}} - \mathcal{S}_{123} \sin{\Omega_{123}} \cos{i_{123}}
.
\label{Ulast}
\end{eqnarray}

Although I could simplify the above
expressions slightly by considering the plane of one of the orbits to be the reference plane,
the complete expressions can be more easily used in conjunction with given sets of orbital parameter
data.

\subsubsection{Expressions for $\psi$}

The following equations are slightly cumbersome, but are
useful for direct computation and do 
illustrate how even circularity does not allow 
for simple functions of $\psi$. Time appears in multiple
locations in each of the equations, but all explicitly,
as opposed to implicitly through $\Pi$ and $r$ as in the eccentric
case. For one star and two planets, I obtain

\[
\cos{\psi_{\rm 1S2P}^{(\rm i)}} =
\left[
\mathcal{C}_{12} \mathcal{C}_{13} + \mathcal{S}_{12} \mathcal{S}_{13} \cos{i_{12}} \cos{i_{13}}
\right]
\cos{\left[\Omega_{12} - \Omega_{13}\right]}
\]

\[
\ \ \ \ \ \ \ \ \ \ \ \ \, + \left[
\mathcal{C}_{12} \mathcal{S}_{13} \cos{i_{13}}  - \mathcal{S}_{12} \mathcal{C}_{13} \cos{i_{12}} 
\right]
\sin{\left[\Omega_{12} - \Omega_{13}\right]}
\]

\begin{equation}
\ \ \ \ \ \ \ \ \ \ \ \ \, +
\mathcal{S}_{12}  \mathcal{S}_{13} \sin{i_{12}}\sin{i_{13}}
.
\label{incpsi1st}
\end{equation}

For two stars and one planet, the mass of both stars is introduced as usual
and I obtain

\[
\cos{\psi_{\rm 2S1P}^{(\rm i)}} = 
\left[ 
\frac
{1}
{\left(M_1 + M_2\right) r_{13,{\rm 2S1P}}^{(\rm i)}}
\right]
\]

\[
\times \bigg[
M_1 a_{12} \left(\mathcal{P}_{12}^2 + \mathcal{U}_{12}^2 \right)
\]

\[
+
\left(M_1 + M_2\right) a_{123} \left(\mathcal{P}_{12} \mathcal{P}_{123} + \mathcal{U}_{12} \mathcal{U}_{123}  \right)
\]

\[
+ \mathcal{S}_{12} \sin{i_{12}} 
\left[
M_1 a_{12} \mathcal{S}_{12} \sin{i_{12}} +
\left(M_1 + M_2\right) a_{123} \mathcal{S}_{123} \sin{i_{123}}
\right]
\bigg]
,
\]

\begin{equation}
\end{equation}

\noindent{}where

\[
\left(r_{13,{\rm 2S1P}}^{(\rm i)}\right)^2
=
a_{123}^2
+
\frac{M_{1}^2 a_{12}^2}
     {\left(M_1 + M_2\right)^2}
     + 
\frac{2M_1 a_{12} a_{123}}{M_1 + M_2}
\]

\[
\times \bigg[
\left(
\mathcal{C}_{12} \mathcal{C}_{123} + \mathcal{S}_{12} \mathcal{S}_{123} \cos{i_{12}} \cos{i_{123}}
\right)
\cos{\left[\Omega_{12} - \Omega_{123}\right]}
\]

\[
\ \ + \left(
\mathcal{C}_{12} \mathcal{S}_{123} \cos{i_{123}}  - \mathcal{S}_{12} \mathcal{C}_{123} \cos{i_{12}} 
\right)
\sin{\left[\Omega_{12} - \Omega_{123}\right]}
\]

\begin{equation}
\ \ +
\mathcal{S}_{12}  \mathcal{S}_{123} \sin{i_{12}}\sin{i_{123}}
\bigg]
.
\end{equation}

The importance of inclination for the cases which include a moon are 
highlighted by Earth-Moon-Sun eclipses, where inclination plays a large
role in determining whether a system is in transit. For the moon
occulter,

\[
\cos{\psi_{\rm 1M-MO}^{(\rm i)}} =
\frac{1}{r_{12,{\rm 1M-MO}}^{(\rm i)}}
\bigg[-a_{23} \mathcal{C}_{13} \mathcal{C}_{23} \cos{\left(\Omega_{13} - \Omega_{23} \right)}
+a_{13}
\]

\[
\ \ \ - a_{23} \mathcal{S}_{13} \mathcal{S}_{23} \left(\cos{i_{13}} \cos{i_{23}} \cos{\left[\Omega_{13} - \Omega_{23}\right]} + \sin{i_{13}} \sin{i_{23}} \right)
\]

\[
\ \ \ + a_{23} \sin{\left[\Omega_{13} - \Omega_{23} \right]}
\left(  
\mathcal{C}_{23} \mathcal{S}_{13} \cos{i_{13}} - \mathcal{C}_{13} \mathcal{S}_{23} \cos{i_{23}}
\right)
\bigg]
,
\]

\begin{equation}
\end{equation}

\noindent{}with

\[
\left(r_{12,{\rm 1M-MO}}^{(\rm i)}\right)^2 =
a_{13}^2 + a_{23}^2 
\]

\[
\ \ \
+ 2a_{13} a_{23}
\bigg[-\mathcal{C}_{13} \mathcal{C}_{23} \cos{\left(\Omega_{13} - \Omega_{23} \right)}
\]

\[
\ \ \ - \mathcal{S}_{13} \mathcal{S}_{23} \left(\cos{i_{13}} \cos{i_{23}} \cos{\left[\Omega_{13} - \Omega_{23}\right]} + \sin{i_{13}} \sin{i_{23}} \right)
\]

\begin{equation}
\ \ \ + \sin{\left[\Omega_{13} - \Omega_{23} \right]}
\left(  
\mathcal{C}_{23} \mathcal{S}_{13} \cos{i_{13}} - \mathcal{C}_{13} \mathcal{S}_{23} \cos{i_{23}}
\right)
\bigg]
,
\end{equation}

\noindent{}whereas for the planet occulter,

\[
\cos{\psi_{\rm 1M-PO}^{(\rm i)}} =
\frac{1}{r_{13,{\rm 1M-PO}}^{(\rm i)}}
\bigg[a_{23} \mathcal{C}_{12} \mathcal{C}_{23} \cos{\left(\Omega_{12} - \Omega_{23} \right)}
+a_{12}
\]

\[
\ \ \ + a_{23} \mathcal{S}_{12} \mathcal{S}_{23} \left(\cos{i_{12}} \cos{i_{23}} \cos{\left[\Omega_{12} - \Omega_{23}\right]} + \sin{i_{12}} \sin{i_{23}} \right)
\]

\[
\ \ \ + a_{23} \sin{\left[\Omega_{12} - \Omega_{23} \right]}
\left(  
-\mathcal{C}_{23} \mathcal{S}_{12} \cos{i_{12}} + \mathcal{C}_{12} \mathcal{S}_{23} \cos{i_{23}}
\right)
\bigg]
,
\]

\begin{equation}
\label{incpsilast}
\end{equation}

\noindent{}with

\[
\left(r_{13,{\rm 1M-PO}}^{(\rm i)}\right)^2 =
a_{12}^2 + a_{23}^2 
\]

\[
\ \ \ 
+ 2a_{12} a_{23}
\bigg[\mathcal{C}_{12} \mathcal{C}_{23} \cos{\left(\Omega_{12} - \Omega_{23} \right)}
\]

\[
\ \ \ + \mathcal{S}_{12} \mathcal{S}_{23} \left(\cos{i_{12}} \cos{i_{23}} \cos{\left[\Omega_{12} - \Omega_{23}\right]} + \sin{i_{12}} \sin{i_{23}} \right)
\]

\begin{equation}
\ \ \ + \sin{\left[\Omega_{12} - \Omega_{23} \right]}
\left(  
-\mathcal{C}_{23} \mathcal{S}_{12} \cos{i_{12}} + \mathcal{C}_{12} \mathcal{S}_{23} \cos{i_{23}}
\right)
\bigg]
.
\end{equation}

\subsubsection{Solving for time}

For circular orbits, Kepler's equation need not be solved in 
order to determine transit durations and frequencies.
All which is required is an implicit solution for time in the relations
between $\psi$ and $\psi_{\blacklozenge}$ (and $\psi_{\diamondsuit}$).
Then the answer in terms of transit durations and frequencies is immediate.

\subsection{Circular, coplanar orbits}

In the simplest, but often representative, case of circular coplanar orbits,
the equations are simpler, and for the 1S2P case, I obtain explicit closed-form solutions.

\subsubsection{Expressions for $\psi$}

I find

\begin{equation}
\psi_{\rm 1S2P}^{(\rm cc)} = \psi_{\rm 1S2P}^{(\rm e)}
\end{equation}

\noindent{}and

\[
\cos{\psi_{\rm 2S1P}^{(\rm cc)}}  =  
\left(
\frac{1}
{r_{13,{\rm 1S2P}}^{({\rm cc})}}
\right)
\bigg[
M_1 a_{12} + \left(M_1 + M_2\right) 
\]

\[
\ \ \ \ \ \ \ \ \ \ \times
a_{123} \cos{\left[\mathfrak{n}_{12} \left(t - \tau_{12} \right) - \mathfrak{n}_{123} \left(t - \tau_{123} \right)  \right]}
\bigg]
,
\]

\begin{equation}
\end{equation}

\noindent{}with

\[
\left( r_{13,{\rm 1S2P}}^{({\rm cc})} \right)^2 =
M_{1}^2 a_{12}^2 + \left(M_1 + M_2\right)^2 a_{123}^2
\]

\[
+ 2 a_{12} a_{123} M_1 \left(M_1 + M_2\right)
\cos{\left[\mathfrak{n}_{12} \left(t - \tau_{12} \right) - \mathfrak{n}_{123} \left(t - \tau_{123} \right)  \right]}
.
\]

\begin{equation}
\end{equation}

\noindent{}For the moon occulter,

\[
\cos{\psi_{\rm 1M-MO}^{(\rm cc)}}  =
\]

\begin{equation}
\frac
{a_{13} - a_{23} \cos{\left[\mathfrak{n}_{13} \left(t - \tau_{13} \right) - \mathfrak{n}_{23} \left(t - \tau_{23} \right)  \right]}  }
{\sqrt{a_{13}^2 + a_{23}^2 - 2 a_{13}a_{23}\cos{\left[\mathfrak{n}_{13} \left(t - \tau_{13} \right) - \mathfrak{n}_{23} \left(t - \tau_{23} \right)  \right]}} }
,
\end{equation}

\noindent{}whereas for the planet occulter,

\[
\cos{\psi_{\rm 1M-PO}^{(\rm cc)}}  =
\]

\begin{equation}
\frac
{a_{12} + a_{23} \cos{\left[\mathfrak{n}_{12} \left(t - \tau_{12} \right) - \mathfrak{n}_{23} \left(t - \tau_{23} \right)  \right]}  }
{\sqrt{a_{12}^2 + a_{23}^2 + 2 a_{12}a_{23}\cos{\left[\mathfrak{n}_{12} \left(t - \tau_{12} \right) - \mathfrak{n}_{23} \left(t - \tau_{23} \right)  \right]}} }
.
\end{equation}

\subsubsection{Solving for time}

Solving for time does not require the solution
of Kepler's equation, and time appears
in the above expressions for $\psi$ in just a single
cosine argument. However, the time dependencies of
$\psi_{\blacklozenge}$ and $\psi_{\diamondsuit}$ are too complex
for explicit solutions, except in the 1S2P case.

In the 1S2P case,
$r_{12}(t) = a_{12}$ and $r_{13}(t) = a_{13}$.
Consequently umbral and antumbral transits begin according
to equation (\ref{trancrit}) such that
the $p$th transit starts at the following time 

\begin{equation}
\mathfrak{s}_{{\rm 1S2P}}^{(p)} = \frac{\mathfrak{n}_{12} \tau_{12} - \mathfrak{n}_{13}\tau_{13}
+ 2 \pi \left(p-1\right) - \left|\psi_{\blacklozenge}\right|}
{\mathfrak{n}_{12} - \mathfrak{n}_{13}}
\label{sta1S2P}
\end{equation}

\noindent{}and ends at

\begin{equation}
\mathfrak{e}_{{\rm 1S2P}}^{(p)} = \frac{\mathfrak{n}_{12} \tau_{12} - \mathfrak{n}_{13}\tau_{13}
+ 2 \pi \left(p-1\right) + \left|\psi_{\blacklozenge}\right|}
{\mathfrak{n}_{12} - \mathfrak{n}_{13}}
\label{end1S2P}
\end{equation}

\noindent{}The duration of each transit is then

\begin{equation}
\mathfrak{d}_{{\rm 1S2P}}^{(p)} =
\mathfrak{e}_{{\rm 1S2P}}^{(p)} - \mathfrak{s}_{{\rm 1S2P}}^{(p)}
= 
\frac
{2 \left|\psi_{\blacklozenge}\right|}
{\mathfrak{n}_{12} - \mathfrak{n}_{13}}
,
\label{dur1S2P}
\end{equation}

\noindent{}the duration of ingress and egress is

\begin{equation}
\mathfrak{D}_{{\rm 1S2P}}^{(p)} =
\frac
{\left|\psi_{\blacklozenge}\right| - \left|\psi_{\diamondsuit}\right|}
{\mathfrak{n}_{12} - \mathfrak{n}_{13}}
\label{ing1S2P}
\end{equation}

\noindent{}and the frequency of transits is given by

\begin{equation}
\mathfrak{f}_{\rm 1S2P} 
= 
\mathfrak{s}_{{\rm 1S2P}}^{(p+1)} - \mathfrak{s}_{{\rm 1S2P}}^{(p)}
=  \frac{2\pi}{\mathfrak{n}_{12} - \mathfrak{n}_{13}}
.
\label{fre1S2P}
\end{equation}

\section{Partial eclipses}

\begin{figure}
\ \ \ \ \ \ \ \ \ \ \ \ \ \ \ 
\includegraphics[width=5cm]{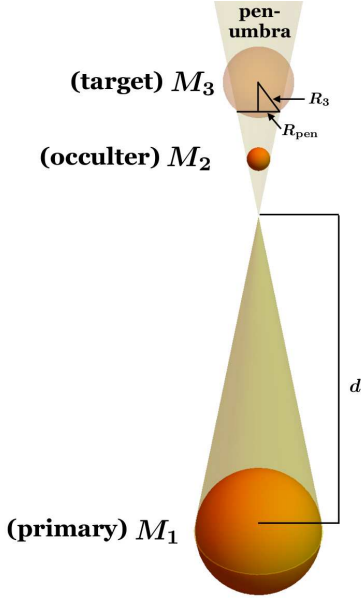}   
\caption{
{\rev 
The radiation cone producing the penumbral shadow, with the target
at syzygy. Comparison with Fig. \ref{difftarget} illustrates the difference
in geometry with the umbral and antumbral cases. This figure is a 
reproduction of figure 2 of Paper I.
}
} 
\label{cart2}
\end{figure}

The entire paper so far has focused on total and annular eclipses. 
However, another type of eclipse -- a partial eclipse -- always 
accompanies a total or annular eclipse. A partial eclipse is
formed through the internal tangent lines (as opposed to the external
tangent lines) between the primary and occulter {\rev (see Fig. \ref{cart2},
which is reproduced from figure 2 of Paper I)}.

The geometry of partial eclipses is akin to that of annular eclipses
because the target intersects the upper nappe of the radiation cone.
The similarities between these two types of eclipses of syzygy were
highlighted in Appendix B of Paper I. There it was revealed that 
several of the quantities for the penumbral cone could be reproduced
by replacing $R_2$ with $-R_2$ in the corresponding antumbral cone.

I use this symmetry here to great effect. Rather than rework all of the
results so far with this different cone, all I need to do is define

\begin{equation}
\Delta \equiv \frac{R_1 + R_2}{r_{12}}
,
\label{Deltaequ}
\end{equation}

\noindent{}and then replace $\delta$ with $\Delta$ in all of the
relevant equations.

\subsection{Condition to be in transit}

Replacing $\delta$ with $\Delta$ in the equation of the intersection
(equations \ref{inteq}-\ref{Jinteq}) is valid, as can be shown by
establishing the equation of the penumbral radiation cone and expressing
it as a quadric. Hence, the
conditions to be in transit and fully engulfed in the shadow 
(equations \ref{trancrit} and \ref{engulfed}) are
the same except for $\delta$ being replaced by $\Delta$.

Whether or not the system is in transit with respect to the
penumbral shadow is of particular interest because observers
on the target will first start to see the primary being
obscured during a partial (not total nor annular) eclipse.

\subsection{Shape of the shadow}

For the penumbral case, the ranks of the matrices in
equations (\ref{rankA}) and (\ref{Bmatrix}) are still three
and one, respectively. 
Consequently,
the penumbral shadow is always a parabolic cylinder, like
both the umbral and antumbral shadows.

\subsection{Size of the shadow}

In the expressions for the size of the antumbral shadow, $\delta$ appears
through $\mathcal{F}$ (equation \ref{calFeq}), $\mathcal{G}^{\vee}$ (equation \ref{calGeqvee})
and $\omega^{\vee}$ (equations \ref{omegaeq}-\ref{omegaveeeq}). Replacing
$\delta$ with $\Delta$ accounts for all of the differences between the penumbral
and antumbral shadows.

\subsection{Time evolution}

The starting and ending times of transits for the penumbral shadow will be
different from the antumbral or umbral cases because the penumbral shadow is much
larger.

The difference in these times can be computed explicitly for the circular, coplanar
case of two planets and one star. In order to compute $\mathfrak{s}$, $\mathfrak{e}$ and $\mathfrak{d}$ for the penumbral case,
just replace $\delta$ with $\Delta$. Then, for example, an observer on a planetary target
who just sees the primary start to be obscured by another planet must wait for a time equal to

\begin{equation}
\frac
{
\left|\cos{\psi_{\blacklozenge}}\left(\delta \rightarrow \Delta \right)\right|
-  
\left|\cos{\psi_{\blacklozenge}}\left(\delta \right)\right|
}
{\mathfrak{n}_{12}-\mathfrak{n}_{13}}
,
\end{equation}

\noindent{}before being able to witness the start of the total or annular eclipse.

\section{External view of a syzygy}

Having considered the three-body case in some detail, I now finish
the main text of the paper by exploring an extension to the four-body
case under the same formalism. In order to keep this analysis manageable, 
throughout the section at least
the primary, occulter and target are assumed to be in syzygy.
My analysis here could be particularly relevant to mutual occultation events
between exoplanets as viewed from Earth, as recently described by
\cite{lugetal2017}, or, for example, to observing both Phobos and Deimos from the
surface of Mars when both moons are in syzygy with the Sun.
 
A key initial question is, does the occulter or the target block the primary's 
light from reaching the external body? The answer depends on the relative sizes and distances
of the bodies.
I denote the external body as the \#4 body, 
and $r_{14}$ as the centre-to-centre 
distance between that body and the primary. 

\subsection{Co-linear syzygy}

\begin{figure}
\ \ \ \ \ \ \ \ \ \ \ \ \ \ \ \ \ \ \ \ \
\includegraphics[height=8.0cm]{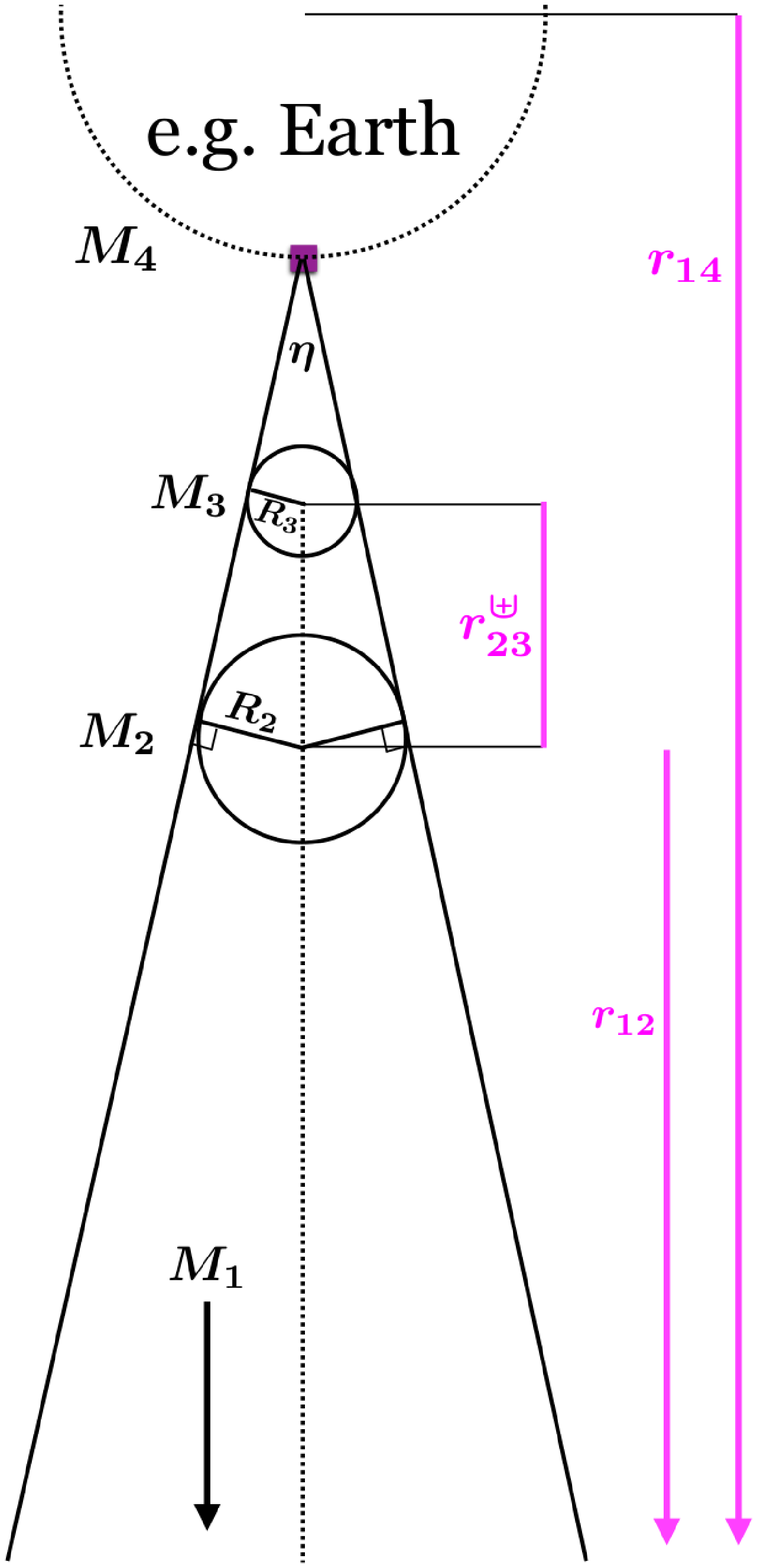}
\caption{
A snapshot of an observer (square) standing on an object ($M_4$; such as the Earth) which is
external to but co-linear with a syzygy {\rev (e.g. two exoplanets as $M_2$ and $M_3$ and
their host star as $M_1$, or two solar system planets $M_2$ and $M_3$ with the Sun as $M_1$)}. 
Shown in this diagram is the limiting
case in which both the occulter and target produce the same-sized shadow on the disk
of the primary in sky. For $r_{23} > r_{23}^{\uplus}$, the target would produce the shadow,
whereas for $r_{23} < r_{23}^{\uplus}$, the occulter would produce the shadow. 
Equations (\ref{ruplusratio}-\ref{etasha}) are derived from this diagram.
}
\label{EarthLim1}
\end{figure}

If the observer on the external body is co-linear with the syzygy, 
then Fig. \ref{EarthLim1}
reveals (given a fixed $r_{12}$) the critical distance
$r_{23}^{\uplus}$ beyond which ($r_{23} > r_{23}^{\uplus}$)
the target, rather than the occulter, is responsible for blocking
the primary's starlight. The geometry of similar triangles illustrates

\begin{equation}
\frac{R_2}{r_{14} - R_4 - r_{12}}
= \frac{R_3}{r_{14} - R_4 - r_{12} - r_{23}^{\uplus}}
\label{ruplusratio}
\end{equation}

\noindent{}which reduces to

\begin{equation}
r_{23}^{\uplus} = \frac{R_2 - R_3}{R_2} \left(r_{14} - R_4 - r_{12}\right)
,
\label{ruplus}
\end{equation}

\noindent{}an expression which is independent of $R_1$.

Let the angular diameter that the observer (or observatory) sees be $\eta$
such that the angular diameter of the shadow on the primary is $\eta_{\rm sha}$.
Then

\[
\eta_{\rm sha} = 2 \, {\rm min}\left[\sin^{-1}\left(\frac{R_3}{r_{14}-r_{13}-R_4}\right),
  \ \sin^{-1}\left(\frac{R_1}{r_{14}-R_4}\right)\right],
\]

\[
\ \ \ \ r_{23} \ge r_{23}^{\uplus}
\]

\[
  \eta_{\rm sha} = 2 \, {\rm min}\left[\sin^{-1}\left(\frac{R_2}{r_{14}-r_{12}-R_4}\right),
    \ \sin^{-1}\left(\frac{R_1}{r_{14}-R_4}\right)\right],
  \]

\begin{equation}  
  \ \ \ \ r_{23} < r_{23}^{\uplus}.
\label{etasha}
\end{equation}

\subsection{Offset syzygy}

Now consider the case when the external body is not co-linear with the syzygy
-- but still co-planar with it -- and offset from the syzygy by a perpendicular
distance $k>0$. Note that
in this case, the distance $r_{14}$ is not parallel to the syzygy. 

\subsubsection{No eclipse}

\begin{figure}
\includegraphics[height=8.0cm]{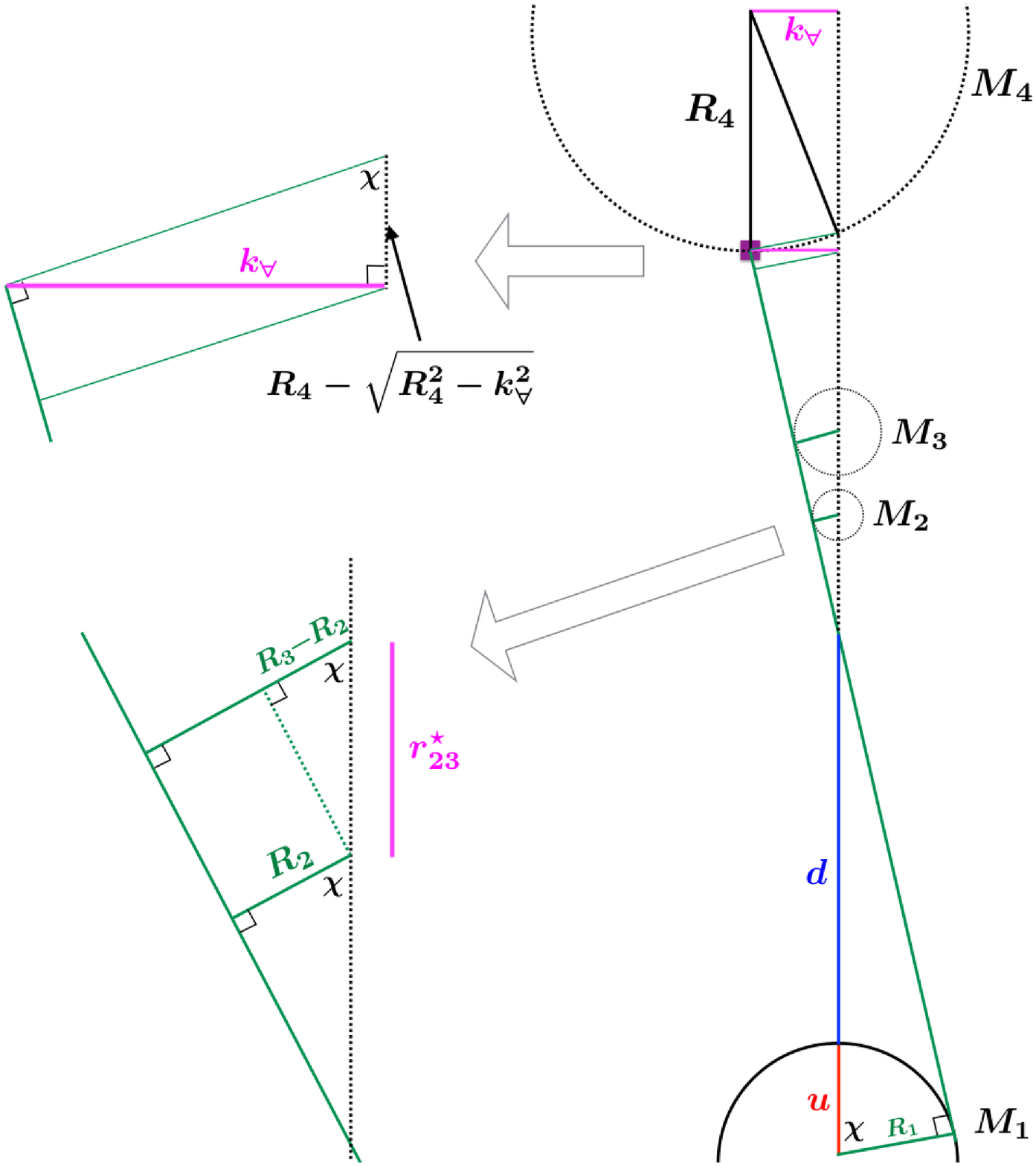}
\caption{
A snapshot of an observer (square) standing on an object ($M_4$) which 
is external to, offset from, and coplanar
with a syzygy. This diagram illustrates
the limiting case where both the occulter and target remain just undetectable 
(at $k = k_{\forall}$ and $r_{23} = r_{23}^*$).
Equations (\ref{earthfrac1}-\ref{kall}) are derived from this diagram.
}
\label{EarthLim2}
\end{figure}

I first consider the limiting value $k_{\forall}$ for which an observer would 
not see any eclipse at all (neither from the occulter, target, nor both). Fig. 
\ref{EarthLim2} illustrates the geometry for this limiting value. The images
on the left side of the figure are zoomed-in, angularly exaggerated
portions of the overall geometry.
I find

\[
\cos{\chi} = 
\frac{R_1}{d+u} 
= \frac{R_{3}-R_2}{r_{23}^*}
= \frac{R_4 - \sqrt{R_{4}^2-k_{\forall}^2}}{\sqrt{\left(R_4 - \sqrt{R_{4}^2-k_{\forall}^2}\right)^2+k_{\forall}^2}}
\]

\begin{equation}
\label{earthfrac1}
\end{equation}

\noindent{}which gives

\begin{equation}
k_{\forall} = \frac{2 R_4 \left(R_1 + R_2\right)}{r_{12}^2}
\sqrt{r_{12}^2 - \left(R_1 + R_2\right)^2}
\label{kall0}
\end{equation}

\begin{equation}
\ \ \ \  = \frac{2 R_4 \left(R_2 - R_{3}^*\right)}{r_{23}^2}
\sqrt{r_{23}^2 - \left(R_2 - R_{3}^*\right)^2}
.
\label{kall}
\end{equation}

\noindent{}Here, $d$ and $u$ are the penumbral equivalents to $h$ and $n$ from Paper I, and $r_{23}^*$ is the critical
value of $r_{23}$ beyond which the target is engulfed in the penumbral shadow (equation B8 of Paper I).
I conclude here that when $k \ge k_{\forall}$ at $R_3 = R_{3}^*$, or when $R_3 \le R_{3}^*$ at $k = k_{\forall}$, then the observer cannot see any type of eclipse.

\subsubsection{Some eclipse}

\begin{figure}
\includegraphics[height=8.0cm]{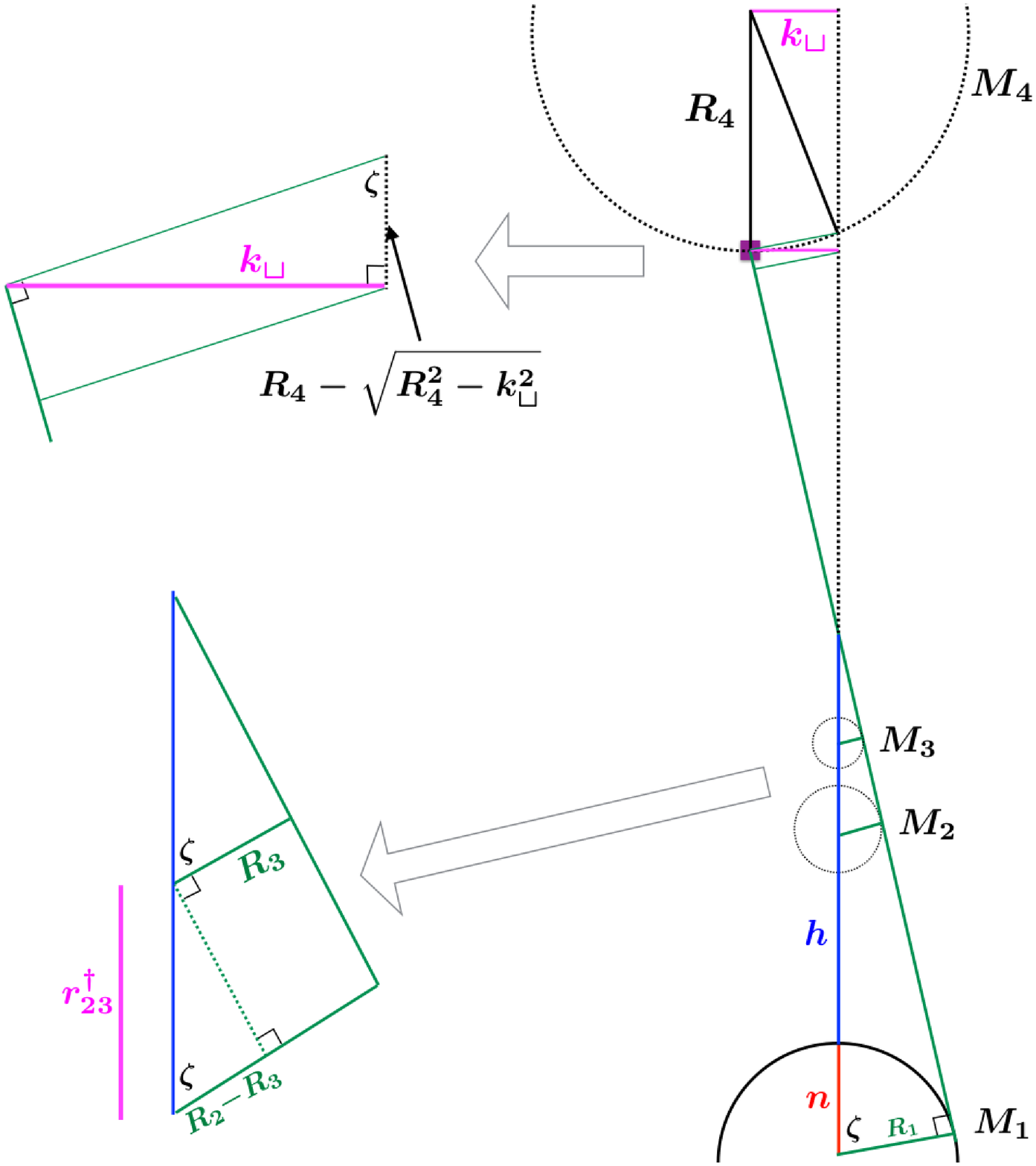}
\caption{
A snapshot of an observer (square) standing on an object ($M_4$) which is
external to, offset from, and coplanar with an syzygy. This diagram illustrates
the limiting case where both the occulter and target appear just fully inside of the
disc on the side of the offset (at $k = k_{\sqcup}$ and $r_{23} = r_{23}^{\dagger}$). 
Equations (\ref{earthfrac2}-\ref{ksqcup2}) are derived from this diagram.
}
\label{EarthLim3}
\end{figure}

\begin{figure}
\includegraphics[height=8.0cm]{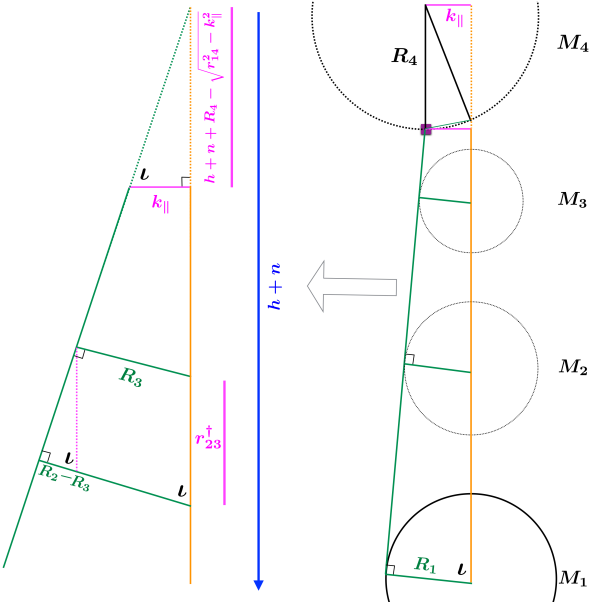}
\caption{
A snapshot of an observer (square) standing on an object ($M_4$) which is
external to, offset from, and coplanar with an syzygy. This diagram illustrates
the limiting case where both the occulter and target appear just fully inside of the
disc on the side opposite the offset (at $k = k_{\parallel}$ and $r_{23} = r_{23}^{\dagger}$). 
Equations (\ref{earthfrac3}-\ref{kparallel}) are derived from this diagram.
}
\label{EarthLim4}
\end{figure}

In contrast, when $k < k_{\forall}$ at $R_3 = R_{3}^*$, or when 
$R_3 > R_{3}^*$ at $k = k_{\forall}$,
then the observer will see some
type of eclipse. Here, I compute two limiting values, 
when the target and occulter are entirely within the field of view on the side 
of the offset (Fig. \ref{EarthLim3}) and on the opposite side 
(Fig. \ref{EarthLim4}). I denote these limiting values
as $k_{\sqcup}$ and $k_{\parallel}$, respectively, and note that the
later can occur only when $k < R_1$ and either $k < R_2$ or $k < R_3$.

The geometry from Fig. \ref{EarthLim3} reveals

\[
\cos{\zeta} = 
\frac{R_1}{h+n} 
= \frac{R_2-R_3}{r_{23}^{\dagger}}
= \frac{R_4 - \sqrt{R_{4}^2-k_{\sqcup}^2}}{\sqrt{\left(R_4 - \sqrt{R_{4}^2-k_{\sqcup}^2}\right)^2+k_{\sqcup}^2}}
\]

\begin{equation}
\label{earthfrac2}
\end{equation}

\noindent{}which yields

\begin{equation}
k_{\sqcup} = \frac{2 R_4 \left(R_1 - R_2\right)}{r_{12}^2}
\sqrt{r_{12}^2 - \left(R_1 - R_2\right)^2}
\label{ksqcup}
\end{equation}

\begin{equation}
\ \ \ \  = \frac{2 R_4 \left(R_2 - R_{3}^{\dagger}\right)}{r_{23}^2}
\sqrt{r_{23}^2 - \left(R_2 - R_{3}^{\dagger}\right)^2}
\label{ksqcup2}
\end{equation}

\noindent{}and the geometry from Fig. \ref{EarthLim4} reveals

\[
\cos{\iota} = 
\frac{R_1}{h+n} 
= \frac{R_2-R_3}{r_{23}^{\dagger}}
\]

\begin{equation}
\ \ \ \ \ \ \ = \frac{k_{\parallel}}{\sqrt{\left(h + n +  R_4 - \sqrt{r_{14}^2-k_{\parallel}^2}\right)^2+k_{\parallel}^2}}
\label{earthfrac3}
\end{equation}

\noindent{}giving

\[
k_{\parallel}^2 = \left(h+n+R_4\right)^2 \left[\left(\frac{R_1}{h+n} \right)^2 - 2 \left(\frac{R_1}{h+n} \right)^4 \right]
\]

\[
+ \left(\frac{R_1r_{14}}{h+n} \right)^2 - \frac{2R_{1}^2 \left(h + n + R_4\right)}{\left(h+n\right)^4}
\bigg\lbrace \left[R_{1}^2 - \left(h+n\right)^2 \right]
\]

\begin{equation}
\times 
  \left[\left(h+n\right)^2 \left(R_{1}^2 - r_{14}^2\right) 
            +     R_4 R_{1}^2 \left[2 \left(h+n\right) + R_4\right] \right]
\bigg\rbrace^{\frac{1}{2}}
.
\label{kparallel}
\end{equation}

Consequently, the side of the primary which is in the same direction of the offset will
be obscured if $k_{\sqcup} < k < k_{\forall}$. In order for the side of the primary
opposite the offset to be obscured, then $k < k_{\parallel}$.

\subsubsection{Angular diameters}

For an offset syzygy, the angular diameters of the primary, occulter and target could all be computed
in similar ways, by analogy with the offset observer from Fig. A3 of Paper I.
The left panel of Fig. \ref{EarthLim5} illustrates the relevant geometry for an arbitrary value
of $k$, and yields

\begin{equation}
\eta_{14} = 2 \sin^{-1} \left[\frac{R_1}{\sqrt{k^2 + \left(\sqrt{r_{14}^2 - k^2} - R_4\right)^2}} \right]
.
\label{eta14arbk}
\end{equation}

\noindent{}Similar formulae hold for $\eta_{24}$ and $\eta_{34}$. 

\subsubsection{Double annular eclipses}

Much trickier is the prospect of
computing $\eta_{\rm sha}$ by combining the angular diameters of the three objects
in syzygy when they overlap in the sky.
I do not delve into the details of various configurations, except for the interesting
case when both the occulter and target are simultaneously 
visible as distinct discs at some point during eclipse. This situation is, in effect,
a double annular eclipse. 

\begin{figure}
\includegraphics[height=8.0cm]{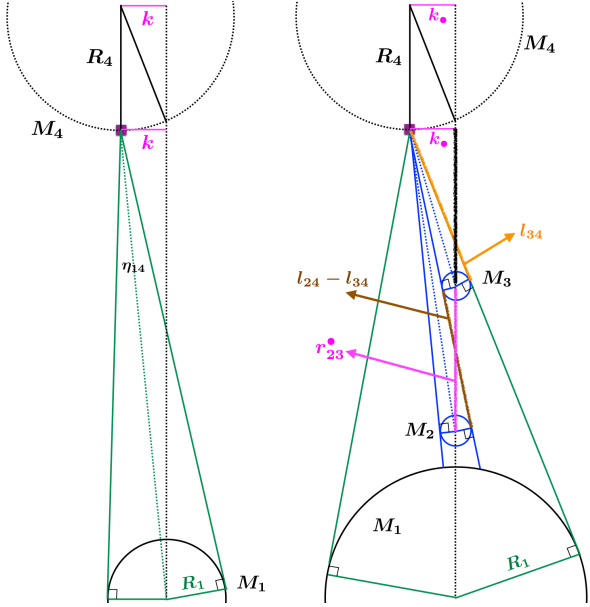}
\caption{
A snapshot of an observer (square) standing on an object ($M_4$) which is
external to, offset from, but coplanar with an syzygy. The left-hand panel
illustrates how the angular diameter for one of the objects (the primary here)
can be derived. The right-hand panel illustrates
the limiting case where both the occulter and target appear 
entirely, simultaneously and non-overlapping on the disc of the primary in the
sky (for $k \ge k_{\bullet}$ and $r_{23} \ge r_{23}^{\bullet}$).
Equations (\ref{eta14arbk}-\ref{pythag}) are derived from this diagram.
}
\label{EarthLim5}
\end{figure}

To determine limiting values of $k$ and $r_{23}$ (denoted as $k_{\bullet}$ and $r_{23}^{\bullet}$) 
which can produce such a configuration,
consider the right panel of Fig. \ref{EarthLim5}. First I helpfully define
the length of the line extending from the observer to either tangent point of 
any of the bodies in syzygy as

\begin{eqnarray}
l_{14} &\equiv& \sqrt{\left(\sqrt{r_{14}^2 - k_{\bullet}^2} - R_4 \right)^2 + k_{\bullet}^2 - R_{1}^2}
,
\label{l14eq}
\\
l_{24} &\equiv& \sqrt{\left[\sqrt{r_{14}^2 - k_{\bullet}^2} - R_4  - r_{12} \right]^2 + k_{\bullet}^2 - R_{2}^2 }
,
\\
l_{34} &\equiv& \sqrt{\left[\sqrt{r_{14}^2 - k_{\bullet}^2} - R_4 - r_{13} \right]^2 + k_{\bullet}^2 - R_{3}^2}
.
\label{l34eq}
\end{eqnarray}

Next, consider the two triangles formed by the orange, black, blue and lower horizontal
magenta lines
such that the bottom triangle partially intersects $M_3$.
The lengths of the orange line is $l_{34}$, the magenta line is $k_{\bullet}$ and the blue
line is $R_3$. The lower part of the orange line (in the bottom triangle) is then given by

\begin{equation}
l_{34}^{(\rm lower)} = \left(\frac{R_3}{k_{\bullet}^2 - R_{3}^2} \right)
                    \left[k_{\bullet} \sqrt{l_{34}^2 + R_{3}^2 - k_{\bullet}^2} - R_3 l_{34} \right]
.
\label{l34low}                    
\end{equation}

\noindent{}Now I obtain a relation based on similar triangles: one of which being the bottom of the
aforementioned triangles, and the other being the downward extension of that triangle to the primary. That comparison yields

\begin{equation}
  \frac{l_{34}^{(\rm lower)}}{R_3} = \frac{l_{14} - l_{34} + l_{34}^{(\rm lower)}}{R_1}
  ,
\label{l34final}
\end{equation}

\noindent{}which is an implicit equation for $k_{\bullet}$, and for which $k_{\bullet}$ can be numerically determined. 

In order to relate
$k_{\bullet}$ to $r_{23}^{\bullet}$, consider the brown line, which has length $l_{24} - l_{34}$, and the two
triangles formed by its intersection with the vertical magenta line of length $r_{23}^{\bullet}$.
The upper part of this magenta line has length $\left[R_3 r_{23}^{\bullet} /\left(R_2 + R_3\right)\right]$
and the upper part of the brown line has length $\left[R_3\left(l_{24} - l_{34}\right)/\left(R_2 + R_3\right)\right]$.
The Pythagorean theorem then gives the following relation:

\begin{equation}
\left(r_{23}^{\bullet}\right)^2 =
\left(R_2 + R_3\right)^2 
+ \left(l_{24} - l_{34}\right)^2
.
\label{pythag}
\end{equation}

\subsubsection{Observing solar system syzygys}

Sometimes, the Sun, Mercury, Venus and Earth will all be in syzygy.
In this case, equation (\ref{ruplus}) reveals that $r_{23}^{\uplus} < 0$,
which indicates that Venus -- and not Mercury -- will block the Sun's light
from a viewer on Earth. If the Earth is coplanar to but
not co-linear with the syzygy
formed by the Sun, Mercury and Venus, then I consider the quantities
$k_{\forall}$, $k_{\sqcup}$ and $k_{\parallel}$ from equations (\ref{kall}),
(\ref{ksqcup}) and (\ref{kparallel}). I find $k_{\parallel} > R_1$,
and so the situation in Fig. \ref{EarthLim4} cannot occur.
Also, $k_{\forall}$~$=$~$153.56$ km and $k_{\sqcup} = 152.49$ km,
meaning that an observer on Earth would have to be within
$153.56$ km of the syzygy in order to see it, and can see just one
side of the Sun being blocked only at an offset which is in-between
152.49 km and 153.56 km.

\subsubsection{Observing extrasolar system syzygys}

In extrasolar planetary systems like Kepler-47 (featured in Paper I), 
with two stars {\rev (Kepler-47A and Kepler-47B)} on a tight orbit 
and two planets {\rev (Kepler-47ABb and Kepler-47ABc)}, syzygys might be common
amongst three, or even all four bodies in the system. In a four-body syzygy, 
an observatory on the outer planet {\rev (Kepler-47ABc)} will see light blocked from
the primary due to the companion star, and not due to the inner planet, because
$r_{23}^{\uplus} > r_{23}$. If the outer planet is offset from but still coplanar
with the syzygy of the other
three bodies\footnote{{\rev As of 29 October 2018, the inclination constraints on both planets
from the Exoplanet Data Explorer (at exoplanets.org) are still consistent with the possibility of coplanarity.}}, 
then $k_{\forall}$~$=$~$4270$~km and $k_{\sqcup}$~$=$~$2000$~km.
If $k$~$<$~$k_{\forall}$, then the observatory will detect the syzygy.
If $k_{\sqcup}$~$<$~$k$~$<$~$k_{\forall}$, then the observatory will see obscuration of the
side of the primary that is on the same side of the offset.
This wide range is due to the large (stellar) size of the occulter.
Because $k_{\parallel} > R_1$, the situation in Fig. \ref{EarthLim4}
cannot hold. What about a person standing on Earth observing
Kepler-47A, Kepler-47B and Kepler-47ABb when they are in syzygy?
If the Earth would be coplanar with this syzygy, then
$k_{\forall}$~$=$~$929$~km and $k_{\sqcup}$~$=$~$434$~km.

\section{Summary}

I have derived user-friendly algebraic relations and criteria 
for eclipses, transits and occultations, having been inspired by the foundational 
geometry of {\rev \cite{verbre2017}}. These relations may be useful for a variety of purposes for transit
identification, in the planning and
analysis of observations, and for an understanding of the shadowing conditions on 
the surfaces and atmospheres of planets and moons; Figs. \ref{bubbles1} and \ref{bubbles2} contain 
flowcharts for quick use. Because nearly all 
formulae are independent of perspective, 
they may be applied to Earth-based
observatories, Solar system-based observatories (on land and in space), and 
observatories within extrasolar systems.

My focus was on three-body systems which include one star, one planet, and
one other body which could be a star, planet or moon. The only major assumptions
I made about these bodies were that they are all spheres, that light rays do not bend,
and for orbital motion cases, that the orbits are fixed. I considered both
snapshots of these systems (Sections 3-7) and their motion
(Section 8, and Appendices A and B) in the context of umbral, antumbral and penumbral (Section 9)
shadows. I also illustrated how some of these results may be extended to four bodies,
with an observer situated on the fourth external body, such as the Earth (Section 10).

\phantom{This text will be invisible}

For snapshots in time, specific results include 

\phantom{This text will be invisible}

{\bf (i)} A criterion to determine if a given system architecture is in or out of transit (equation \ref{trancrit}), 

{\bf (ii)} If in transit, whether the target will be fully engulfed in the shadow (equation \ref{engulfed}),

{\bf (iii)} Whether a specific location on the target is in transit (equation \ref{trancritloc}), 

{\bf (iv)} Whether or not the shadow is umbral or antumbral (equation \ref{newantot});
a penumbral shadow will always accompany both, 

{\bf (v)} The size of the umbral shadow (equation \ref{Lumbfinal}) or antumbral shadow 
(equation \ref{Lantfinal}),

{\bf (vi)} The minimum shadow size (equations \ref{minshaumb}-\ref{minshaant}) and 
maximum shadow size (equations \ref{maxshaumb}-\ref{maxshaant}), and

{\rev {\bf (vii)} A transformation (from equation \ref{deltaeq} to equation \ref{Deltaequ}) which enables
one to obtain snapshot results for penumbral cones without having to perform derivations similar to
those carried out for the umbral and antumbral cones.}

\phantom{This text will be invisible}

\noindent{}Of more mathematical interest is 

\phantom{This text will be invisible}

{\bf (viii)} That the shadow is always a parabolic cylinder (Section 4), 

{\bf (ix)} The equation of the intersection of the radiation cone and spherical 
target in Cartesian coordinates (equations \ref{inteq}-\ref{TheKint}) and 
in radial coordinates (equation \ref{radialeq}), and

{\bf (x)} The transformed version of this equation in standard form (equation \ref{stanform}).

\phantom{This text will be invisible}

For time evolution, specific results include 

\phantom{This text will be invisible}

{\bf (i)} Start times, end times, durations and frequencies
of transits (including ingresses and egresses) for two planets
on circular coplanar orbits around a star
(equations \ref{sta1S2P}-\ref{fre1S2P}), 

{\bf (ii)} The angles $\psi$ which allow
for a single implicit computation of start times, end times, durations and frequencies of transits
for circular arbitrarily inclined orbits (equations \ref{incpsi1st}-\ref{incpsilast}), and 

{\bf (iii)} The angles $\psi$ which allow for implicit computation of transit properties for arbitrarily eccentric, coplanar 
orbits (equations \ref{phisimp}-\ref{psimoonPO}).

\phantom{This text will be invisible}

For the special extended case of an observer of an exo-syzygy, specific results include

\phantom{This text will be invisible}

{\bf (i)} The critical distance beyond which the target, rather than the occulter, blocks
the primary's starlight (equation \ref{ruplus}),

{\bf (ii)} The maximum allowable transverse distance of the observer (perpendicular to the syzygy;
equations \ref{kall0}-\ref{kall}),

{\bf (iii)} The angular diameters in the sky of the shadows on the primary (equations \ref{etasha} 
and \ref{eta14arbk}), and

{\bf (iv)} The transverse distance of an observer who is able to see a double annular eclipse
(equations \ref{l14eq}-\ref{pythag}).

\section*{Acknowledgements}

{\rev I thank the referee for their careful reading of the manuscript and their thoughtful and astute comments, which have led to clear 
improvements.} I gratefully acknowledge the 
support of the STFC via an Ernest Rutherford Fellowship (grant ST/P003850/1).

\appendix

\section{Cartesian elements for time evolutions}

In this appendix I provide explicit expressions for Cartesian elements in terms
of orbital elements and time for all of the architectures 
seen in Figs. \ref{bubbles1} and \ref{bubbles2}. The goal in every case is to derive
expressions in terms of $x_{12}, y_{12}, z_{12}, x_{13}, y_{13},$
and $z_{13}$ in terms of the masses and radii of the three bodies, and
the orbital parameters from Table \ref{Tabassum}. These computations
are performed through equations (\ref{r23}-\ref{zr}).
The following equations represent useful references to expedite determinations
of transit characteristics.

\subsection{Arbitrarily eccentric, coplanar orbits}

I can impose coplanarity by assuming
$i=w=\Omega=0$ for each {\rev orbit}. If a user wishes the plane of motion
to not coincide with the $z$ plane, then they can impose the
appropriate rotation on the below expressions.

\subsubsection{One star and two planets (1S2P)}


\begin{equation}
x_{12,\ {\rm 1S2P}}^{\rm (e)} = 
\frac
{a_{12} \left(1-e_{12}^2\right)}
{1 + e_{12} \cos{\Pi_{12}(t)}}
\cos{\left[\mathfrak{n}_{12} \left(t - \tau_{12}\right)  \right]}
,
\label{x12twoplan}
\end{equation}

\begin{equation}
y_{12,\ {\rm 1S2P}}^{\rm (e)} = 
\frac
{a_{12} \left(1-e_{12}^2\right)}
{1 + e_{12} \cos{\Pi_{12}(t)}}
\sin{\left[\mathfrak{n}_{12} \left(t - \tau_{12}\right)  \right]}
,
\end{equation}

\begin{equation}
z_{12,\ {\rm 1S2P}}^{\rm (e)} = 0
,
\label{z12twoplan}
\end{equation}

\begin{equation}
x_{13,\ {\rm 1S2P}}^{\rm (e)} = 
\frac
{a_{13} \left(1-e_{13}^2\right)}
{1 + e_{13} \cos{\Pi_{13}(t)}}
\cos{\left[\mathfrak{n}_{13} \left(t - \tau_{13}\right)  \right]}
,
\end{equation}

\begin{equation}
y_{13,\ {\rm 1S2P}}^{\rm (e)} = 
\frac
{a_{13} \left(1-e_{13}^2\right)}
{1 + e_{13} \cos{\Pi_{13}(t)}}
\sin{\left[\mathfrak{n}_{13} \left(t - \tau_{13}\right)  \right]}
,
\end{equation}

\begin{equation}
z_{13,\ {\rm 1S2P}}^{\rm (e)} = 0
.
\end{equation}

\subsubsection{Two stars and one planet (2S1P)}

\begin{equation}
x_{12,\ {\rm 2S1P}}^{\rm (e)} = x_{12,\ {\rm 1S2P}}^{\rm (e)}
,
\end{equation}

\begin{equation}
y_{12,\ {\rm 2S1P}}^{\rm (e)} = y_{12,\ {\rm 1S2P}}^{\rm (e)}
,
\end{equation}

\begin{equation}
z_{12,\ {\rm 2S1P}}^{\rm (e)} = 0
,
\end{equation}

\[
x_{13,\ {\rm 2S1P}}^{\rm (e)} = 
\frac
{a_{123} \left(1-e_{123}^2\right)}
{1 + e_{123} \cos{\Pi_{123}(t)}}
\cos{\left[\mathfrak{n}_{123} \left(t - \tau_{123}\right)  \right]}
\]

\[
\ \ \ \ \ \ +
\left(\frac{M_1}{M_1 + M_2} \right) x_{12, \ {\rm 1S2P} }^{\rm (e)}
,
\]

\begin{equation}
\end{equation}

\[
y_{13,\ {\rm 2S1P}}^{\rm (e)} = 
\frac
{a_{123} \left(1-e_{123}^2\right)}
{1 + e_{123} \cos{\Pi_{123}(t)}}
\sin{\left[\mathfrak{n}_{123} \left(t - \tau_{123}\right)  \right]}
\]

\[
\ \ \ \ \ \ +
\left(\frac{M_1}{M_1 + M_2} \right) y_{12, \ {\rm 1S2P} }^{\rm (e)}
,
\]

\begin{equation}
\end{equation}

\begin{equation}
z_{13,\ {\rm 2S1P}}^{\rm (e)} = 0
.
\end{equation}

\subsubsection{One moon, with moon occulter (1M\textendash MO)}

\[
x_{12,\ {\rm 1M-MO}}^{\rm (e)} = 
x_{13, \ {\rm 1S2P} }^{\rm (e)}
\]

\begin{equation}
\ \ \ \ \ \ \ \ \ \ -
\frac
{a_{23} \left(1-e_{23}^2\right)}
{1 + e_{23} \cos{\Pi_{23}(t)}}
\cos{\left[\mathfrak{n}_{23} \left(t - \tau_{23}\right)  \right]}
,
\end{equation}

\[
y_{12,\ {\rm 1M-MO}}^{\rm (e)} = 
y_{13, \ {\rm 1S2P} }^{\rm (e)}
\]

\begin{equation}
\ \ \ \ \ \ \ \ \ \ -
\frac
{a_{23} \left(1-e_{23}^2\right)}
{1 + e_{23} \cos{\Pi_{23}(t)}}
\sin{\left[\mathfrak{n}_{23} \left(t - \tau_{23}\right)  \right]}
,
\end{equation}

\begin{equation}
z_{12,\ {\rm 1M-MO}}^{\rm (e)} = 0
,
\end{equation}

\begin{equation}
x_{13,\ {\rm 1M-MO}}^{\rm (e)} = x_{13,\ {\rm 1S2P}}^{\rm (e)}
,
\end{equation}

\begin{equation}
y_{13,\ {\rm 1M-MO}}^{\rm (e)} = y_{13,\ {\rm 1S2P}}^{\rm (e)}
,
\end{equation}

\begin{equation}
z_{13,\ {\rm 1M-MO}}^{\rm (e)} = 0
.
\end{equation}

\subsubsection{One moon, with planet occulter (1M\textendash PO)}

\begin{equation}
x_{12,\ {\rm 1M-PO}}^{\rm (e)} = x_{12,\ {\rm 1S2P}}^{\rm (e)}
,
\end{equation}

\begin{equation}
y_{12,\ {\rm 1M-PO}}^{\rm (e)} = y_{12,\ {\rm 1S2P}}^{\rm (e)}
,
\end{equation}

\begin{equation}
z_{12,\ {\rm 1M-PO}}^{\rm (e)} = 0
,
\end{equation}

\[
x_{13,\ {\rm 1M-PO}}^{\rm (e)} = 
x_{12, \ {\rm 1S2P} }^{\rm (e)}
\]

\begin{equation}
\ \ \ \ \ \ \ \ \ \ +
\frac
{a_{23} \left(1-e_{23}^2\right)}
{1 + e_{23} \cos{\Pi_{23}(t)}}
\cos{\left[\mathfrak{n}_{23} \left(t - \tau_{23}\right)  \right]}
,
\end{equation}

\[
y_{13,\ {\rm 1M-PO}}^{\rm (e)} = 
y_{12, \ {\rm 1S2P} }^{\rm (e)}
\]

\begin{equation}
\ \ \ \ \ \ \ \ \ \ +
\frac
{a_{23} \left(1-e_{23}^2\right)}
{1 + e_{23} \cos{\Pi_{23}(t)}}
\sin{\left[\mathfrak{n}_{23} \left(t - \tau_{23}\right)  \right]}
,
\end{equation}

\begin{equation}
z_{13,\ {\rm 1M-PO}}^{\rm (e)} = 0
.
\end{equation}

\subsection{Circular, arbitrarily inclined orbits}

Circular orbits imply $r(t)=a$, and hence eliminates a time
dependence that is present with eccentric orbits. Instead the
time dependence arises within the orbital angles
as follows:

\subsubsection{One star and two planets (1S2P)}

\begin{equation}
x_{12, \ {\rm 1S2P}}^{\rm (i)} = a_{12} \left[\mathcal{C}_{12} \cos{\Omega_{12}}  
                       - \mathcal{S}_{12} \cos{i_{12}} \sin{\Omega_{12}}  \right]
,
\end{equation}

\begin{equation}
y_{12, \ {\rm 1S2P}}^{\rm (i)} = a_{12} \left[\mathcal{C}_{12} \sin{\Omega_{12}}  
                       + \mathcal{S}_{12} \cos{i_{12}} \cos{\Omega_{12}}  \right]
,
\end{equation}

\begin{equation}
z_{12, \ {\rm 1S2P}}^{\rm (i)} = a_{12} \mathcal{S}_{12} \sin{i_{12}}
,
\end{equation}

\begin{equation}
x_{13, \ {\rm 1S2P}}^{\rm (i)} = a_{13} \left[\mathcal{C}_{13} \cos{\Omega_{13}}  
                       - \mathcal{S}_{13} \cos{i_{13}} \sin{\Omega_{13}}  \right]
,
\end{equation}

\begin{equation}
y_{13, \ {\rm 1S2P}}^{\rm (i)} = a_{13} \left[\mathcal{C}_{13} \sin{\Omega_{13}}  
                       + \mathcal{S}_{13} \cos{i_{13}} \cos{\Omega_{13}}  \right]
,
\end{equation}

\begin{equation}
z_{13, \ {\rm 1S2P}}^{\rm (i)} = a_{13} \mathcal{S}_{13} \sin{i_{13}}
.
\end{equation}

\[
\]

\subsubsection{Two stars and one planet (2S1P)}

\[
\]

\begin{equation}
x_{12, \ {\rm 2S1P}}^{\rm (i)} = x_{12, \ {\rm 1S2P}}^{\rm (i)}
,
\end{equation}

\begin{equation}
y_{12, \ {\rm 2S1P}}^{\rm (i)} = y_{12, \ {\rm 1S2P}}^{\rm (i)}
,
\end{equation}

\begin{equation}
z_{12, \ {\rm 2S1P}}^{\rm (i)} = z_{12, \ {\rm 1S2P}}^{\rm (i)}
,
\end{equation}

\[
x_{13, \ {\rm 2S1P}}^{\rm (i)} = a_{123} 
\left[ 
\mathcal{C}_{123} \cos{\Omega_{123}} 
-
\mathcal{S}_{123} \cos{i_{123}} \sin{\Omega_{123}}
\right]
\]

\begin{equation}
\ \ \ \
+
\left(\frac{M_1}{M_1 + M_2} \right) x_{12, \ {\rm 1S2P}}^{\rm (i)}
,
\end{equation}

\[
y_{13, \ {\rm 2S1P}}^{\rm (i)} = a_{123} 
\left[ 
\mathcal{C}_{123} \sin{\Omega_{123}} 
+
\mathcal{S}_{123} \cos{i_{123}} \cos{\Omega_{123}}
\right]
\]

\begin{equation}
\ \ \ \
+
\left(\frac{M_1}{M_1 + M_2} \right) y_{12, \ {\rm 1S2P}}^{\rm (i)}
,
\end{equation}

\begin{equation}
z_{13, \ {\rm 2S1P}}^{\rm (i)} = a_{123} \mathcal{S}_{123} \sin{i_{123}} 
                            +
    \left(\frac{M_1}{M_1 + M_2} \right) z_{12, \ {\rm 1S2P}}^{\rm (i)}
.
\end{equation}

\[
\]

\subsubsection{One moon, with moon occulter  (1M\textendash MO)}

\[
\]

\[
x_{12, \ {\rm 1M-MO}}^{\rm (i)} = x_{13, \ {\rm 1S2P}}^{\rm (i)} 
\]

\begin{equation}
\ \ \ \ - a_{23} \left[\mathcal{C}_{23} \cos{\Omega_{23}} - \mathcal{S}_{23} \cos{i_{23}} \sin{\Omega_{23}}  \right]
,
\end{equation}

\[
y_{12, \ {\rm 1M-MO}}^{\rm (i)} = y_{13, \ {\rm 1S2P}}^{\rm (i)} 
\]

\begin{equation}
\ \ \ \ - a_{23} \left[\mathcal{C}_{23} \sin{\Omega_{23}} + \mathcal{S}_{23} \cos{i_{23}} \cos{\Omega_{23}}  \right]
,
\end{equation}

\begin{equation}
z_{12, \ {\rm 1M-MO}}^{\rm (i)} = z_{13, \ {\rm 1S2P}}^{\rm (i)} - a_{23} \mathcal{S}_{23} \sin{i_{23}}
,
\end{equation}

\begin{equation}
x_{13, \ {\rm 1M-MO}}^{\rm (i)} = x_{13, \ {\rm 1S2P}}^{\rm (i)}
,
\end{equation}

\begin{equation}
y_{13, \ {\rm 1M-MO}}^{\rm (i)} = y_{13, \ {\rm 1S2P}}^{\rm (i)}
,
\end{equation}

\begin{equation}
z_{13, \ {\rm 1M-MO}}^{\rm (i)} = z_{13, \ {\rm 1S2P}}^{\rm (i)}
.
\end{equation}

\subsubsection{One moon, with planet occulter (1M\textendash PO)}

\begin{equation}
x_{12, \ {\rm 1M-PO}}^{\rm (i)} = x_{12, \ {\rm 1S2P}}^{\rm (i)}
,
\end{equation}

\begin{equation}
y_{12, \ {\rm 1M-PO}}^{\rm (i)} = y_{12, \ {\rm 1S2P}}^{\rm (i)}
,
\end{equation}

\begin{equation}
z_{12, \ {\rm 1M-PO}}^{\rm (i)} = z_{12, \ {\rm 1S2P}}^{\rm (i)}
,
\end{equation}

\[
x_{13, \ {\rm 1M-PO}}^{\rm (i)} = x_{12, \ {\rm 1S2P}}^{\rm (i)} 
\]

\begin{equation}
\ \ \ \ + a_{23} \left[\mathcal{C}_{23} \cos{\Omega_{23}} - \mathcal{S}_{23} \cos{i_{23}} \sin{\Omega_{23}}  \right]
,
\end{equation}

\[
y_{13, \ {\rm 1M-PO}}^{\rm (i)} = y_{12, \ {\rm 1S2P}}^{\rm (i)} 
\]

\begin{equation}
\ \ \ \ + a_{23} \left[\mathcal{C}_{23} \sin{\Omega_{23}} + \mathcal{S}_{23} \cos{i_{23}} \cos{\Omega_{23}}  \right]
,
\end{equation}

\begin{equation}
z_{13, \ {\rm 1M-PO}}^{\rm (i)} = z_{12, \ {\rm 1S2P}}^{\rm (i)} + a_{23} \mathcal{S}_{23} \sin{i_{23}}
.
\end{equation}

\subsection{Circular, coplanar orbits}

In the simplest case, the equations are the most compact {\rev of the three example geometries considered}.

\subsubsection{One star and two planets (1S2P)}

\begin{equation}
x_{12, \ {\rm 1S2P}}^{\rm (cc)} = a_{12} \cos{\left[\mathfrak{n}_{12} \left(t - \tau_{12} \right) \right]} 
,
\end{equation}

\begin{equation}
y_{12, \ {\rm 1S2P}}^{\rm (cc)} = a_{12} \sin{\left[\mathfrak{n}_{12} \left(t - \tau_{12} \right) \right]} 
,
\end{equation}

\begin{equation}
z_{12, \ {\rm 1S2P}}^{\rm (cc)} = 0
,
\end{equation}

\begin{equation}
x_{13, \ {\rm 1S2P}}^{\rm (cc)} = a_{13} \cos{\left[\mathfrak{n}_{13} \left(t - \tau_{13} \right) \right]} 
,
\end{equation}

\begin{equation}
y_{13, \ {\rm 1S2P}}^{\rm (cc)} = a_{13} \sin{\left[\mathfrak{n}_{13} \left(t - \tau_{13} \right) \right]} 
,
\end{equation}

\begin{equation}
z_{13, \ {\rm 1S2P}}^{\rm (cc)} = 0
.
\end{equation}

\subsubsection{Two stars and one planet (2S1P)}

\begin{equation}
x_{12, \ {\rm 2S1P}}^{\rm (cc)} = x_{12, \ {\rm 1S2P}}^{\rm (cc)}
,
\end{equation}

\begin{equation}
y_{12, \ {\rm 2S1P}}^{\rm (cc)} = y_{12, \ {\rm 1S2P}}^{\rm (cc)}
,
\end{equation}

\begin{equation}
z_{12, \ {\rm 2S1P}}^{\rm (cc)} = 0
,
\end{equation}

\[
x_{13, \ {\rm 2S1P}}^{\rm (cc)} = a_{123} \cos{\left[\mathfrak{n}_{123} \left(t - \tau_{123}\right)\right]}
\]

\[
\ \ \ \ \ \ \
+ \left(\frac{M_1}{M_1 + M_2} \right) x_{12, \ {\rm 1S2P}}^{\rm (cc)}
,
\]

\begin{equation}
\end{equation}

\[
y_{13, \ {\rm 2S1P}}^{\rm (cc)} = a_{123} \sin{\left[\mathfrak{n}_{123} \left(t - \tau_{123}\right)\right]}
\]
                            
\[
\ \ \ \ \ \ \
+ \left(\frac{M_1}{M_1 + M_2} \right) y_{12, \ {\rm 1S2P}}^{\rm (cc)}
,
\]

\begin{equation}
\end{equation}

\begin{equation}
z_{13, \ {\rm 2S1P}}^{\rm (cc)} = 0
.
\end{equation}

\subsubsection{One moon, with moon occulter (1M\textendash MO) }

\begin{equation}
x_{12, \ {\rm 1M-MO}}^{\rm (cc)} = x_{13, \ {\rm 1S2P}}^{\rm (cc)}
                              - a_{23} \cos{\left[\mathfrak{n}_{23} \left(t - \tau_{23}\right)\right]}
,
\end{equation}

\begin{equation}
y_{12, \ {\rm 1M-MO}}^{\rm (cc)} = y_{13, \ {\rm 1S2P}}^{\rm (cc)}
                              - a_{23} \sin{\left[\mathfrak{n}_{23} \left(t - \tau_{23}\right)\right]}
,
\end{equation}

\begin{equation}
z_{12, \ {\rm 1M-MO}}^{\rm (cc)} = 0
,
\end{equation}

\begin{equation}
x_{13, \ {\rm 1M-MO}}^{\rm (cc)} = x_{13, \ {\rm 1S2P}}^{\rm (cc)}
,
\end{equation}

\begin{equation}
y_{13, \ {\rm 1M-MO}}^{\rm (cc)} = y_{13, \ {\rm 1S2P}}^{\rm (cc)}
,
\end{equation}

\begin{equation}
z_{13, \ {\rm 1M-MO}}^{\rm (cc)} = 0
.
\end{equation}

\subsubsection{One moon, with planet occulter  (1M\textendash PO)}

\begin{equation}
x_{12, \ {\rm 1M-PO}}^{\rm (cc)} = x_{12, \ {\rm 1S2P}}^{\rm (cc)}
,
\end{equation}

\begin{equation}
y_{12, \ {\rm 1M-PO}}^{\rm (cc)} = y_{12, \ {\rm 1S2P}}^{\rm (cc)}
,
\end{equation}

\begin{equation}
z_{12, \ {\rm 1M-PO}}^{\rm (cc)} = 0
,
\end{equation}

\begin{equation}
x_{13, \ {\rm 1M-PO}}^{\rm (cc)} = x_{12, \ {\rm 1S2P}}^{\rm (cc)} + a_{23} \cos{\left[\mathfrak{n}_{23} \left(t - \tau_{23}\right)\right]}
,
\end{equation}

\begin{equation}
y_{13, \ {\rm 1M-PO}}^{\rm (cc)} = y_{12, \ {\rm 1S2P}}^{\rm (cc)} + a_{23} \sin{\left[\mathfrak{n}_{23} \left(t - \tau_{23}\right)\right]}
,
\end{equation}

\begin{equation}
z_{13, \ {\rm 1M-PO}}^{\rm (cc)} = 0
.
\end{equation}

\section{Goodness of static orbit approximation}

{\rev

When deriving relations for orbital motion in Section 8, I assumed that the bodies
move along static orbits. In reality, mutual three-body gravitational effects perturb these orbits, as well
as altering the speed of the bodies which traverse these orbits. Hence, the applicability
of my relations for motion depend on (i) the architecture considered, (ii) the accuracy sought, and (iii)
the timescale over which the results are to be generated.
 
In this appendix, I provide some quantitative context for {\revrev this applicability}. I compute the extent of the departure between the static and perturbed cases for just a few representative architectures. I do so by performing multi-body numerical simulations with point-mass bodies. Their outcomes generate $(x,y,z)$ locations of an object at a series of times, {\revrev and} I calculate the distance from these locations to the locations analytically obtained from the formulae in Section 8 and Appendix A at those same times. This time-dependent distance is henceforth denoted as {\it deviation}. The maximum deviation would then represent approximately the longest axis of the initial orbit.

For my numerical simulations, I used the Bulirsch-Stoer integrator from the {\it Mercury} integration package \citep{chambers1999} with an accuracy tolerance of $10^{-12}$. This integrator utilises a variable timestep and so is adaptable to all of the architectures considered here, but does suffer from floating point round-off error. Throughout I adopted $1M_{\odot}$ stars and set the output time interval to be at most a few per cent of the orbital period of the smallest orbit, but not small enough to resolve variations during an individual transit. I ran the simulations for no longer than a few human lifetimes, but sufficiently long enough to sample tens, hundreds or thousands of transits depending on architecture. I started with the orbital elements in Table \ref{Tabassum} and subsequently converted them into Cartesian elements for input into the code.

For consistency, the numerical constants which I inserted into the analytics needed to be equivalent to those in the code. Hence, I adopted {\it Mercury's} now slightly-outdated value for the astronomical unit ($149597870000$~m), its value for Solar mass ($1.9891 \times 10^{30}$ kg), its definition of year ($365.25$ days) and its value for $\mathfrak{G}$ -- obtained through the code representation of {\tt K2} -- of $6.67198422296 \times 10^{-11}$ N$\cdot$kg$^{-2}$$\cdot$m$^2$.

\subsection{The 1S2P case}

Two planets orbiting one star may be subject to a variety of secular and mean motion resonances. These resonances are encountered at particular planet separations and orbital angles. Generally, however, the larger the initial separation, the smaller the magnitude of the perturbation, particularly on timescales of tens or hundreds of orbits.

I present deviations for a variety of two-planet, one-star, circular, coplanar (cc) case instances in Fig. \ref{1S2Pnbody}. The curves in the figure are not solid lines, but rather contain oscillations at different scales and so appear fuzzy. 

In all instances, the inner planet was initially located at 1 au and the two planets were of equal mass. The {\revrev gray} curve (third from top) represents a type of fiducial case for close orbits, with $M_2 = M_3 = 10^{-6}M_1$ and $a_{13}/a_{12} = 2$ (initially). The red curves then keep this same initial semimajor axis ratio but sample different masses: $M_2 = M_3 = 10^{-4}M_1$ (top curve), $M_2 = M_3 = 10^{-8}M_1$ (third from bottom curve), and $M_2 = M_3 = 10^{-10}M_1$ (bottom curve). The blue curves keep $M_2 = M_3 = 10^{-6}M_1$ but vary the initial mutual separations as $a_{13}/a_{12} = 1.5$ (second from top curve), $3$ (fourth from top curve), $5$ (fourth from bottom curve) and $10$ (second from bottom curve).

These curves do not sample the entire range of known separations. For example, the semimajor axis ratio of Neptune and Mercury is about 78. On the lower end, semimajor axis ratios much smaller than 1.5 are subject to dynamical instabilities depending on the planet masses and other orbital parameters.

\begin{figure}
\includegraphics[width=8.5cm, height=6.5cm]{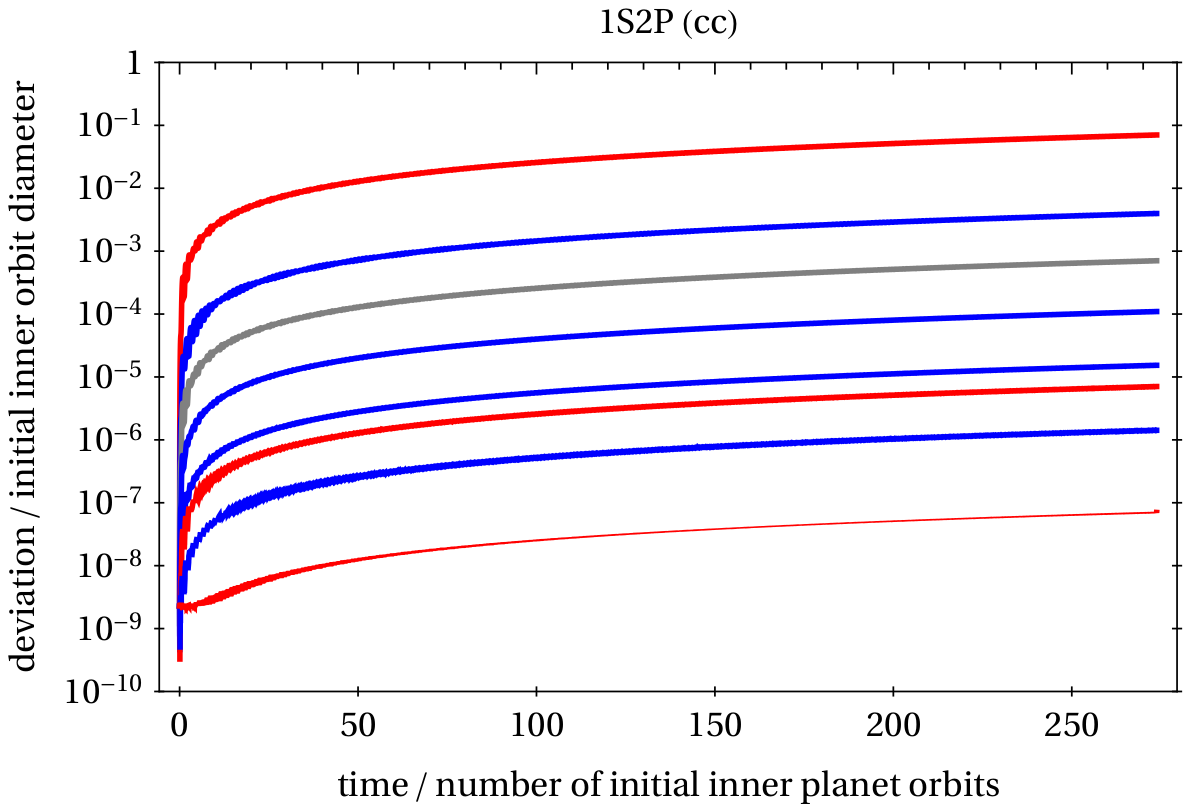}
\caption{
{\rev
For the 1S2P circular, coplanar (cc) case, shown is the separation between the location of an inner planet within a gravitational three-body simulation and the location of an inner planet whose time evolution is dictated by the analytical formulae from Section 8 and Appendix A. Here, $M_1 = M_{\odot}$ and $a_{12}=1$ au. The blue and {\revrev gray} curves all share $M_2 = M_3 = 10^{-6}M_1$, with, from top to bottom, initial $a_{13}/a_{12} = 1.5, 2.0, 3.0, 5.0$ and $10.0$.  The red and {\revrev gray} curves all share $a_{13}/a_{12} = 2.0$, with, from top to bottom,  $M_2/M_1 = M_3/M_1 = 10^{-4}, 10^{-6}, 10^{-8}$ and $10^{-10}$.
}
}
\label{1S2Pnbody}
\end{figure}

\subsection{The 2S1P case}

\begin{figure}
\centering
\includegraphics[width=8.5cm, height=6.5cm]{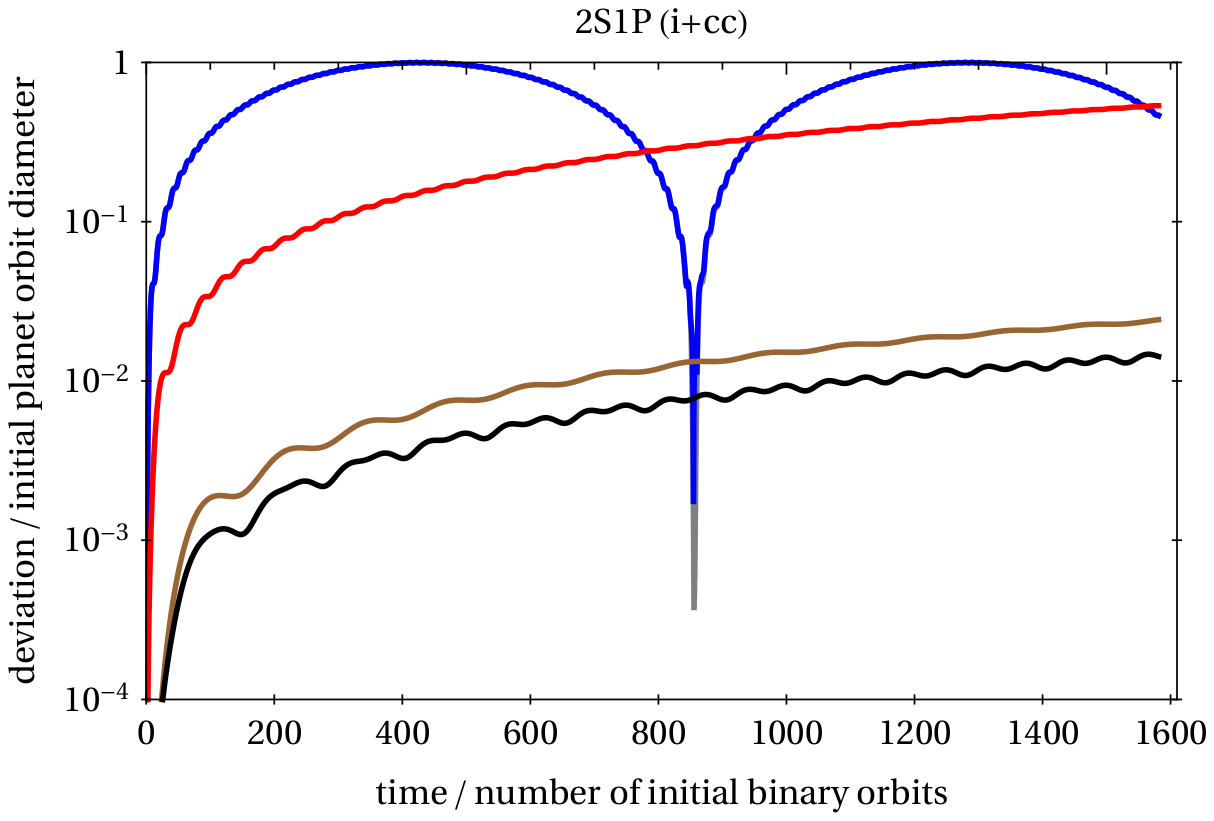}
\phantom{}
\includegraphics[width=8.5cm, height=6.5cm]{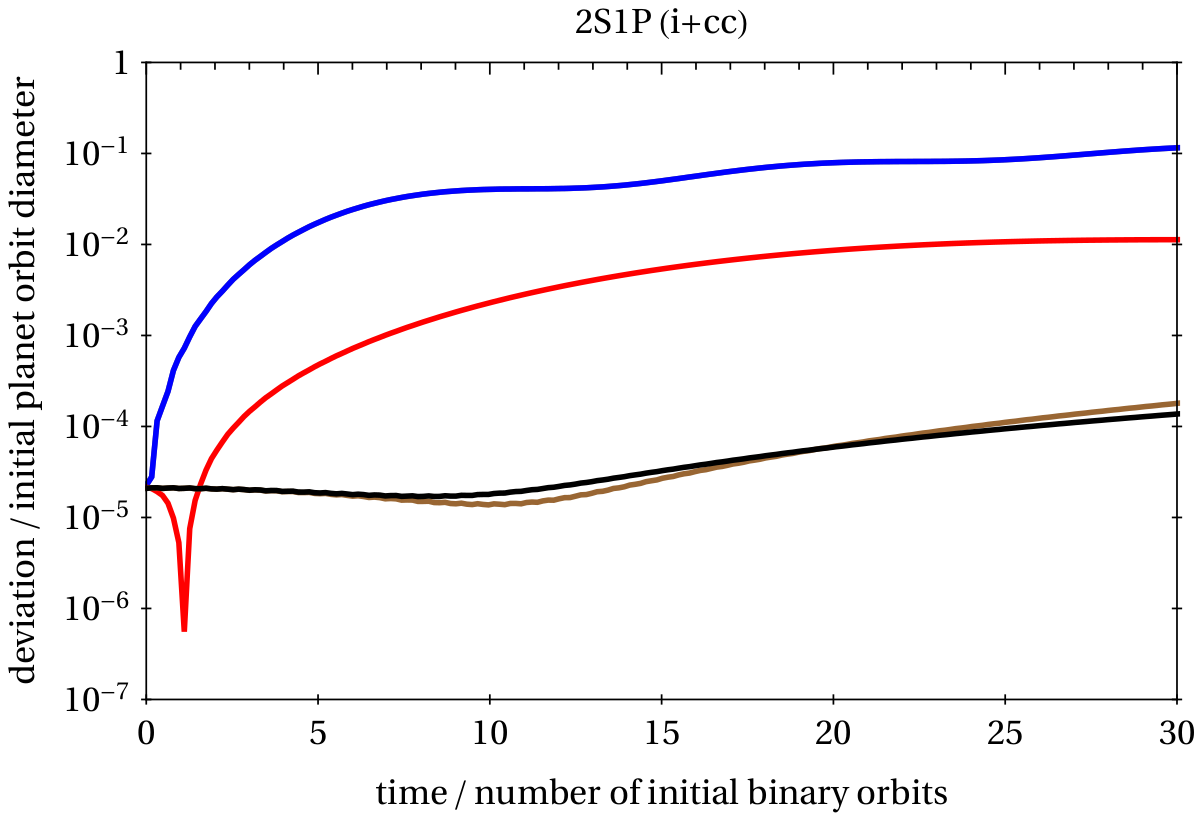}
\caption{
Deviations like those in Fig. \ref{1S2Pnbody}, except for the 2S1P case.
{\revrev The circular, coplanar case is given by the gray curves (which are mostly
under the blue curves), and the circular,
inclined case is given by the other curves.} 
The bottom plot is a zoom-in of the top plot for the first 30 binary orbits.
Here, $M_1 = M_2 = M_{\odot}$, $M_3 = 10^{-6} M_{1}$, and $a_{12} = 0.2$ au, {\revrev $i_{12} = 0^{\circ}$}. 
The {\revrev five} curves from top to bottom correspond 
to ($a_{123} = 1$ au, $i_{123} = 0^{\circ}$; gray),  ($a_{123} = 1$ au, $i_{123} = 1^{\circ}$; blue), ($a_{123} = 2$ au, $i_{123} = 1^{\circ}$; red), ($a_{123} = 5$ au, $i_{123} = 1^{\circ}$; brown), and ($a_{123} = 5$ au, $i_{123} = 50^{\circ}$; black).
}
\label{2S1Pnbody}
\end{figure}

2S1P architectures are more prone to the effects of precession than in the 1S2P case. {\revrev For eccentric and/or inclined orbits,} both stars can quickly precess the outer orbit's argument of pericentre and quickly regress the outer orbit's longitude of ascending node. {\revrev Even in the circular, coplanar case, there will be short-term variations of the osculating elements, and in a more pronounced way than in the 1S2P circular coplanar case. Quick orbital changes explain why circumbinary exoplanets like} Kepler-413 ABb disappeared and re-appeared during the original {\it Kepler} mission \citep{kosetal2014}.

Fig. \ref{2S1Pnbody} presents {\revrev five} curves of deviations, all of which feature a planet mass of $M_3 = 10^{-6} M_2 = 10^{-6} M_1$ and an initial inner binary semimajor axis of $a_{12} = 0.2$ au. Both orbits were initially circular, and in all cases $i_{12} = 0^{\circ}$, $\Omega_{12} = \Omega_{123} = 0^{\circ}$. The outer orbit initial semimajor axis and inclination for the curves, from top to bottom, were respectively, ($a_{123} = 1$ au, $i_{123} = 0^{\circ}$ for gray), ($a_{123} = 1$ au, $i_{123} = 1^{\circ}$ for blue), ($a_{123} = 2$ au, $i_{123} = 1^{\circ}$ for red), ($a_{123} = 5$ au, $i_{123} = 1^{\circ}$ for brown), and ($a_{123} = 5$ au, $i_{123} = 50^{\circ}$ for black). The bottom plot is a zoom-in of the top plot for the first 30 binary eclipses. {\revrev The blue and gray curves are nearly coincident, except for the trough or kink in the top plot.}

The initially 1 au planet achieved a maximum deviation after about 400 initial binary orbits, and then repeated this behaviour in a cyclical pattern. Alternatively, the initially 2 au planet did not reach maximum deviation until after 1600 initial binary orbits, demonstrating a strong dependence of deviation on $a_{123}$. In contrast, the dependence on inclination is relatively weak, as shown by the {\revrev blue, brown, gray and black curves. The differences in deviation are easily discernable between the $i_{123} = 1^{\circ}$ and $i_{123} = 50^{\circ}$ cases, at least after several tens of binary orbits. The difference in deviation between the $i_{123} = 0^{\circ}$ and $i_{123} = 1^{\circ}$ cases, however, is discernable only at the kink. This kink represents the location where the orbit has precessed through a complete revolution, such that the deviation returns close to zero for a relatively short time. The deviation stretches closer to zero in the coplanar case (gray curve) than in the inclined case (blue curve). The inclination differences ultimately indicate that the orbital changes cannot be dominated by nodal precession.}

These dependencies have been well-studied analytically in the limiting cases of $M_3 = 0$ and for averaged, or secular, orbits. In these limits, perturbation theory can be used to approximate the time evolution of $\Omega_{123}$ as a series (see Eq. 208 of \citealt*{veras2014}), the leading term of which is similar to leading term of the time evolution of $w_{123}$ (see Eq. 212 of \citealt*{veras2014}).  An alternate expression for this leading term, expressed as a precession timescale, can be found in Eq. 15 of \cite{martin2017a}, having originated from the studies of \cite{schneider1994}, \cite{farlas2010} and \cite{dooblu2011}.

The availability of analytic approximations for secular precession timescales, combined with the periodic behaviour of the curves in Fig. \ref{2S1Pnbody} (apparent in the top two curves of the top plot only), potentially allows one to provide a correction term to the analytical treatment in this paper. Doing so may reduce deviations, particularly for the planets with the smallest orbits.

\subsection{The 1M-MO case}

A moon with a planet host which is close to its parent star might be subject to similarly 
fast orbit precession (see Eq. 3 of \citealt*{martin2017b}, and \citealt*{mardling2010}),
leading to large deviations on short timescales.  In order to explore this possibility within
the context of my 1M-MO case, I considered a (giant) planet with mass $M_3 = 10^{-3} M_{\odot}$
which hosts a moon of mass $M_2 = 10^{-8} M_{\odot}$ on an inclined circular orbit with
respect to $i_{23} = 0^{\circ}$. I chose
a large initial satellite distance of $a_{23} = 10^6$~km in order to test an extreme-case scenario of least
accuracy. This distance still however
lies within the Hill sphere of all sampled semimajor axes (described below), meaning that the moon
orbits the planet and not the star. On the other extreme, a moon like Phobos
resides just 6000 km from Mars, with resulting deviations which may be significantly lower
(see Eq. 3 of \citealt*{martin2017b}).

Figure \ref{1MMOnbody} illustrates the results with three pairs of curves. The three pairs from top
to bottom correspond to initial values of $a_{13} = 0.5, 1.0$ and $5.0$ au. The blue curves illustrate the case
of an initial value of $i_{13} = 1^{\circ}$ whereas the red curves showcase an initial value of $i_{13} = 50^{\circ}$. The 
deviations again illustrate a weak dependence on inclination and a stronger dependence on $a_{13}$.

\subsection{Regions of validity}

The plots in this Appendix suggest that the static orbit approximation cannot be used on timescales at or
beyond which a curve reaches unity, unless correction terms are applied. 
Consider the top (blue) curve in Fig. \ref{2S1Pnbody},
which {\revrev reaches unity} after about 400 binary orbits. {\revrev Although this case is representative
of most currently known circumbinary exoplanets, as more distant exoplanets are discovered orbiting two stars,
the fixed-orbit approximation will become more useful.}

Up until what time is the static orbit approximation then applicable? The answer depends on the user's
motivation, the accuracy they seek, the deviation of the other bodies in the system, and how all deviations are broken
down into $\left(x,y,z\right)$ components. For example, for those wishing to compute a single transit duration 
{\revrev -- which is a small fraction of one orbit--} the
resulting accuracy can vary by at least eight orders of magnitude depending on which of the small sample of architectures
considered here is chosen.

In general, however, some trends have emerged. The static orbit approximation
is better suited to the 1S2P case than the 2S1P and 1M-MO cases. Also, distant transits tend to reduce deviations. 
More precisely, large mutual separations of three bodies will improve the static orbit approximation. 
Overall, the static orbit approximation is better used to approximate transit durations and frequencies
rather than predict actual instances of eclipses for sufficiently
massive objects residing within about 10 au of their parent star or stars.

\begin{figure}
\includegraphics[width=8.5cm, height=6.5cm]{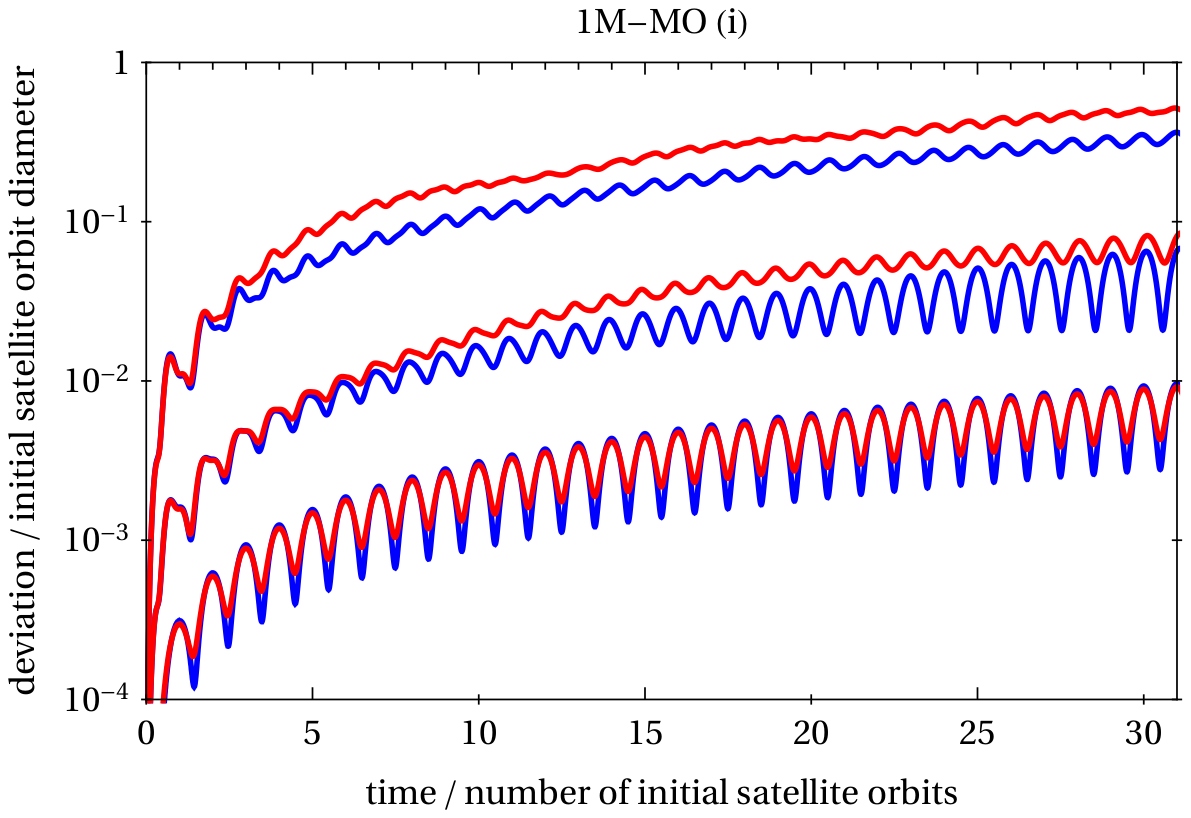}
\caption{
{\rev
Deviations like those in Figs. \ref{1S2Pnbody} and \ref{2S1Pnbody}, except for the 1M-MO inclined (i) case. 
Here, $M_1 = M_{\odot}, M_2 = 10^{-8} M_{\odot}$, $M_{3} = 10^{-3} M_{\odot}$, and 
$a_{23} = 10^6$ km, {\revrev $i_{23} = 0^{\circ}$}. The top, middle and bottom pairs of curves correspond to $a_{13} = 0.5, 1.0$ and
$5.0$ au. The blue and red curves respectively correspond to initial values of $i_{13} = 1^{\circ}$ and $50^{\circ}$.
}
}
\label{1MMOnbody}
\end{figure}

}

\label{lastpage}
\end{document}